\documentclass[journal]{IEEEtran}

\ifCLASSINFOpdf
\else
\fi

\usepackage[usenames,dvipsnames]{xcolor}
\definecolor{TUMBlue}{RGB}{0,101,189} %
\definecolor{TUMBlueDark}{RGB}{0,82,147} %
\definecolor{TUMBlueLight}{RGB}{152,198,234} %
\definecolor{TUMBlueMedium}{RGB}{100,160,200} %
\definecolor{TUMIvory}{RGB}{218,215,203} %
\definecolor{TUMGreen}{RGB}{162,173,0} %
\definecolor{TUMGray}{gray}{0.6} %
\definecolor{TUMOrange}{RGB}{227,114,34} %
\definecolor{TUMGreenDark}{RGB}{0,124,48} %
\definecolor{TUMRed}{RGB}{196,7,27} %
\definecolor{TUMPink}{RGB}{181,92,165}
\definecolor{TUMPinkDark}{RGB}{155,70,141}

\usepackage{amsmath,amssymb,amsfonts,tikz,pgfplots,bm}
\usepackage{cite,booktabs,tabularx}
\usepackage{algorithmic}
\usepackage{graphicx}
\usepackage{textcomp}
\usepackage{verbatim}
\usepackage{amsmath}
\usepackage[ruled,vlined,linesnumbered,titlenumbered]{algorithm2e}
\usepackage{nicefrac}
\usepackage[inline,shortlabels]{enumitem}
\usepackage{array}
\usepackage{balance}
\usepackage{multirow}
\usepackage{orcidlink}
\usepackage[capitalise]{cleveref}
\usepackage{float}
\usepackage{lineno}

\usepackage{soul}

\renewcommand{\baselinestretch}{1.0}

\newcommand{\cC}{{\cal C}}
\newcommand{\cD}{{\cal D}}
\newcommand{\cE}{{\cal E}}

\newcommand{\cO}{{\cal O}}

\newcommand{\cR}{{\cal R}}
\newcommand{\cS}{{\cal S}}
\newcommand{\cT}{{\cal T}}

\newcommand{\cW}{{\cal W}}

\newcommand{\bfd}{{\boldsymbol d}}
\newcommand{\bfe}{{\boldsymbol e}}

\newcommand{\bfp}{{\boldsymbol p}}
\newcommand{\bfq}{{\boldsymbol q}}

\newcommand{\bfu}{{\boldsymbol u}}
\newcommand{\bfv}{{\boldsymbol v}}
\newcommand{\bfw}{{\boldsymbol w}}
\newcommand{\bfx}{{\boldsymbol x}}
\newcommand{\bfy}{{\boldsymbol y}}
\newcommand{\bfz}{{\boldsymbol z}}

\newcommand{\bfB}{{\boldsymbol B}}
\newcommand{\bfC}{{\boldsymbol C}}
\newcommand{\bfD}{{\boldsymbol D}}

\newcommand{\bfF}{{\boldsymbol F}}

\newcommand{\bfH}{{\boldsymbol H}}

\newcommand{\bfP}{{\boldsymbol P}}

\newcommand{\bfT}{{\boldsymbol T}}

\newcommand{\bfX}{{\boldsymbol X}}
\newcommand{\bfY}{{\boldsymbol Y}}
\newcommand{\bfZ}{{\boldsymbol Z}}

\usetikzlibrary{plotmarks}
\usepgfplotslibrary{colormaps}
\usetikzlibrary{decorations.pathreplacing, calc, intersections, pgfplots.groupplots}
\usetikzlibrary{arrows, shapes, shapes.multipart, shapes.geometric, trees, positioning,
  backgrounds, fit, matrix}
\tikzstyle{block} = [rectangle, draw, text centered, rounded corners, minimum height=1.5em,fill=black!5!]

\newcommand{\lormid}{\scalebox{0.75}{$\mathrel{\raisebox{1.0pt}{$\mathbf{\mid}$}}$}}

\tikzstyle{round} = [circle, draw, text centered, minimum height=1.5em, fill=black!5!]

\usepackage{ifthen}
\newif\ifprintcomments
\printcommentstrue                  %

\newcommand{\floor}[1]{\lfloor{#1}\rfloor}

\DeclareMathOperator*{\argmax}{argmax} %

\newcommand{\dnaA}{\ensuremath{\mathsf{A}}}
\newcommand{\dnaC}{\ensuremath{\mathsf{C}}}
\newcommand{\dnaG}{\ensuremath{\mathsf{G}}}
\newcommand{\dnaT}{\ensuremath{\mathsf{T}}}
\newcommand{\alphDNA}{\ensuremath{\Sigma}_\mathrm{DNA}}

\newcommand{\evspac}{\ensuremath{\cE}}

\newcommand{\Ins}{{\tt{{Ins}}}}
\newcommand{\Del}{{\tt{{Del}}}}
\newcommand{\Sub}{{\tt{{Sub}}}}
\newcommand{\Tra}{{\tt{Tra}}}

\newcommand{\kmerChfo}{\ensuremath{\textsf{KMER}_k^\mathrm{f}}}
\newcommand{\kmerChba}{\ensuremath{\textsf{KMER}_k^\mathrm{b}}}
\newcommand{\kmerCh}{\ensuremath{\textsf{KMER}_k}}
\newcommand{\kmer}{\ensuremath{k\text{mer}}}
\newcommand{\datasetFo}{\ensuremath{\cD^\mathrm{f}}}
\newcommand{\datasetBa}{\ensuremath{\cD^\mathrm{b}}}

\newcommand{\accbc}{acc-BC}
\newcommand{\fastbc}{fast-BC}

\newcommand{\pI}{\ensuremath{p_\mathrm{I}}}
\newcommand{\pD}{\ensuremath{p_\mathrm{D}}}
\newcommand{\pS}{\ensuremath{p_\mathrm{S}}}

\newcommand{\multidraw}{sampling}
\newcommand{\Multidraw}{Sampling}
\newcommand{\chM}{\ensuremath{M}}
\newcommand{\chN}{\ensuremath{N}}
\newcommand{\chNfo}{\ensuremath{N^{\mathrm{f}}}}
\newcommand{\chNba}{\ensuremath{N^{\mathrm{b}}}}

\newcommand{\chL}{\ensuremath{L}}
\newcommand{\chLout}{\ensuremath{L'}}
\newcommand{\chc}{\ensuremath{c}}

\newcommand{\chX}{\ensuremath{\bfX}}
\newcommand{\chY}{\ensuremath{\bfY}}
\newcommand{\chYfo}{\ensuremath{\bfY^{\mathrm{f}}}}
\newcommand{\chYba}{\ensuremath{\bfY^{\mathrm{b}}}}

\newcommand{\chq}{\ensuremath{4}}
\newcommand{\drawdist}{\ensuremath{\bfP^\mathrm{d}}}

\newcommand{\trchM}{\ensuremath{A}}%
\newcommand{\trchY}{\ensuremath{\chY^\mathrm{t}}}

\newcommand{\Rtot}{\ensuremath{R}}
\newcommand{\field}{\ensuremath{\mathbb{F}}}
\newcommand{\primer}{\ensuremath{\bfp}}
\newcommand{\primL}{\ensuremath{\chL_\mathrm{p}}}

\newcommand{\Rtheo}{\ensuremath{R_\text{th}}}

\newcommand{\leno}{\ensuremath{n_{\mathsf{o}}}}
\newcommand{\dimo}{\ensuremath{k_{\mathsf{o}}}}
\newcommand{\Rout}{\ensuremath{R_\mathrm{o}}}

\newcommand{\scmem}{\ensuremath{m_\mathrm{sc}}}
\newcommand{\sccoup}{\ensuremath{o_\mathrm{sc}}}
\newcommand{\prot}{\ensuremath{\bfB}}
\newcommand{\protn}{\ensuremath{n_\mathsf{p}}}
\newcommand{\protr}{\ensuremath{r_\mathsf{p}}}
\newcommand{\protlift}{\ensuremath{Q_\mathsf{p}}}
\newcommand{\scprot}{\ensuremath{\prot_\mathrm{sc}}}
\newcommand{\bleno}{\ensuremath{L_{\mathsf{o}}}}

\newcommand{\rbcjr}{\ensuremath{r_{\bfd}}}
\newcommand{\avMIR}{\ensuremath{\overline{R}_{\text{MI}}}}
\newcommand{\qout}{\ensuremath{q_{\mathrm{out}}}}
\newcommand{\pout}{\ensuremath{p_{\mathrm{out}}}}
\newcommand{\mi}{\ensuremath{\mathrm{I}}}

\newcommand{\Rtrace}{\ensuremath{R_{\text{trace}}}}

\newcommand{\nix}{\ensuremath{n_{\mathsf{ix}}}}
\newcommand{\kix}{\ensuremath{k_{\mathsf{ix}}}}

\newcommand{\Rix}{\ensuremath{R_\mathrm{ix}}}
\newcommand{\bfii}[1]{\ensuremath{\mathrm{\mathbf{ind}}(#1)}}
\newcommand{\ii}[1]{\ensuremath{\mathrm{ind}(#1)}}
\newcommand{\outq}{\ensuremath{4}}

\newcommand{\Rin}{\ensuremath{R_\mathrm{i}}}
\newcommand{\nin}{\ensuremath{n_{\mathsf{i}}}}
\newcommand{\kin}{\ensuremath{k_{\mathsf{i}}}}

\newcommand{\bfwi}[1]{\ensuremath{\bfw^{(#1)}}}
\newcommand{\bfvi}[1]{\ensuremath{\bfv^{(#1)}}}
\newcommand{\lseq}{\ensuremath{L}}%

\newcommand{\outC}{\ensuremath{\bfC}}
\newcommand{\outS}{\ensuremath{\cS}}
\newcommand{\outW}{\ensuremath{\cW}}
\newcommand{\threshClus}{\ensuremath{\delta_\mathrm{C}}}
\newcommand{\threshIx}{\ensuremath{\delta_\mathrm{I}}}
\newcommand{\distClus}{\ensuremath{d_\mathrm{C}}}

\newcommand{\pmfconst}{\ensuremath{\gamma_\mathrm{c}}}

\newcommand{\poprRe}{\ensuremath{\cR}}
\newcommand{\poprTop}{\ensuremath{\cT}}
\newcommand{\distPost}{\ensuremath{d_\mathrm{P}}}

\newcommand{\rc}{\ensuremath{c_{\mathrm{r}}}} %
\newcommand{\wc}{\ensuremath{c_{\mathrm{w}}}}   %

\newcommand{\entr}{\ensuremath{\mathrm{H}}}

\newcommand{\expect}{\ensuremath{\mathbb{E}}}

\begin{document}
\title{An End-to-End Coding Scheme for DNA-Based Data Storage With Nanopore-Sequenced Reads}

\author{Lorenz~Welter$^{\orcidlink{0000-0002-5135-6731}}$,
        Roman~Sokolovskii$^{\orcidlink{0000-0002-3156-5923}}$,
        Thomas~Heinis$^{\orcidlink{0000-0002-7470-2123}}$,
        Antonia~Wachter-Zeh$^{\orcidlink{0000-0002-5174-1947}}$,\\
        Eirik~Rosnes$^{\orcidlink{0000-0001-8236-6601
        }}$,
        and~Alexandre~Graell~i~Amat$^{\orcidlink{0000-0002-5725-869X}}$
\thanks{Earlier versions of this paper were presented in part
at the 2023 IEEE Information Theory Workshop (ITW) in \cite{multidrawConCat_ITW_2023} and \cite{IssamEirikAlex_AIRkmer_23}.}%
\thanks{Corresponding author: Lorenz Welter (\texttt{lorenz.welter@tum.de}).}
\thanks{Lorenz Welter and Antonia Wachter-Zeh are with the
Institute for Communications Engineering, Technical University of Munich,
80333 Munich, Germany.}%
\thanks{Roman Sokolovskii and Thomas Heinis are with the
Department of Computing, Imperial College London,
London SW7 2AZ, United Kingdom.}%
\thanks{Eirik Rosnes is with Simula UiB, 5006 Bergen, Norway.}%
\thanks{Alexandre Graell i Amat is with the Department of Electrical Engineering,
Chalmers University of Technology, 41296 Gothenburg, Sweden.}%
\thanks{Funded by the European Union through projects NEO (101115317), MoSS (101058035), DNAMIC (101115389), and DiDAX (101115134). Views and opinions expressed are however those of the author(s) only and do not necessarily reflect those of the European Union or the European Research Council Executive Agency. Neither the European Union nor the granting authority can be held responsible for them.}
\thanks{The work of Alexandre Graell i Amat was financially supported by the Swedish Research Council (VR) under grant 2020-03687.
}
}

\markboth{}%
{}

\maketitle

\begin{abstract}
We consider error-correcting coding for deoxyribonucleic acid (DNA)-based storage using nanopore sequencing. We model the DNA storage channel as a sampling noise channel where the input data is chunked into $\chM$ short DNA strands, which are copied a random number of times, and the channel outputs a random selection of $\chN$ noisy DNA strands. The retrieved DNA reads are prone to strand-dependent insertion, deletion, and substitution (IDS) errors. We construct an index-based concatenated coding scheme consisting of the concatenation of an outer code, an index code, and an inner code. We further propose a low-complexity (linear in $\chN$) maximum a posteriori probability decoder that takes into account the strand-dependent IDS errors and the randomness of the drawing to infer symbolwise a posteriori probabilities for the outer decoder. We present Monte-Carlo simulations for information-outage probabilities and frame error rates for different channel setups on experimental data. We finally evaluate the overall system performance using the read/write cost trade-off. A powerful combination of tailored channel modeling and soft information processing allows us to achieve excellent performance even with error-prone nanopore-sequenced reads outperforming state-of-the-art schemes.%
\end{abstract}

\IEEEpeerreviewmaketitle

\section{Introduction}

\IEEEPARstart{I}{n} the quest for efficient and durable data storage solutions, deoxyribonucleic acid (DNA)-based data storage has emerged as a promising frontier.
The idea of DNA-based data storage dates back to R. Feynman's speech as early as 1959~\cite{feynman1959} and was first demonstrated experimentally in 1988 in the \textit{Microvenus} artwork~\cite{davis1996microvenus}.
After some fundamental small-scale experiments (e.g., \cite{clelland1999hiding,leier2000cryptography,bancroft2001long,gibson2010creation}), several larger-scale experiments (e.g., \cite{church_next-generation_2012,goldman_towards_2013,grass_robust_2015,yazdi_rewritable_2015,erlich_dna_2017,yazdi_portable_2017,organick_random_2018,antkowiak_low_2020,bornholt_dna-based_2016,Lopez2019,Anavy2019composite,stanford_magneticDNAstorage_2023}) showed the viability of DNA as a medium for high-density archival data storage.
Initially, research focused on general proofs-of-concept, whereas nowadays, the focus has shifted to experimental and theoretical investigation of particular aspects of the storage solution.

Typically, a DNA-based storage system consists of three components: 
\begin{enumerate*}[label=\arabic*)]
\item \emph{Synthesis:} Data is translated to several $\chM$ short DNA strands of length $\chL$ (typically in the range of $100$--$200$ nucleotides (nt)), each by chaining the four DNA nucleotides (Adenine (\dnaA), Cytosine (\dnaC), Guanine (\dnaG), and Thymine (\dnaT)). 
\item \emph{Storage:} The DNA strands are stored spatially unordered within an appropriate environment. Multiple copies of each strand are directly stored, and the number of copies depends on the applied synthesis method.
\item \emph{Sequencing:} Before reading, the DNA strands are typically copied by polymerase chain reaction (PCR). Then, a random subset of strands of cardinality $\chN$ is drawn from the storage pool, and a sequencing device retrieves the data of the drawn DNA strands, also called DNA reads; this work focuses specifically on \textit{Oxford Nanopore} sequencing.
\end{enumerate*}
For data consisting of multiple files, current methods for random file access, e.g., \cite{organick_random_2018}, utilize different primer sequences attached to the file-specific DNA strands and use primer-targeted PCR amplification for the readout. 

Naturally, this storage process poses interesting information-theoretic problems. 
In addition to the impairments mentioned above and inherent redundancy opportunities, within all three stages, the DNA strands are prone to insertion, deletion, and substitution (IDS) errors \cite{heckel_characterization_2019}.
For a detailed overview of information- and coding-theoretic contributions to DNA-based storage, we refer the reader to~\cite{HeckelShomorony_Foundations_2022}. In the following, we focus on the most relevant literature to efficiently highlight our contribution.

Theoretical research on the DNA storage channel started with the work \cite{heckel_fundamental_ISIT_2017}, where the authors derived the capacity of the DNA storage channel in the noise-free case. The authors showed that the unordered nature of the storage and the random subset selection induce a nonrecurring capacity loss even %
in the absence of errors. Follow-up works included substitution errors during the storage process \cite{lenz_upperboundcapDNA_2019, lenz_achieving_2020} and multiple drawing distributions \cite{lenz_noisyDrawChan_2023} in their capacity analysis. These capacity results were recently extended to any memoryless error channel within the DNA storage channel model \cite{weinberger_dna_2022} for a uniform drawing distribution. Interestingly, \cite{weinberger_dna_2022} showed that the decrease in theoretical decoding error probability is dominated by the number of stored DNA strands $\chM$, and not by the total number $\chM\!\cdot\!\chL$ of stored nucleotides. This implies that at the receiver side, reordering the DNA strands and recovering the data of erased, i.e., not drawn, strands is a crucial point. This is also underlined by its follow-up work \cite{weinberger_codedindex_2022}, where the author investigates a concatenated coding scheme based on random coding arguments with a coded index for strand reordering and additional low-complexity decoding, such as decoding strands individually before regrouping. Furthermore, noteworthy from the primarily theoretical perspective
are the code constructions for the DNA storage channel with erasure noise \cite{levick2022achieving} and substitution noise \cite{lenz_concatachieve_2020, polarDNAshuffle2023haghighat}.

For tackling IDS errors in single-sequence transmission, the schemes most relevant to our work were introduced in \cite{davey_reliable_2001} and \cite{ratzer_marker_2005}. These works model the independent and identically distributed (i.i.d.) IDS channel together with an error-correcting code as a hidden Markov process. On that basis, the presented decoding method was later improved by \cite{briffa_improved_2010, buttigieg_improved_2015}. Further, \cite{banerjee_sequential_2022} investigated sequential decoding techniques, and \cite{maarouf2023finite} examined finite blocklength phenomena. 

As the DNA storage channel naturally provides multiple copies of a stored strand on the receiver's side, decoding is connected with the \textit{trace reconstruction} problem \cite{carrillo1988multiple,levenshtein2001efficient,batu2004reconstructing,holenstein2008trace,peres2017average,holden2020subpolynomial,Srinivasavaradhan_2021,sakogawa2020symbolwise,press2020hedges,sabary2024reconstruction}.
Independently, \cite{Srinivasavaradhan2021TrellisBMA,Maarouf2022ConcatenatedCF} analyzed the coded trace reconstruction channel under i.i.d. IDS errors extending ideas of \cite{davey_reliable_2001} by means of achievable information rates (AIRs), where the latter one also presented frame error rates (FERs) when using concatenated codes. These works show significant gains by leveraging multiple noisy copies of the same transmit sequence, even with sub-optimal decoding techniques with linear complexity in the number of traces. These schemes have been extended to non-i.i.d. errors, explicitly targeting nanopore read channels, by a preliminary version of this paper \cite{IssamEirikAlex_AIRkmer_23} and \cite{hamoum_conf_aplanmer_2021,frenchKmer_belaid_2023}. Other modeling and coding approaches focusing on the nanopore channel have been investigated in \cite{mao2018models,hulett2021coding,mcbain2022finite,banerjee2023error,concatViterbo2024}. The work \cite{chandak_overcoming_2020} also focuses on non-i.i.d. errors as the decoder is evaluated on experimental data, which naturally contains non-i.i.d. errors, and is based on a deep-learning algorithm together with Viterbi decoding. 

\textbf{Our contribution.} This work advances the research on reliable DNA-based data storage with \emph{Oxford Nanopore sequencing} by incorporating more realistic modeling based on experimental data and introducing a flexible end-to-end/decoding solution designed for efficient error-free data retrieval. 
We model the DNA storage channel explicitly, including strand-dependent IDS errors%
, narrowing the modeling gap to the actual DNA storage channel. We call this model the \emph{\multidraw\ KMER channel}. 

We employ an index-based concatenated coding scheme and provide random access through primer-targeted PCR. In our proposed coding scheme, the information for one file is encoded by an outer code to provide overall error protection, mainly tackling unresolved errors and strand erasures of the data. Subsequently, the resulting codeword is split into $\chM$ data blocks. In each data block, its index information is embedded after being encoded by an index code, whereas the data block itself is encoded by another (inner) code.  
Both codes are specially tailored to tackle IDS errors. The collection of $\chM$ strands is then transmitted via the DNA storage channel incorporating strand erasures, multiple read copies, and strand-dependent IDS noise. 

At the receiver side, each noisy strand is decoded separately by a maximum a posteriori (MAP) decoder tailored explicitly to the nanopore channel, outputting symbolwise a posteriori probabilities (APPs). This low-complexity strategy initially ignores the possible multi-copy gain; however, to exploit this gain, we later combine the APPs according to the subsequent clustering and index decoding steps. We optionally cluster the received strands based on the obtained APPs. An index decision is made for each cluster by jointly considering the received strands within a cluster. Accurate clustering is computationally expensive and scales polynomially in the number of reads $\chN$ \cite{rashtchian_clustering_2017}. Hence, instead of complete clustering after the individual strand decoding, we leverage the soft output of our index decisions by grouping only reads with a clear index decision. Subsequently, we assign the remaining reads by comparing them to a constant number of the pre-grouped reads. By that strategy, we keep the decoding complexity to scale linearly in the number of total reads $\chN$ but benefit from an additional grouping performance gain. 

We analyze our proposed scheme in terms of outage and FERs on experimental data. Further, we provide comparisons based on the read/write cost trade-off to other experiments and coding setups in the literature. The created random dataset is available at \cite{roman2023experimentaldata}.

To achieve good performance, our scheme relies on channel estimations obtained from a random dataset. Moreover, given the targeted system parameters, mainly the number of stored strands $\chM$ and the targeted coverage $c$, we can use the random dataset to optimize the code components and decoding parameters.

We summarize our main contributions as follows.
\begin{itemize}
    \item Competitive and practical coding scheme for an end-to-end DNA-based data storage system.
    \item Flexible design of the scheme, which can be optimized to various DNA storage channel parameters.
    \item Incorporating strand-dependent IDS noise in the channel model and the decoder.
    \item Efficient decoding with complexity scaling only linearly in the number of total reads $\chN$.
    \item Comprehensive analysis of our scheme, providing valuable insights for the DNA storage channel.
    \item Verification of the correctness of our scheme on experimental data.
    \item Experimental random dataset with nanopore-sequenced DNA reads.
\end{itemize}

We remark that our coding approach is also applicable to storage methods without nanopore sequencing. For example, strand-dependent errors may happen as well during the synthesis process. Nonetheless, the channel model is inspired by the physical process of nanopore sequencing, where the electrical signal registered at the nanopore is influenced not only by the nucleotide translocating directly through it, but also by a window of surrounding nucleotides. Further, the optimization of our proposed system, i.e., channel estimation and decoder optimization, is based specifically on nanopore data.

The remainder of the paper is organized as follows. In \cref{sec:system-model}, we introduce the considered DNA storage channel model. We present our coding scheme in \cref{sec:coding-scheme} and the decoding procedure in \cref{sec:decoding}. Since we have conducted laboratory experiments with random data, we describe our experimental setup in \cref{sec:dataset}. The optimization procedure is laid out in \cref{sec:codeopt}. Then, we present our results in \cref{sec:simresults}, including the simulated end-to-end performance of the optimized scheme for an example target scenario. Finally, we draw some conclusions in~\cref{sec:conclusion}.

\section{{DNA} Storage Channel Model}\label{sec:system-model}

For one file, we consider the practical scenario in which a long data sequence is divided into short DNA strands, which are individually synthesized, and the PCR amplification and sequencing processes produce a random number of noisy reads for each short DNA strand. This channel is called the \emph{sampling noisy} channel.

In the following, we focus first on the single-strand noise channel by which each DNA strand is impaired during the whole storage process. After that, we discuss the sampling nature of our channel model in detail.

\subsection{KMER Channel Model}\label{sec:kmer-model}

We consider a variant of the state-based memory-$k$ nanopore channel model introduced in \cite{frenchKmer_belaid_2023}, which we abbreviate simply as the \kmerCh\ channel. (We describe the modifications we introduced to the model in~\cite{frenchKmer_belaid_2023} below.) The essential characteristic of this channel model is that it incorporates memory effects for errors within DNA strands---i.e., it forgoes the common assumption that errors are i.i.d. Introducing memory effects in the channel model allows us to match more closely the empirically observed behavior of nanopore sequencing, e.g., a bias toward higher deletion rates in homopolymers~\cite{del21troubles}. In particular, the occurrence of an error event (insertion, deletion, or substitution)---and hence the error-free transmission event as well---depends in the model on the surrounding $k$ input symbols and on the previous event (insertion \Ins, deletion \Del, substitution \Sub, or transmission \Tra). The \kmerCh\  channel we use in this work is depicted schematically in \cref{fig:kmer-channel} and introduced formally below.

Let $\bfx = (x_1, \dots, x_\chL)$, $x_t \in \alphDNA = \{\dnaA, \dnaC, \dnaG, \dnaT\}$, be the DNA strand to be transmitted and $\bfy = (y_1, \dots, y_{\chLout})$  the output DNA strand, where due to insertions and deletions the length of the output strand $\chLout$ is random and depends on the channel realizations. We define $\kmer_t = (x_{t-\floor{\frac{k}{2}}}, \dots, x_t, \dots, x_{t+\floor{\frac{k}{2}}})$ to be the surrounding \kmer\ at time instance $t \in \{\floor{\frac{k}{2}}+1, \dots, \chL - \floor{\frac{k}{2}}\}$. For any other possible instance $t$, we define the $\kmer_t$ to be the longest possible but symmetric substring surrounding $x_t$, e.g., for $k > 1$, $\kmer_1 = (x_1)$, $\kmer_2 = (x_1,x_2,x_3)$, etc. Let $e \in \{\Ins, \Del, \Sub, \Tra\}$ denote the previous channel event. Note that the total number of channel events is at least $\chL$ depending on the number of insertions.

The input to the channel can be seen as a queue in which symbols $x_t$ are successively enqueued for transmission. The output strand $\bfy$ is generated as follows.

Before $x_1$ is enqueued, we initialize $\bfy$ to an empty vector and set $e$ to the special event $e = \tt{Beg}$, which stands for the beginning event. Then, for each input symbol $x_t$, the following four events may occur.
\begin{enumerate}
    \item A uniformly at random symbol $a \in \alphDNA$ is inserted with probability $p(\Ins | \kmer_t, e)$. In this case, $\bfy$ is concatenated with symbol $a$ as $\bfy \leftarrow (\bfy , a)$, $x_t$ remains in the queue, and $e$ is updated as $e \leftarrow \Ins$.
    \item The symbol $x_t$ is deleted with probability $p(\Del | \kmer_t, e)$. Hence, $\bfy$ remains unchanged. We set $e \leftarrow \Del$ and $x_{t+1}$ is enqueued.
    \item With probability $p(\Sub | \kmer_t, e)$ the symbol $x_t$ is substituted by $a' \in \alphDNA \setminus \{x_t\}$ selected with probability $p( a' | x_t, \Sub)$ (independent of $\kmer_t$ and $e$). We set $\bfy \leftarrow (\bfy , a')$ and assign $e \leftarrow \Sub$. Subsequently, $x_{t+1}$ is enqueued.
    \item The symbol $x_t$ is transmitted with probability $p(\Tra | \kmer_t, e)$. The output is set to $\bfy \leftarrow (\bfy , x_t)$, $e$ is updated by $e \leftarrow \Tra$, and $x_{t+1}$ is enqueued.
\end{enumerate}
After $x_{\chL}$ leaves the queue, we obtain the output strand $\bfy$, of length $\chLout$. One important aspect is that the symbols in $\bfy$ are no longer synchronized with $\bfx$. Hence, $y_t$ could result from transmitting a symbol $x_{t^\prime}$ with $t^\prime \neq t$.

The main differences between the model described above and the model originally introduced in~\cite{frenchKmer_belaid_2023} are:
\begin{enumerate}
\item The definition of the window for the \kmer\ dependencies: Instead of conditioning on $k$ previously transmitted symbols, we condition on $k$ \textit{surrounding} symbols, in more realistic correspondence with the physical nature of the nanopore sequencing process. 
\item Modeling of the insertions:  The model in~\cite{frenchKmer_belaid_2023} assumes that a deletion must follow (a burst of)  insertions. In contrast, we allow for insertions to be followed by any channel event, see Fig.~\ref{fig:kmer-channel}.
\end{enumerate}
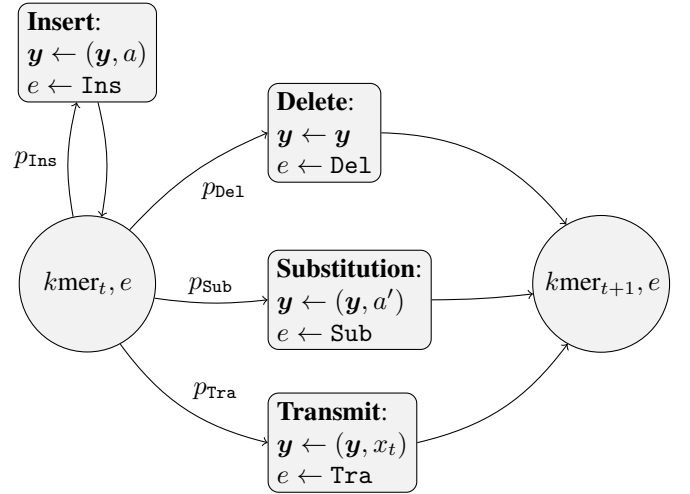
\begin{figure}[t]
	\centering
    \begin{tikzpicture}
		\node[round, text width = 1.5cm] (xi) {$\kmer_t, e$};
		
		\node[block, above=1.5cm of xi, align=left] (ins) {{\bf Insert}: \\ $\bfy \leftarrow (\bfy,a)$ \\ $e \leftarrow \Ins$};
		
		\node[block, above right=.7cm and 1.75cm of xi, align=left] (del) {{\bf Delete}: \\ $\bfy \leftarrow \bfy$ \\ $e \leftarrow \Del$};
		
		\node[block, below right=-1.1cm and 1.75cm of xi, align=left] (sub) {{\bf Substitution}: \\ $\bfy \leftarrow (\bfy,a')$ \\ $e \leftarrow \Sub$};

        \node[block, below right=.8cm and 1.75cm of xi, align=left] (tra) {{\bf Transmit}: \\ $\bfy \leftarrow (\bfy,x_t)$ \\ $e \leftarrow \Tra$};

		\node[round, text width = 1.5cm, right=5.0cm of xi] (xip1) {$\kmer_{t+1}, e$};

		\draw[->] (xi) to [bend left=12] node[left] {$p_\Ins$} (ins);
		\draw[->] (ins) to [bend left=12] (xi);
		
		\draw[->] (xi) to [bend left=12] node[below right] {$p_\Del$} (del.west);
        \draw[->] (del.east) to [bend left=25] (xip1);

        \draw[->] (xi) to [bend right=7] node[above] {$p_\Sub$} (sub.west);
        \draw[->] (sub.east) to [bend right=3] (xip1);

        \draw[->] (xi) to [bend right=20] node[above right] {$p_\Tra$} (tra.west);
        \draw[->] (tra.east) to [bend right=20] (xip1);

	\end{tikzpicture}
	\vspace{-1ex}
	\caption{Adapted state-based memory-$k$ nanopore (\kmerCh) channel inspired from \cite{frenchKmer_belaid_2023}. We have $\kmer_t = (x_{t-\floor{\frac{k}{2}}}, \dots, x_t, \dots, x_{t+\floor{\frac{k}{2}}})$ and abbreviate $p_\Ins = p(\Ins | \kmer_t , e)$, $p_\Del = p(\Del | \kmer_t , e)$, $p_\Sub = p(\Sub | \kmer_t , e)$, and $p_\Tra = p(\Tra | \kmer_t , e)$.}
	\label{fig:kmer-channel}
\end{figure}

\subsection{\Multidraw\ KMER Channel Model}

The data sequence is divided into $\chM$ short DNA strands $\bfx_1,\ldots,\bfx_\chM$, each of length $\chL$, which are synthesized and stored multiple times in a \emph{pool}. We define the input list $\chX \triangleq ( \bfx_1, \dots, \bfx_\chM )$. 
 We assume that the overall storage process, i.e., synthesis, PCR amplification, and nanopore sequencing, generates $\chNfo$ forward and $\chNba$ backward reads from the strands in the pool, where each read is a noisy (forward) or noisy reverse-complemented (backward) copy of an input strand $\bfx_i$. A reverse-complement $\bfx^\mathrm{b} \in \alphDNA^{\chL}$ of a strand $\bfx^\mathrm{f}$ is defined as $\bfx^\mathrm{b} = (\bar{x}_\chL^\mathrm{f},\ldots,\bar{x}_1^\mathrm{f})$, where $\bar{x} \in \alphDNA$ denotes the complement of the symbol $x$ with the respective complement relation $\dnaA$ -- $\dnaT$ and $\dnaC$ -- $\dnaG$ (Watson-Crick pairing). The idea of forward and backward reads originates from the fact that DNA is a double-helix formed by two complementary, directional, and base-wise strands twisted around each other \cite{milenkovic_2ndStru_2005}. 
 
 Consequently, the number of total strands $\chN$ is described as
 \begin{equation*}
     \chN = \chNfo + \chNba = \chc \chM\,,
 \end{equation*}
for a constant $\chc \in \mathbb{R}^{+}$. In the literature, the factor $c$ is called the \emph{coverage depth}.

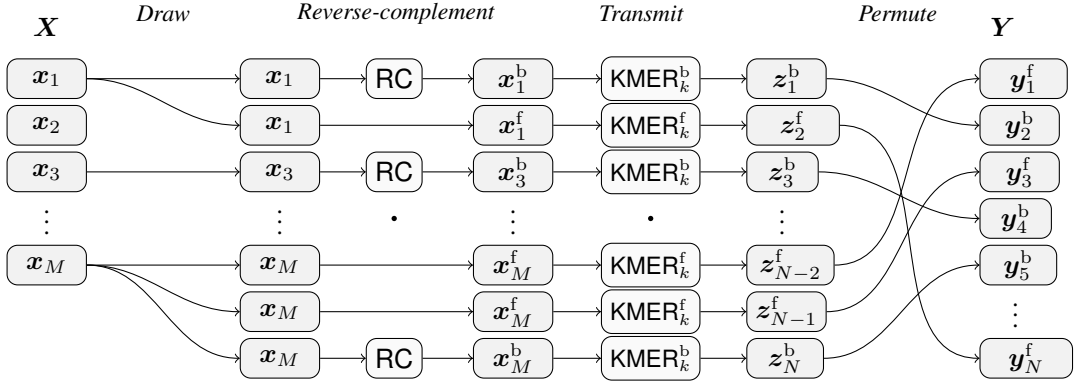
\begin{figure*}[t]
	\centering
    			\begin{tikzpicture}
			
			\def\xydist{0.62};
			\def\xxyydist{3.1};
			\def\yydist{0.62};
			
			\def\bwidth{3.0em};

			\tikzstyle{seqblock} = [rectangle, draw, text centered, rounded corners, minimum height=1.5em,fill=black!5!, minimum width=\bwidth]
			\tikzset{
				zblock/.style = {
					minimum width= {#1}em
					},
				rectangle, draw, text centered, rounded corners, minimum height=1.5em,fill=black!5!
			}
			\tikzstyle{idsblock} = [rectangle, draw, text centered, rounded corners, minimum height=1.5em,fill=black!2!, minimum width=0.5*\bwidth]

			\foreach \ii in {1,...,3}
				{\node[seqblock] (x\ii) at (-1*\xxyydist,-1*\ii*\xydist) {$\bfx_\ii$};	}
				
			\node (x4) at (-1*\xxyydist,-1*3.9*\xydist) {$\vdots$};
				
			\node[seqblock] (x5) at (-1*\xxyydist,-1*5*\xydist) {$\bfx_\chM$};

            \foreach \ii in {1,...,2}
				{\node[seqblock] (x1\ii) at (0*\xxyydist,-1*\ii*\xydist) {$\bfx_1$};	}

            \foreach \ii in {1}
				{\node[seqblock] (x2\ii) at (0*\xxyydist,-2*\xydist-1*\ii*\xydist) {$\bfx_3$};	}
    
			\node (x4) at (0*\xxyydist,-3.9*\xydist) {$\vdots$};

            \foreach \ii in {1,...,3}
				{\node[seqblock] (x5\ii) at (0*\xxyydist,-4*\xydist-1*\ii*\xydist) {$\bfx_\chM$};	}

            \foreach \ii/\jj in {1/b,2/f}
				{\node[seqblock, inner sep=0pt] (xrc1\ii) at (1*\xxyydist,-1*\ii*\xydist) {$\bfx^\mathrm{\jj}_1$};	}

            \foreach \ii/\jj in {1/b}
				{\node[seqblock, inner sep=0pt] (xrc2\ii) at (1*\xxyydist,-2*\xydist-1*\ii*\xydist) {$\bfx^\mathrm{\jj}_3$};	}
    
			\node (xrc4) at (1*\xxyydist,-3.9*\xydist) {$\vdots$};

            \foreach \ii/\jj in {1/f,2/f,3/b}
				{\node[seqblock, inner sep=0pt] (xrc5\ii) at (1*\xxyydist,-4*\xydist-1*\ii*\xydist) {$\bfx^\mathrm{\jj}_\chM$};	}

			\node[rectangle, draw, text centered, rounded corners, minimum height=1.5em,fill=black!5!, minimum width=\bwidth, anchor=west, inner sep=0] (z1) at (2.00*\xxyydist,-1*1*\yydist) {$\bfz_1^\mathrm{b}$};
			\node[rectangle, draw, text centered, rounded corners, minimum height=1.5em,fill=black!5!, minimum width=\bwidth+0.5em, anchor=west, inner sep=0] (z2) at (2.00*\xxyydist,-1*2*\yydist) {$\bfz_2^\mathrm{f}$};
			\node[rectangle, draw, text centered, rounded corners, minimum height=1.5em,fill=black!5!, minimum width=\bwidth-0.3em, anchor=west, inner sep=0] (z3) at (2.00*\xxyydist,-1*3*\yydist) {$\bfz_3^\mathrm{b}$};
			\node[rectangle, draw, text centered, rounded corners, minimum height=1.5em,fill=black!5!, minimum width=\bwidth+0.3em, anchor=west, inner sep=0] (z4) at (2.00*\xxyydist,-1*5*\yydist) {$\bfz_{\chN-2}^\mathrm{f}$};
			\node[rectangle, draw, text centered, rounded corners, minimum height=1.5em,fill=black!5!, minimum width=\bwidth+0.0em, anchor=west, inner sep=0] (z5) at (2.00*\xxyydist,-1*6*\yydist) {$\bfz_{\chN-1}^\mathrm{f}$};
			\node[anchor=west] (z6) at (2.1*\xxyydist,-1*3.9*\yydist) {$\vdots$};
			\node[rectangle, draw, text centered, rounded corners, minimum height=1.5em,fill=black!5!, minimum width=\bwidth-0.1em, anchor=west, inner sep=0] (z7) at (2.00*\xxyydist,-1*7*\yydist) {$\bfz_\chN^\mathrm{b}$};

            \node[rectangle, draw, text centered, rounded corners, minimum height=1.5em,fill=black!5!, minimum width=\bwidth+0.3em, anchor=west, inner sep=0] (y1) at (3.00*\xxyydist,-1*1*\yydist) {$\bfy_1^\mathrm{f}$};
			\node[rectangle, draw, text centered, rounded corners, minimum height=1.5em,fill=black!5!, minimum width=\bwidth, anchor=west, inner sep=0] (y2) at (3.00*\xxyydist,-1*2*\yydist) {$\bfy_2^\mathrm{b}$};
			\node[rectangle, draw, text centered, rounded corners, minimum height=1.5em,fill=black!5!, minimum width=\bwidth+0.0em, anchor=west, inner sep=0] (y3) at (3.00*\xxyydist,-1*3*\yydist) {$\bfy_3^\mathrm{f}$};
			\node[rectangle, draw, text centered, rounded corners, minimum height=1.5em,fill=black!5!, minimum width=\bwidth-0.3em, anchor=west, inner sep=0] (y4) at (3.00*\xxyydist,-1*4*\yydist) {$\bfy_4^\mathrm{b}$};
			\node[rectangle, draw, text centered, rounded corners, minimum height=1.5em,fill=black!5!, minimum width=\bwidth-0.1em, anchor=west, inner sep=0] (y5) at (3.00*\xxyydist,-1*5*\yydist) {$\bfy_5^\mathrm{b}$};
			\node[anchor=west] (y6) at (3.1*\xxyydist,-1*5.9*\yydist) {$\vdots$};
			\node[rectangle, draw, text centered, rounded corners, minimum height=1.5em,fill=black!5!, minimum width=\bwidth+0.5em, anchor=west, inner sep=0] (y7) at (3.00*\xxyydist,-1*7*\yydist) {$\bfy_\chN^\mathrm{f}$};

            \draw[->] (x1.east) to[out=0,in=-180] node[midway,right,inner sep=2pt] {} (x11.west);
			\draw[->] (x1.east) to[out=0,in=-180] node[midway,right,inner sep=2pt] {} (x12.west);
			\draw[->] (x3.east) to[out=0,in=-180] node[midway,right,inner sep=2pt] {} (x21.west);
			\draw[->] (x5.east) to[out=0,in=-180] node[midway,right,inner sep=2pt] {} (x51.west);
			\draw[->] (x5.east) to[out=0,in=-180] node[midway,right,inner sep=2pt] {} (x52.west);
			\draw[->] (x5.east) to[out=0,in=-180] node[midway,right,inner sep=2pt] {} (x53.west);

            \node[idsblock] (rc1) at (0.49*\xxyydist,-1*1*\yydist) {\textsf{RC}};
            \node[idsblock] (rc2) at (0.49*\xxyydist,-3*1*\yydist) {\textsf{RC}};
            \node[idsblock] (rc3) at (0.49*\xxyydist,-7*1*\yydist) {\textsf{RC}};
            \draw[->] (x11.east) -- (rc1.west);     \draw[->] (rc1.east) -- (xrc11.west);
            \draw[->] (x21.east) -- (rc2.west);     \draw[->] (rc2.east) -- (xrc21.west);
            \draw[->] (x53.east) -- (rc3.west);     \draw[->] (rc3.east) -- (xrc53.west);           
            \draw[->] (x12.east) -- (xrc12.west);
            \draw[->] (x51.east) -- (xrc51.west);
            \draw[->] (x52.east) -- (xrc52.west);
            \node[circle, fill=black, minimum size=2pt,
				inner sep=0pt, outer sep=0pt] at (0.49*\xxyydist,-4*\yydist) {};

            \node[idsblock] (ids1) at (1.59*\xxyydist,-1*1*\yydist) {\small $\kmerChba$}; %
			\draw[->] (xrc11.east) -- (ids1.west) ;
			\draw[->] (ids1.east) -- (z1.west) ;

            \node[idsblock] (ids2) at (1.59*\xxyydist,-2*1*\yydist) {\small $\kmerChfo$};
            \draw[->] (xrc12.east) -- (ids2.west) ;
			\draw[->] (ids2.east) -- (z2.west) ;

            \node[idsblock] (ids3) at (1.59*\xxyydist,-3*1*\yydist) {\small $\kmerChba$};
            \draw[->] (xrc21.east) -- (ids3.west) ;
			\draw[->] (ids3.east) -- (z3.west) ;

            \node[idsblock] (ids4) at (1.59*\xxyydist,-5*1*\yydist) {\small $\kmerChfo$};
            \draw[->] (xrc51.east) -- (ids4.west) ;
			\draw[->] (ids4.east) -- (z4.west) ;

            \node[idsblock] (ids5) at (1.59*\xxyydist,-6*1*\yydist) {\small $\kmerChfo$};
            \draw[->] (xrc52.east) -- (ids5.west) ;
			\draw[->] (ids5.east) -- (z5.west) ;

            \node[idsblock] (ids6) at (1.59*\xxyydist,-7*1*\yydist) {\small $\kmerChba$};
            \draw[->] (xrc53.east) -- (ids6.west) ;
			\draw[->] (ids6.east) -- (z7.west) ;

			\foreach \ii in {4}
			{
				\node[circle, fill=black, minimum size=2pt,
				inner sep=0pt, outer sep=0pt] at (1.59*\xxyydist,-1*\ii*\yydist) {};
			}

			\draw[->] (z1.east) to[out=0,in=-180] node[midway,right,inner sep=2pt] {} (y2.west);
			\draw[->] (z2.east) to[out=0,in=-180] node[midway,right,inner sep=2pt] {} (y7.west);
			\draw[->] (z3.east) to[out=0,in=-180] node[midway,right,inner sep=2pt] {} (y4.west);
			\draw[->] (z4.east) to[out=0,in=-180] node[midway,right,inner sep=2pt] {} (y1.west);
			\draw[->] (z5.east) to[out=0,in=-180] node[midway,right,inner sep=2pt] {} (y3.west);
			\draw[->] (z7.east) to[out=0,in=-180] node[midway,right,inner sep=2pt] {} (y5.west);

			\node (x0) at (-1*\xxyydist,0.1*\xydist) {$\bfX$};			
			\node (z0) at (3.1*\xxyydist,0.1*\xydist) {$\chY$};
			\node (samp) at (-0.5*\xxyydist,0.4*\yydist) {\small \textit{Draw}};
            \node (rc) at (0.5*\xxyydist,0.4*\yydist) {\small \textit{Reverse-complement}};
             \node (perm) at (2.65*\xxyydist,0.4*\yydist) {\small \textit{Permute}};
			\node (trans) at (1.55*\xxyydist,0.4*\yydist) {\small \textit{Transmit}};
			\end{tikzpicture}
	\vspace{-1ex}
	\caption{\Multidraw\ \textsf{KMER} channel model. First, each input strand gets copied randomly. Second, the strands get reverse-complemented (\textsf{RC}) according to a Bernoulli RV and get consequently labeled as forward or backward reads (see $(\cdot)^\mathrm{f}$/$(\cdot)^\mathrm{b}$). Third, the resulting strands are passed through independent \kmerChfo/\kmerChba\ channels. Fourth, the strands are randomly permuted.}
	\label{fig:multidraw-channel}
\end{figure*}

The channel between the input to be synthesized, $\chX$, and the output of the sequencing process, $\chY$, can be split into forward and backward reads as $\chY = [\chYfo, \chYba]$. 
In the following, we describe the channel model in four phases.
\begin{enumerate}
    \item \emph{Draw:} $\chN$ draws are performed from the list $\bfX$ at random with replacement according to the fixed drawing probabilities $\drawdist=(p^\mathrm{d}_1, \ldots, p^\mathrm{d}_\chM)$ for each strand $\bfx_i$ with $\sum_i p^\mathrm{d}_i = 1$. Let $D_i$ be the random variable (RV) corresponding to the number of times strand $\bfx_i$ is drawn and define the random vector $\bfD \triangleq (D_1,\dots,D_{\chM} )$, with $\sum_i D_i = N$, and $Q_d $ the RV giving the number of sequences which are drawn $d$ times. Then, the random draw vector $\bfD = (D_1,\dots,D_\chM)$ is multinomially distributed as $\bfD \sim \mathsf{Multinom}(\chM, \chN, \bfP^\mathrm{d})$.
    \item \emph{Reverse-complement:} Each drawn strand is randomly reverse-complemented according to a Bernoulli RV $B \sim \mathsf{Ber}(p^\mathrm{rc})$ and labeled accordingly as a forward or backward read.
    \item \emph{Transmit:} The drawn strands are transmitted through independent channels, where the corresponding forward \kmerChfo\ or backward \kmerChba\ channel is applied. We denote by $\bfz_j^\mathrm{f/b}$ the output of the $j$-th \kmerChfo/\kmerChba\ channel, $j\in \{1,\dots,\chN\}$, and define $\bfZ \triangleq (\bfz_1^\mathrm{f},\dots,\bfz_{\chNfo}^\mathrm{f}, \bfz_1^\mathrm{b},\dots,\bfz_{\chNba}^\mathrm{b})$ of total length $\chN$.  
    \item \emph{Permute:} To obtain the final output $\chY = [\chYfo, \chYba]$, the list $\bfZ$ is permuted uniformly at random and subsequently split into forward reads $\chYfo$ and backward reads $\chYba$ according to their respective labels.

\end{enumerate}
We illustrate the channel model in \cref{fig:multidraw-channel}.\footnote{In  \cite{weinberger_dna_2022, lenz_upperboundcapDNA_2019, HeckelShomorony_Foundations_2022}, the \emph{draw} and \emph{permute} steps are considered as a joint process referred to as \emph{sampling}. We split the step for the sake of presentation; our channel model is essentially the same, except that here we consider the noise channel as a \kmerCh\ channel, which induces non-i.i.d. IDS errors and an arbitrary drawing distribution.}
We define the channel parameter $\beta = \frac{\log_2(\chM)}{\chL}$ that relates the total number and the length of the strands and can be interpreted as the penalty, in bit/nt, that the channel induces by the loss of order during the \textit{permute} step.

\subsection{Extended Trace Reconstruction Channel Model} \label{sec:trace-recon-model}

This channel model simplifies the \multidraw\ channel model by fixing the drawing step and removing the permutation step. In particular, let the input be $\chM$ strands of length $\chL$. Each input strand is drawn exactly $\trchM$ times. The event of reverse-complementing follows the same Bernoulli RV $B \sim \mathsf{Ber}(p^\mathrm{rc})$. By analyzing this simplified channel, we can gain insights into the behavior of different codes in the presence of IDS errors and perform optimization procedures for the \multidraw\ channel (see \cref{sec:codeopt}). We use the terminology \emph{extended trace reconstruction channel} to highlight the close relationship to the simple \emph{trace reconstruction channel}, e.g., as defined in \cite{holenstein2008trace} (for only deletion noise).

\section{Index-Based Coding Scheme}\label{sec:coding-scheme}

We propose an index-based concatenated coding scheme consisting of three codes: an outer spatially-coupled low-density parity-check (SC-LDPC) code, providing overall protection to the data and against non-drawn strands; an index code, which counteracts the loss of order in the presence of IDS errors; and an inner code whose primary goal is to maintain synchronization.

The information $\bfu \in \field _{\chq} ^{\dimo}$ is encoded by an  $[\leno,\dimo,\sccoup,\scmem]_{\chq}$ SC-LDPC code  of output length $\leno$, input length $\dimo$, coupling length $\sccoup$, and coupling memory $\scmem$ over the finite field $\field_{\chq}$ \cite{felstrom1999time, lentmaier_SCLDPC_2010}. Note that there is a one-to-one mapping of the field $\field_{4} = \field_{2^2}$ to the  DNA alphabet $\alphDNA = \{\dnaA,\dnaC,\dnaG,\dnaT\}$. We interpret the alphabet $\alphDNA$ as a field $\field_4$ to allow linear operations over vectors in the DNA alphabet. The resulting codeword is subsequently interleaved uniformly at random forming $\bfw = (w_1,\ldots,w_{\leno}) \in \field_{\chq} ^{\leno}$. We split $\bfw$ into $\chM$ equal-length data blocks as $\bfw = (\bfwi{1}, \ldots , \bfwi{\chM})$ such that $\bfwi{i} = (w^{(i)}_1,\ldots,w^{(i)}_{\bleno}) \in \field_{\chq}^{\bleno}$,   $\bleno = \frac{\leno}{\chM}$. Independently, the data blocks $\bfwi{i}$ are encoded by an $[\nin,\kin]_\chq$ inner code over $\field_\chq$ of output length $\nin$, input length $\kin$, and rate $\Rin=\frac{\kin}{\nin}$, generating an inner data codeword of length ${\bleno}\Rin^{-1}$. We consider the marker-repeat (MR) codes presented in \cite{Srinivasavaradhan2021TrellisBMA, inoue_adaptivemarker2012} for the inner code, where the $\kin$-th input symbol is repeated once such that $\nin = \kin + 1$ and $\Rin=1-\frac{1}{\nin}$. The MR codes have shown good performance in high-rate and low-error regimes for i.i.d. trace reconstruction channels using real DNA strands \cite{Srinivasavaradhan2021TrellisBMA}.\footnote{We consider a slight variant where we place the redundancy symbols evenly spaced within the data regions of the strand. This is due to the termination effects of our decoder at the beginning and end of the strands; hence, we obtain equal distributed synchronization capabilities along the strand.} Further, each index $i$ is encoded by an $[\nix,\kix]_\chq$ index code over $\field_{\chq}$ of even output length $\nix$ and even input length $\kix \geq \log_\chq (\chM)$ to an index codeword. Our decoding technique provides higher error protection at the beginning and at the end of the DNA strands. Thus, we split the index codeword in half and insert the two halves at the beginning and the end of the inner codeword of the respective data block $\bfwi{i}$. Subsequently, a random offset sequence is added to the generated strand. It supports the synchronization capability of the index and inner codes and ensures that the nucleotides of the stored DNA strands are uniformly distributed over $\Sigma_4$. Finally, the fixed file-specific front and back primer sequences $\primer_1, \primer_2 \in \alphDNA^{\primL}$ are attached, resulting in the final encoded output strand $\bfx_i$, of length $\lseq =  \bleno \Rin^{-1} + \nix + 2 \primL$. Note that the primer sequences and the random offset are known to the decoder. The final codeword list is then described by $\bfX = (\bfx_1,\dots,\bfx_\chM)$.

The overall code rate, or information density (without primers), is 
\begin{align*}
    \Rtot =  \Rout \Rin \left(2 - \frac{\beta}{\Rix}\right)\,, \quad \text{\footnotesize[bit/nt]}
\end{align*}
with individual rates $\Rout = 2 \frac{\dimo}{\leno}$, $\Rix = 2\frac{\kix}{\nix}$, and recall $\Rin = 2\frac{\kin}{\nin}$, all normalized to the unit bit/nt.

For decoding, it is convenient to introduce the following equivalent interpretation of the primer, index, and inner encoding as a joint process. A final encoded output strand $\bfx_i$ consists of segments encoded with different codes. This allows us to perform decoding on a single trellis, which we call the \emph{joint-code} trellis.\footnote{We emphasize that the \emph{joint-code} does not include the outer code since we encode/decode the outer code separately.} We transform the index integer $i$ to an index vector $\bfii{i} = \left(\ii{i}_1, \ldots, \ii{i}_{\kix}\right) \in \field_\chq^{\kix}$. This index vector is split in half into $\mathrm{\mathbf{ix}}_1^{(i)}$ and $\mathrm{\mathbf{ix}}_2^{(i)}$, and placed together with the corresponding primer sequence $\primer_1$, $\primer_2$ at the beginning and end of the vector $\bfwi{i}$ resulting in the vector $\bfvi{i}=(v^{(i)}_1,\ldots,v^{(i)}_{\kix+\bleno+2\primL})$, of length $\kix+\bleno+2\primL$. Moreover, the  $[\nix, \kix]_\chq$ index code is generated by a serial concatenation of $a$ equal $[\frac{\nix}{a},\frac{\kix}{a}]_\chq$ codes, where $a$ is a multiple of 2 and divides both $\kix$ and $\nix$. Subsequently, the vector $\bfvi{i}$ is encoded at the primer positions with a rate-one code, at the index positions by the $[\frac{\nix}{a},\frac{\kix}{a}]_\chq$ codes, and symbolwise by the inner MR code at the positions corresponding to $\bfwi{i}$.

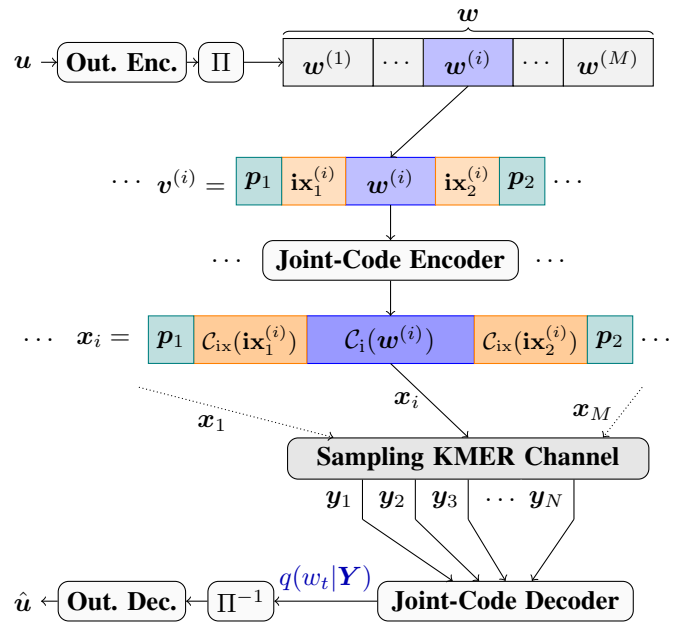
\begin{figure}[t]
	\centering
	\begin{tikzpicture}
    \def\xxdist{5em};
	\def\yydist{3.0em};
	\def\bwidth{3.4em}
    \def\blockshift{2.35}

	\tikzstyle{seqblock} = [rectangle, draw, text centered, minimum height=1.8em,fill=black!5!, minimum width=\bwidth, outer sep=0]
 \tikzstyle{seqblockempty} = [rectangle, draw, text centered, minimum height=1.8em,fill=black!5!, minimum width=0.5*\bwidth, outer sep=0]
    \tikzstyle{seqixblock} = [rectangle, draw, text centered, minimum height=1.8em,fill=black!1!, minimum width=0.2*\bwidth, outer sep=0]
			\tikzstyle{coding} = [rectangle, draw, text centered, rounded corners, minimum height=1.5em,fill=black!2!, minimum width=0.5*\bwidth]

		\node (u) at (0,0.5*\yydist) {$\bfu$};
		\node[coding] (outenc) at (0.75*\xxdist, 0.5*\yydist) {\bf Out. Enc.};

        \node[coding] (interenc) at (1.5*\xxdist, 0.5*\yydist) {\bf $\Pi$};

        \node[seqblock, fill=blue!25!white, text opacity=1] (wi) at (\blockshift*\xxdist+1.5*\bwidth,0.5*\yydist) {$\bfwi{i}$};
        \node[seqblockempty,anchor=east] (w2) at ($(wi.west)$) {\small $\dots$}; 
         \node[seqblockempty,anchor=west] (w3) at ($(wi.east)$) {\small $\dots$};
  \node[seqblock,anchor=east] (w1) at ($(w2.west)$) {$\bfwi{1}$};
   \node[seqblock, anchor=west] (w4) at ($(w3.east)$) {$\bfwi{\chM}$};

     \node[seqblock, draw=blue, fill=blue!25!white, text opacity=1] (ixseq) at (0.75*\blockshift*\xxdist+1.5*\bwidth,-1*\yydist) {$\bfwi{i}$};
    \node[seqixblock, anchor=east, draw=orange, fill=orange!25!white, text opacity=1] (ixseq1) at ($(ixseq.west)$) {\small $\mathrm{\mathbf{ix}}_1^{(i)}$};
    \node[seqixblock, anchor=east, draw=teal, fill=teal!25!white, text opacity=1] (ixseq2) at ($(ixseq1.west)$) { $\primer_1$};
     \node at ($(ixseq2.west)+(-0.5*\bwidth,0)$) { $\bfv^{(i)} = $};
    \node at ($(ixseq2.west)+(-1.2*\bwidth,0)$) { $\dots$};
    \node[seqixblock, anchor=west, draw=orange, fill=orange!25!white, text opacity=1] (ixseq3) at ($(ixseq.east)$) {\small $\mathrm{\mathbf{ix}}_2^{(i)}$};
    \node[seqixblock, anchor=west, draw=teal, fill=teal!25!white, text opacity=1] (ixseq4) at ($(ixseq3.east)$) { $\primer_2$};
    \node at ($(ixseq4.east)+(0.28*\bwidth,0)$) { $\dots$};

		\node[coding, text width=9em] (enci) at (0.75*\blockshift*\xxdist+1.5*\bwidth,-2*\yydist) {\bf Joint-Code Encoder};
	       \node (enc1) at (0.75*\blockshift*\xxdist-0.3*\bwidth,-2*\yydist) {$\dots$};
        \node (enc2) at (0.75*\blockshift*\xxdist+3.3*\bwidth,-2*\yydist) {$\dots$};

     \node[seqblock, draw=blue, fill=blue!40!white, text opacity=1] (ENCixseq) at (0.75*\blockshift*\xxdist+1.5*\bwidth,-3*\yydist) {\hspace{2.5ex}$\cC_{\mathrm{i}}(\bfwi{i})$ \hspace{1.0ex} \vphantom{o}};
    \node[seqixblock, anchor=east, draw=orange, fill=orange!40!white, text opacity=1] (ENCixseq1) at ($(ENCixseq.west)$) {\small $\cC_{\mathrm{ix}}(\mathrm{\mathbf{ix}}_1^{(i)})$};
    \node[seqixblock, anchor=east, draw=teal, fill=teal!25!white, text opacity=1] (ENCixseq2) at ($(ENCixseq1.west)$) { $\primer_1$};
     \node at ($(ENCixseq2.west)+(-0.5*\bwidth,0)$) { $\bfx_i = $};
    \node at ($(ENCixseq2.west)+(-1.2*\bwidth,0)$) { $\dots$};
    \node[seqixblock, anchor=west, draw=orange, fill=orange!40!white, text opacity=1] (ENCixseq3) at ($(ENCixseq.east)$) {\small $\cC_{\mathrm{ix}}(\mathrm{\mathbf{ix}}_2^{(i)})$};
    \node[seqixblock, anchor=west, draw=teal, fill=teal!25!white, text opacity=1] (ENCixseq4) at ($(ENCixseq3.east)$) { $\primer_2$};
    \node at ($(ENCixseq4.east)+(0.28*\bwidth,0)$) { $\dots$};

		\node[coding, text width=13em, fill=black!10!white, text opacity=1] (channel) at (\blockshift*\xxdist+1.5*\bwidth,-4.5*\yydist) {\bf \Multidraw\ KMER Channel};
		
		\node[coding, text width=9em] (indec) at (\blockshift*\xxdist+1.5*\bwidth+6*\blockshift,-6.3*\yydist) {\bf Joint-Code Decoder};

        \node[coding] (interdec) at (1.65*\xxdist, -6.3*\yydist) {\bf $\Pi^{-1}$};
  
		\node[coding] (outdec) at (0.75*\xxdist, -6.3*\yydist) {\bf Out. Dec.};
		
		\node (uhat) at (0.0, -6.3*\yydist) {$\hat{\bfu}$};
		
		\draw[->] (u.east) -- (outenc.west);
         \draw[->] (outenc.east) -- (interenc.west);
		\draw[->] (interenc.east) -- (w1.west);
		
        \draw[->] (wi.south) -- (ixseq.north);

         \draw[->] (ixseq.south) -- (enci.north);
         \draw[->] (enci.south) -- (ENCixseq.north);

        \draw[->] ($(ENCixseq.south)+(0,0)$) -- ($(channel.north)+(0,0)$) node[midway, left] {$\bfx_i$};
        \draw[->,densely dotted] ($(ENCixseq.south)+(-2.8*\bwidth,-0.3*\yydist)$) -- ($(channel.north)+(-1.5*\bwidth,0)$) node[midway, left,yshift=-4] {$\bfx_1$};
        \draw[->,densely dotted] ($(ENCixseq.south)+(2.8*\bwidth,-0.3*\yydist)$) -- ($(channel.north)+(1.5*\bwidth,0)$) node[midway, left] {$\bfx_{\chM}$};
		
		\draw[-] ($(channel.south)+(-0.8*\xxdist,0)$) -- ($(channel.south)+(-0.8*\xxdist,-0.5*\yydist)$)  node[midway, left] {$\bfy_1$};
		\draw[-] ($(channel.south)+(-0.4*\xxdist,0)$) -- ($(channel.south)+(-0.4*\xxdist,-0.5*\yydist)$)  node[midway, left] {$\bfy_2$};
		\draw[-] ($(channel.south)+(-0.0*\xxdist,0)$) -- ($(channel.south)+(-0.0*\xxdist,-0.5*\yydist)$)  node[midway, left] {$\bfy_3$};
		\draw[draw=white] ($(channel.south)+(0.4*\xxdist,0)$) -- ($(channel.south)+(0.4*\xxdist,-0.5*\yydist)$)  node[midway, xshift=-0.7em] {$\dots$}; 
		\draw[-] ($(channel.south)+(0.8*\xxdist,0)$) -- ($(channel.south)+(0.8*\xxdist,-0.5*\yydist)$)  node[midway, left] {$\bfy_\chN$};
		
		\draw[->] ($(channel.south)+(-0.8*\xxdist,-0.5*\yydist)$) -- ($(indec.north)+(-0.4*\xxdist,0)$);
		\draw[->] ($(channel.south)+(-0.4*\xxdist,-0.5*\yydist)$) -- ($(indec.north)+(-0.2*\xxdist,0)$);
		\draw[->] ($(channel.south)+(-0.0*\xxdist,-0.5*\yydist)$) -- ($(indec.north)+(-0.0*\xxdist,0)$);
		\draw[->] ($(channel.south)+(0.8*\xxdist,-0.5*\yydist)$) -- ($(indec.north)+(0.2*\xxdist,0)$);
		
		\draw[->] (indec.west) -- (interdec.east) node [midway,above,yshift=0] {\textcolor{blue!70!black}{$q(w_t|\chY)$}};

        \draw[->] (interdec.west) -- (outdec.east);
		
		\draw[->] (outdec.west) -- (uhat.east);
		
		\draw [decorate,decoration = {brace}] ($(w1.north west)+(0,0.05*\yydist)$) --  ($(w4.north east)+(0,0.05*\yydist)$) node [midway, above,yshift=0.05*\yydist] {$\bfw$};

\end{tikzpicture}
	\caption{Concatenated coding scheme for  communication over the multi-draw \textsf{KMER} channel. For each block $\bfwi{i}$, the corresponding index vector is split into two parts and inserted at the beginning and the end of the block together with the primers $\primer_1$ and  $\primer_2$. \textit{Joint-code} refers to the combination of primer, index, and inner code. We use $\cC_\mathrm{ix}(\cdot)$ and $\cC_\mathrm{i}(\cdot)$ as the encoding functions for the index code and inner code, respectively. The reads $\bfy_1,\ldots, \bfy_\chN$ are either labeled forward or backward reads. The term $q(w_t|\chY)$ denotes the mismatched symbol APPs computed by the joint-code decoder. Further, the symbols $\Pi$ and $\Pi^{-1}$ represent the random interleaving and its reversal operation, respectively.} 	\label{fig:coding-scheme}
\end{figure}

The coding/decoding scheme is depicted in Fig.~\ref{fig:coding-scheme}.

\section{Reads Decoding}\label{sec:decoding}

The channel introduces three main challenges (see \cref{fig:multidraw-channel}): 
\begin{enumerate*}
    \item The randomness of the drawing may cause an absence of input strands. However, it is also an inherent redundancy source since strands may be drawn multiple times. For unbiased drawing distributions, the coverage effectively controls the interplay of these two effects. As the coverage depth grows, the amount of provided redundancy increases; thus, the probability of absent strands decreases.
    \item The insertion and deletion events during the single-strand transmission lead to a loss of synchronization within the respective strand.
    \item The random permutation effect leads to a loss of order within the input strands.
\end{enumerate*}

\begin{figure}
    \centering
    \input{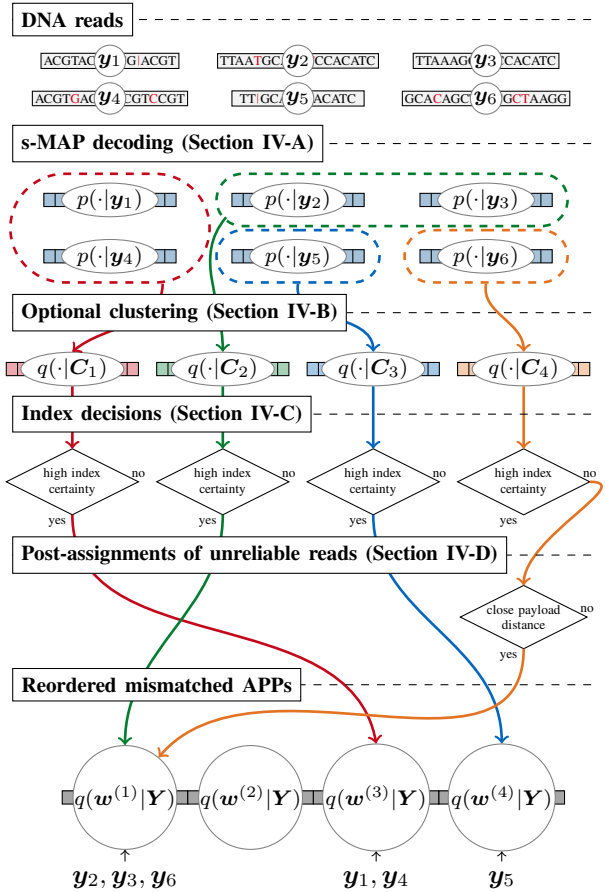}
    \caption{Simplified decoding flow (no outer decoding). For simplicity, we do not differentiate forward or backward reads. The arrows in the last row of the figure indicate the resulting assignments of the DNA reads $\bfy_1,\ldots,\bfy_6$ to the respective outer codeword blocks.}
    \label{fig:decoderflow}
\end{figure}

Optimal decoding would consider these effects jointly to produce an estimate of the encoded information $\hat{\bfu}$ given a list of all received DNA strands $\chY$; such decoding is infeasible in practice. Instead, we propose a sub-optimal decoding scheme based on \emph{mismatched decoding} principles. In essence, we first perform inner strand-level decoding for each read, then group the generated estimations into clusters and reconstruct their ordering, combine estimations within each cluster, and finally feed these consensus estimations to the outer decoder to output an estimate $\hat{\bfu}$. We describe our decoding procedure in detail below, and for a better grasp, a simplified flowchart is depicted in \cref{fig:decoderflow}.

First, we infer APPs using a symbolwise MAP decoder which works on the joint-code (recall combined primer, index, and inner code) and the forward/backward channel trellis by employing the BCJR algorithm for each received strand $\bfy_1,\dots,\bfy_\chN$ independently (Section~\ref{sse:inner-decoding}). The channel labels the received strands at no cost as a forward or backward (i.e., reverse-complemented) read,\footnote{The decoder relies on its knowledge of the primer sequences $\primer_1, \primer_2$ (see Fig.~\ref{fig:coding-scheme}) to distinguish between forward and backward reads.} which follow different channel statistics. Hence, we use separate channel estimations for forward and backward reads in our decoding. Optionally, clustering of the DNA strands using the APPs is performed (Section~\ref{sse:clustering})\textemdash clustering at this stage enables us to use the provided coding gain for the subsequent step of index decisions. For combining the APPs within a cluster, the APPs are symbolwise multiplied, generating the mismatched APPs for one cluster. By making another mismatched (ignoring APP correlations) maximum likelihood decision on the index APPs for a given cluster (Section~\ref{sse:index-decoding}), we assign the block index $\hat{i}$ within $\bfw$ whenever the decoder-assigned probability of $\hat{i}$ is higher than a predefined threshold; otherwise, we keep the reads of the cluster for post-processing. Therefore, assuming optimal clustering, we can benefit from the multi-copy gain of the channel for index estimation. We later attempt to assign the reads that were ruled out based on the index decision threshold by comparing their \textit{payload} distances, i.e., aggregated APP distance, to already-assigned reads (Section~\ref{sec:postclust}). The APPs of the data block belonging to the same index $\hat{i}$ are multiplied, generating the mismatched APPs $q(w_t^{(i)}|\chY)$. These APPs are further processed to mitigate underconfidence/overconfidence bias following ideas from the theory of generalized mutual information. Finally, the overall ordered APPs $q(w_t|\chY)$ are passed to the outer decoder, reversing the interleaving operation and neglecting any possible correlations between the symbolwise APPs. The combination of the outer code and the interleaver is designed to handle \emph{block-fading} effects (i.e., the loss of strands), as described in Section~\ref{sec:codeopt-outer}.

Combining APPs belonging to different strands by symbolwise multiplication is inspired by the well-performing \emph{separate decoding} strategy of \cite{Maarouf2022ConcatenatedCF}. In that work, the authors compared two decoding approaches on the trace reconstruction channel, a computationally intensive but optimal joint decoding approach, and their low-complexity mismatched approach by separately decoding each strand first and multiplying the APPs before handing it to an outer decoder.

With slight abuse of notation, hereafter, \pmfconst\ denotes a normalizing factor such that the corresponding computation generates a valid probability mass function when using mismatched decoding laws. 

\subsection{Symbolwise MAP Decoding of a DNA Read}
\label{sse:inner-decoding}

In this subsection, we consider the case of decoding a single output forward read $\bfy \in \chYfo$ of length $\chL'$ given an input $\bfv$ of length $\kix+\bleno+2\primL$, which includes the information words of the primer sequences, the index, and the data, using the BCJR decoder \cite{BCJR_1974} over the trellis combination of the \kmerChfo\ channel with the joint-code trellis. Processing the backward reads follows similarly by reverse-complementing the chain of codebooks of the joint-code trellis and combining it afterward with the backward channel trellis. We drop the superscript of the sequence $\bfv$, i.e., its index information, due to the permutation effect of the channel. We follow the basic concept for decoding synchronization errors introduced in \cite{davey_reliable_2001} and refined in \cite{briffa_improved_2010} for the i.i.d. IDS channel. Approaches that already adapted this concept for slightly different \kmerCh\ channels have been independently proposed in \cite{IssamEirikAlex_AIRkmer_23} and \cite{frenchKmer_belaid_2023}. 

The goal of the decoder is to perform symbolwise estimation of the input vector $\bfv$ given the channel observation $\bfy$. The APP of a symbol $v_t$ given $\bfy$ is denoted by
\begin{align*}
 p(v_t|\bfy) = \frac{p(v_t,\bfy)}{p(\bfy)}\,,
\end{align*}
where $p(v_t,\bfy)$ is normally obtained by demarginalizing over the joint-code states corresponding to $v_t$. However, insertion and deletion events introduce a synchronization loss that destroys the Markov property of the joint-code trellis, meaning there is no straightforward factorization of $p(v_t,\bfy)$. By introducing a hidden state variable, the so-called \textit{drift}, one can circumvent the problem and restore the Markov property. Hence, the combined channel and joint-code model can be seen as a hidden Markov model (HMM). The drift state $d_t$ is defined as the number of insertions minus the number of deletions that occurred until time $t$. We define the joint state $\sigma_t = (\kmer_t, e_t, d_t)$ of the combined channel and joint-code trellis. Let $\bfx$ be the transmit sequence corresponding to $\bfy$. Denote $\bfx_a^b = (x_a,x_{a+1},\dots, x_b)$ and let $l_t$ denote the number of code symbols at trellis section $t$ and $\chL_{t-1} = \sum _{i=1}^{t-1} l_i$. A transition from time $t$ to $t+1$ corresponds to a transmission of code symbols $\bfx_{\chL_{t-1}+1}^{\chL_{t}}$ corresponding to symbol $v_t$ over the \kmerChfo\ channel. Using the drift variable, the HMM models the emission of $l_t +d_{t+1}-d_t$ symbols depending only on the next state $\sigma_{t+1}$ and the current state $\sigma_t$. Note that within this transmission, there could be intermediate channel states not included in the combined joint-code and channel trellis since it is possible that $l_t \geq 1$. Moreover, due to \kmer\ and event dependency, there is a memory effect of up to $k$ symbols and of one previous channel event. Thus, we say that the HMM, emitting the state sequence $\sigma_0, \dots, \sigma_{\bleno+\kix+2\primL}$, has memory $k$.

Overall, the joint probability $p(\bfy,v_t)$ can be computed by marginalizing the trellis states that contain an outgoing edge labeled with symbol $v_t$ as
\begin{align*}
    p(\bfy,v_t) = \sum_{\sigma,\sigma'} p(\bfy,v_t,\sigma,\sigma')\,,
\end{align*}
where $\sigma$ and $\sigma'$ denote realizations of the RVs $\sigma_t$ and $\sigma_{t+1}$, respectively. Due to the Markov property of the HMM, we can expand the joint probability to
\begin{align*}
	&p(\bfy,v_t,\sigma,\sigma') = \\
	& \underbrace{p\left(\bfy_{1}^{\chL_{t-1}+d}, \sigma \right)}_{\triangleq \alpha_{t-1}(\sigma)} 
 \underbrace{\left( \bfy_{{\chL_{t-1}+d+1}}^{\chL_{t}+d'}, v_t ,\sigma'\big| \sigma \right)}_{\triangleq \gamma _{t}(\sigma,\sigma',v_t)}
	\underbrace{p\left(\bfy_{{\chL_{t}+d'+1}}^{\lseq^\prime}\Big| \sigma'\right)}_{\triangleq \beta_{t}(\sigma')} \\
	&= \alpha_{t-1}(\sigma) \cdot \gamma _{t}(\sigma,\sigma',v_t) \cdot \beta_{t}(\sigma')\, ,
\end{align*}
where $\alpha_{t-1}(\sigma)$ denotes the forward metric, $\gamma _{t}(\sigma,\sigma',v_t)$ the branch metric, and $\beta_{t}(\sigma')$ the backward metric. In line with these derivations, we can deduce the forward and backward recursions as
\begin{align*}
    \alpha_t(\sigma') &= \sum_{\sigma} \sum_{v_t}\alpha_{t-1}(\sigma) \gamma_{t}(\sigma, \sigma',v_t)\, ,  \\
    \beta_{t-1}(\sigma) &= \sum_{\sigma'}  \sum_{v_t} \gamma_{t}(\sigma, \sigma',v_t) \beta_{t}(\sigma')\, ,
\end{align*}
with starting conditions
\begin{align*}
    \alpha_0(\sigma) &= \begin{cases}
		\frac{1}{4} & \text{if } \sigma = (x,\mathtt{Beg}, 0)\\
		0& \text{otherwise}\, ,
	\end{cases}
	\\
	\beta_{{\bleno+\kix+2\primL}}(\sigma) &= \begin{cases}
		\frac{1}{12} & \text{if } \sigma = (x, e, \chLout-\chL)\\
		0& \text{otherwise}\, ,
	\end{cases}
\end{align*}
for any $x \in \alphDNA$ and $e \in \{\Del, \Sub, \Tra\}$, since we cannot have a final state $\Ins$. The branch metric 
\begin{align*}
\gamma _{t}(\sigma,\sigma',v_t) \propto p(v_t) p \left( \bfy_{{\chL_{t-1}+d+1}}^{\chL_{t}+d'}, \sigma'\big|\sigma ,v_t \right)
\end{align*}
is computed by a three-dimensional expansion of the lattice structure introduced in~\cite{bahl_decoding_1975}, similar to~\cite{frenchKmer_belaid_2023}. As described before, the decoding trellis of the joint-code and the channel only considers channel states at time transitions $t$. Hence, the following lattice can be interpreted as an intermediate forward computation over multiple channel states for the same $v_t$. We define the lattice $\bfF$ of dimensions $(l_t + 1) \times (l_t +d_{t+1}-d_t + 1) \times 5$, where the first dimension corresponds to the length (plus one) of the codeword $\dot{\bfx} = (\dot{x}_1,\ldots, \dot{x}_{l_t}) = \bfx_{\chL_{t-1}+1}^{\chL_{t}}$ associated with the symbol $v_t$, the second dimension to the given window $\dot{\bfy} = (\dot{y}_1,\ldots, \dot{y}_{l_t +d_{t+1}-d_t}) = \bfy_{{\chL_{t-1}+d_t+1}}^{\chL_{t}+d_{t+1}}$ according to the considered drift changes (plus one), and the third dimension to the possible channel event states $\evspac = \{\tt{Beg},\Ins, \Del, \Sub, \Tra\}$. The lattice $\bfF$ is initialized as $F_{\iota,\mu,\epsilon} = 0$ for all possible $\iota, \mu, \epsilon$, except we set $F_{0,0,e}=1$ according to the current event state $e$. 
Note that the possible choices of current state $\kmer$ and future state $\kmer'$ depend partly on the choice of $\dot{\bfx}$, or vice versa. 
For increasing $\iota$ and $\mu$, we have the following iterative update rules,
\begin{align*}
    F_{\iota,\mu,\Ins} &= \sum_{\epsilon \in \evspac} \frac{1}{4} \cdot p(\Ins | \kmer_\iota, \epsilon) \cdot F_{\iota,\mu-1,\epsilon} \\ %
    &0 \leq \iota \leq l_t \text{ and } 1 \leq \mu \leq l_t+d_{t+1}-d_t\,,  \\
    F_{\iota,\mu,\Del} &= \sum_{\epsilon \in \evspac} p(\Del | \kmer_\iota, \epsilon) \cdot F_{\iota-1,\mu,\epsilon} \\ %
    &1 \leq \iota \leq l_t \text{ and } 0 \leq \mu \leq l_t+d_{t+1}-d_t\,,  \\
    F_{\iota,\mu,\Sub} &= \sum_{\epsilon \in \evspac} p(\Sub | \kmer_\iota, \epsilon) \!\cdot\! p(\dot{y}_{\mu-1}|\dot{x}_{\iota-1}, \Sub) \cdot F_{\iota-1,\mu-1,\epsilon} \\ %
    &1 \leq \iota \leq l_t \text{ and } 1 \leq \mu \leq l_t+d_{t+1}-d_t\,,  \\
    F_{\iota,\mu,\Tra} &= \sum_{\epsilon \in \evspac} p(\Tra | \kmer_\iota, \epsilon) \cdot F_{\iota-1,\mu-1,\epsilon} \\ %
    &1 \leq \iota \leq l_t \text{ and } 1 \leq \mu \leq l_t+d_{t+1}-d_t\,.  
\end{align*}
As a result, we have that
\begin{align*}
    p \left( \bfy_{{\chL_{t-1}+d+1}}^{\chL_t+d'}, \sigma'\big|\sigma ,v_t \right) = F_{l_t,l_t+d_{t+1}-d_t,e'}\, , 
\end{align*}
where $e'$ is the final event after the transmission of $\dot{\bfx}$ of the information symbol $v_t$. Note that $e'$ cannot be the event $\Ins$ since we model the insertions always before a time change. Further, we consider a substitution matrix but model the insertions as random symbols. 

For complexity reasons, in each trellis section $t$, we limit the maximum and minimum considered drift around its expected value  by some multiple of its total variance according to \cite{davey_reliable_2001},
\begin{align*}
    \chL_t \cdot \frac{\pI-\pD}{1-\pI} \pm 3.5 \cdot \sqrt{\chL \cdot \frac{(1-\pD)(\pD+\pI)}{(1-\pI)^2}}\,,
\end{align*}
where $\pI$ and $\pD$ are the overall average insertion and deletion probabilities. Further, the decoder considers at most $I_\mathrm{max}$ consecutive insertions per channel state transition. Observe that we could have decided to combine multiple symbols/codebooks to possibly further enhance the decoding performance \cite{briffa_improved_2010, buttigieg_improved_2015}, but the decoding complexity is dependent on the cardinality of the created codebook and additionally grows quadratically in the length of the trellis, which renders decoding infeasible for larger trellises.

During decoding, the primer sequences $\primer_1$ and $\primer_2$ are known. We use them to create `soft' boundaries for the strand's index and payload parts. By incorporating the primers into the trellis, we avoid the negative effects of making a hard decision on primer boundaries.

\subsection{Reads Clustering}
\label{sse:clustering}

Typical approaches involve clustering the received strands before performing error-correction decoding. These methods usually use approximation techniques for the Levenshtein distance (minimum number of IDS operations required to change one sequence into another), e.g., $q$-gram distance \cite{antkowiak_low_2020, rashtchian_clustering_2017} to reduce the distance computation cost. We present a clustering approach with complexity $\cO(\chN^2 \lseq)$, which computes pairwise distances in $\cO(\lseq)$ based on the resynchronized soft information the strand-by-strand decoding provides, which we call \emph{payload} distance. We use the aggregated statistical distance between the two APP outputs as a distance metric between two reads $\bfy_{j_1}$ and $\bfy_{j_2}$. In particular, we define
\begin{align*}
    \distClus(\bfy_{j_1}, \bfy_{j_2}) = \sum_{t=\primL+1}^{\kix+\bleno+\primL} \frac{1}{2} \sum _{v_t} \left\vert p(v_t | \bfy_{j_1}) - p(v_t |  \bfy_{j_2}) \right\vert \,, %
\end{align*}
where the second sum is over all possible symbols at time $t$. Two reads are considered to stem from the same DNA input strand if their pairwise distance is lower than a threshold $\threshClus$. For a given setup, this threshold is optimized such that the probability of clustering two reads originating from different DNA input strands is at most $10^{-6}$ based on simulations.

In particular, we follow a hierarchical clustering approach using the single-linkage method. Hence, we start by computing the pairwise distances of the reads and considering each read a single cluster. We iteratively merge clusters with the minimum current cluster distance, where the cluster distance is defined as the minimum distance of any two reads of the two clusters. We stop merging whenever no cluster distance is below the threshold $\threshClus$.

Let the $r$-th cluster, $1 \leq r \leq \chM'$, be formed as $\outC_r \triangleq (\bfy_{r,1}, \dots, \bfy_{r,|\outC_r|})$, where $\bfy_{r,j}$ is the $j$-th strand placed in cluster $\outC_r$ and $|\outC_r|\neq 0$. 
We denote the overall clustering output as the list $\outC = (\outC_1,\dots,\outC_{\chM'})$,  with $1 \leq \chM' \leq \chN $ and $\sum_{r=1}^{\chM'} |\outC_r| = \chN$.

In general, this or other reads clustering approaches enable the system to take advantage of the multi-copy gain introduced by the channel to improve the subsequent step of index recovery of the received strands. The disadvantage of most reads clustering approaches is the associated computing cost for large $\chM$ and $\chN$. We have also considered trading complexity against clustering accuracy similar to \cite{rashtchian_clustering_2017}, where one clusters iteratively random small subsets of reads. Due to either the impracticability for large $\chN$ or non-significant performance gains in the latter for a fixed coverage, we emphasize that this clustering step is optional (there are good performance gains for small $\chN$).

\subsection{Index Decoding} \label{sse:index-decoding}

The index code and recovering the index information of each strand/cluster at the receiver side are the crucial points for tackling the permutation effect and leveraging the multi-copy gain of the channel. 

For all strands $\bfy_{r,j}$ of each cluster $\outC_r$, we obtain the soft information of the indices by extracting
\begin{align*}
  &\bfp\left(\bfii{i}|\bfy_{r,j}\right) \\ 
  & \triangleq \left(p(v_{\primL+1}|\bfy_{r,j}),{\tiny \dots},p(v_{\primL+(\kix/2)}|\bfy_{r,j}), \right.  \\
  & \hspace{6ex} \left. p(v_{\primL+\bleno+(\kix/2)+1}|\bfy_{r,j}),\dots, p(_{\primL+\bleno+\kix}|\bfy_{r,j})\right)\,   
\end{align*}
from the respective strand APPs $p(v_t|\bfy_{r,j})$. We generate mismatched index APPs $\bfq(\bfii{i}|\outC_r)$ for a cluster $\outC_r$ symbolwise. For any $1 \leq \kappa \leq \kix$, we compute
\begin{align*}
    q\left(\ii{i}_\kappa|\outC_r \right) = \pmfconst \cdot \left[ \prod _{j=1} ^{\lvert \outC_r \rvert} p(\ii{i}_\kappa|\bfy_{r,j})\right]^{s(\lvert \outC_r \rvert)}\, ,
\end{align*}
where $s(\cdot) \geq 0$ is a scaling parameter inspired by the theory of generalized mutual information analysis that allows to mitigate overconfidence/underconfidence bias of the mismatched rule and is, in our scenario, dependent on the number of combined traces (see, e.g., \cite{kaplan1993information,ARXIV_Srinivasavaradhan2021TrellisBMA}). 

For all indices $1 \leq i \leq M$, we compute the mismatched APP $q(i|\outC_r)$ using the memoryless mismatched decoding rule, i.e., assuming independence between the index APPs, as 
\begin{align*}
    q(i|\outC_r) = \pmfconst \cdot \prod _{\kappa=1} ^{\kix} q(\ii{i}_\kappa | \outC_r)\,, 
\end{align*}
 inspired by maximum likelihood decoding for memoryless channels. This method naturally provides confidence levels for each cluster belonging to specific indices (if no clustering is performed for each read). We assign all reads of a cluster $\outC_r$ to index $\hat{i}$ only if the maximum confidence is higher than a threshold $\threshIx$ as 
\begin{align*}
    \hat{i}(\outC_r) = \begin{cases} \displaystyle{\argmax_{1\leq i \leq M}}\ q(i|\outC_r) & {\text{ if $\displaystyle{\max_{1\leq i \leq M}}\ q(i|\outC_r) \geq \threshIx$}}  \\
    \mathrm{None} &\text{ otherwise}\,.
    \end{cases}
\end{align*}
Introducing the index reliability threshold $\threshIx$ reduces the negative effect of misassigning reads/clusters to the wrong index. This stems from the idea that read erasures are easier for the outer decoder to handle than wrongly assigned reads, especially in low-coverage scenarios. The threshold $\threshIx$ is optimized to yield the highest average mutual information rate (MIR) (see \cref{sec:codeopt-index}). This strategy can be compared to adding error-detection redundancy like a cyclic-redundancy-check to the index or the whole DNA strand, as employed, e.g., in \cite{stanford_magneticDNAstorage_2023}, but by leveraging the soft-output decoder instead of additional redundancy. 

Rather than discarding the non-assigned reads, we can store them in the set
\begin{align*}
    \poprRe = \left\{ \bfy \in \outC_r : {\displaystyle{\max_{1\leq i \leq M}}\ q(i|\outC_r) < \threshIx}, \ \forall\,  1 \leq r \leq \chM' \right\}
\end{align*}
for post-processing as described in \cref{sec:postclust}. 

We remark that computing $q(i|\outC_r)$ for any $i \in \chM$ is computationally expensive and only feasible for small enough $\chM$ (e.g., we consider $\chM = 30\,589$ feasible, see \cref{sec:simresults}). In our previous work \cite{multidrawConCat_ITW_2023}, we performed hard decisions on the symbolwise index APPs $\bfq(\bfii{i}|\outC_r)$. This method is less expensive but less accurate. Moreover, it does not allow for the use of the confidence threshold $\threshIx$ and the later presented post-processing.

Let $\outS_i$ be the set of reads assigned to index $i$, i.e., 
\begin{align*}
    \outS_i = \left\{ \bfy \in \outC_r : i = \hat{i}(\outC_r), \ \forall\,  1 \leq r \leq \chM' \right\}\,, 
\end{align*}
 where $0 \leq |\outS_i| \leq \chN$ and $1 \leq i \leq \chM$. For all data positions of the $i$-th block, i.e., for symbols of $\bfwi{i}$, we compute the mismatched APPs according to 
\begin{align*}
    q(w^{(i)}_t|\chY) = \pmfconst \cdot \left[ \prod _{\bfy_j \in \outS_i} p(w^{(i)}_t|\bfy_{j}) \right] ^{s\left( \lvert \outS_i \rvert \right)}\,,
\end{align*}
where again $s(\cdot)$ is the scaling parameter for mitigating underconfidence/overconfidence bias and depends on the number of combined traces. Further, we set $q(w^{(i)}_t|\chY) = \frac{1}{\outq}$ (or equally $s(0) = 0$) when $\outS_i = \emptyset$.

\subsection{Post-Assignment of Reads}\label{sec:postclust}

Assigning the reads according to the maximum soft information of the decoded index is a practical way of grouping the reads. Further, filtering according to the reliability of the actual decision using the index threshold $\threshIx$ can be considered a straightforward way of tackling misassignments. However, this method solely focuses on the index part of the strand and hence has the drawback of throwing away reads with possible bad index region quality but good payload quality. 

The general idea of the post-assignment step is to assign the so-far-ignored reads in $\poprRe$ using the \textit{payload} distance to the already-grouped reads. In that way, we can consider the index decoding step as a sort of pre-grouping and use the computed soft information for each block $\bfwi{i}$ as the group representative for additional read assignments. To further reduce computational complexity, we fix the performed comparison to the top index candidates $\poprTop_{\bfy}$ according to the already computed probabilities $q(i|\bfy)$ by the index decoder for each read $\bfy \in \poprRe$, where $\lvert \poprTop_{\bfy} \rvert$ is constant with respect to $\chN$. 

In more detail, recall that $\bfvi{i}$ is the vector associated with block $\bfwi{i}$ plus the corresponding index and primer symbols. For $\bfy \in \poprRe$, we use the distance measure $\distPost(\bfwi{i}, \bfy)$ which is defined similarly as the clustering distance as
\begin{align*}
    \distPost(\bfwi{i}, \bfy) = \sum_{t=\primL+1}^{\kix+\bleno+\primL} \frac{1}{2} \sum _{v^{(i)}_t} \left\vert q\left(v^{(i)}_t | \widetilde{\chY}\right) - q\left(v^{(i)}_t |  \bfy\right) \right\vert \,, %
\end{align*}
where the computed APPs of the pre-grouped reads $\widetilde{\chY}$, i.e., all reads $\bfy \notin \poprRe,$ are denoted as $q\left(v^{(i)}_t | \widetilde{\chY}\right)$. Further, in the following, we formally define the top candidate list $\poprTop_{\bfy}$ for any $\bfy \in \poprRe$. Denote as $\bfT_\bfy = [i_1, \ldots, i_\chM]$ the list of indices sorted according to their soft information, i.e., for all $i_i < i_j$ it holds $q(i_i|\bfy) > q(i_j|\bfy)$. Then, the top candidate list $\poprTop_{\bfy}$ is defined as the first $\lvert \poprTop_{\bfy} \rvert$ elements of $\bfT_\bfy$. The final index assignment of the read $\bfy \in \poprRe$ is then performed as 
\begin{align*}
    \hat{i}(\bfy) = \begin{cases}
        i &\text{ if } \distPost(\bfwi{i}, \bfy) < \threshClus \\
        &\text{ for any } i \in \poprTop_{\bfy} \text{ and } \outS_i \neq \emptyset \\
        \mathrm{None} &\text{ otherwise}\,.
    \end{cases}
\end{align*}
If several candidates fulfill the distance requirement, we select the index $\hat{i}(\bfy)$ with the highest $q(\hat{i}|\bfy)$.

For a specific block index $i$, let $\outW_i$ be the union of the pre-assigned reads $\outS_i$ and the assigned reads in the post-assignment step. We compute the final mismatched APPs for the block $\bfwi{i}$ as
\begin{align*}
    q(w^{(i)}_t|\chY) = \pmfconst \cdot \left[ \prod _{\bfy_j \in \outW_i} p(w^{(i)}_t|\bfy_{j}) \right] ^{s\left( \lvert \outW_i \rvert \right)}\,.
\end{align*}

Note that by this method, solely pre-assigned groups get additional reads assigned. Hence, the primary benefit of this method stems from assigning reads to groups with a small number of pre-assigned reads. On the other hand, already incorrectly pre-assigned groups will be pushed further in the wrong direction. During optimization of the parameter $\threshIx$ (see \cref{sec:codeopt-index}), we have observed that the choice tends to thresholds close to $1$. This directly helps minimize the number of incorrect assignments, hence also attempting to minimize the aforementioned `avalanche' effect.

Further, the cardinality of the top candidate list $\poprTop_{\bfy}$ is a sensitive design parameter that not only controls the complexity of the post-processing but also influences the performance of the method, again in an ambiguous way. Increasing the cardinality does increase the probability that the correct index candidate is in the list, but it also increases the probability of misassignments. We also remark that we have attempted to redo the index decision based on the newly assigned reads, but it does not seem to noticeably affect the system performance. Ultimately, some reads remain unassigned since no suitable representative might be found, e.g., due to empty clusters/no pre-assgined reads or structural errors in the index region. One might try to perform a version of iterative decoding between the outer code and this post-processing method to create representative APP vectors of otherwise-absent blocks $\bfwi{i}$. 

\subsection{Outer Decoding} \label{sec:dec-outer}

We consider an outer decoder that ignores possible correlations between the symbolwise estimates $q(w_t|\chY)$ given by the joint-code MAP decoder. Hence, we apply the mismatched outer decoding metric
\begin{align*}
    q(\bfw | \chY ) = \prod _{i=1}^{\chM} \prod _{t=1} ^{\bleno} q(w_t^{(i)}|\chY)\,
\end{align*}
to finally output an estimate $\hat{\bfu}$ of our input data. Note that the outer decoder is aware of the applied random interleaving at the encoder's stage; hence, the decoder can reverse the operation.

We have summarized the decoding flow in \cref{alg:decoder} for the reader's convenience.

\SetKwComment{Comment}{/* }{ */}
\begin{algorithm}
\caption{Decoding of DNA reads}\label{alg:decoder}
\SetKwFunction{ssdec}{BCJR}
\SetKwData{appy}{$p(\bfv|\bfy_j)$}
\SetKwFunction{cluster}{clusteringAPP}
\SetKwData{clus}{$\outC_1,\ldots,\outC_{\chM'}$}
\SetKwFunction{excind}{extractInd}
\SetKwFunction{excdata}{multiplyData}
\SetKwData{ix}{$\hat{i}$}
\SetKwFunction{outdec}{OuterDecoder}
\SetKwData{infhat}{$\hat{\bfu}$}
\SetKwInOut{Input}{Input}\SetKwInOut{Output}{Output}
\Input{reads $\chY = \left[\chYfo,\chYba \right]$}
\Output{$\hat{\bfu}$}
\Comment{Single strand decoding}
\Comment{(see Sec.~\ref{sse:inner-decoding})}
\For{$\bfy_j \in \left[\chYfo,\chYba \right]$}{
    \If{forward}{
    \appy $\leftarrow$ \ssdec{$\bfy_j$,\kmerChfo}\;
    }
    \Else{
    \appy $\leftarrow$ \ssdec{$\bfy_j$,\kmerChba}\;
    }
}
\Comment{Opt.\ clustering of reads}
\Comment{(see Sec.~\ref{sse:clustering})}
\clus $\leftarrow$ \cluster{$p(\bfv|\bfy_1),\ldots,p(\bfv|\bfy_\chN)$}\;
\Comment{Index decision}
\Comment{(see Sec.~\ref{sse:index-decoding})}
\For{$r \leftarrow 1$ \KwTo $\chM'$}{
    \ix $\leftarrow$ \excind{$q(\bfv|\outC_r)$}\;
    \Comment{Opt.\ index threshold decoding}
    \If{$\displaystyle{\max_{1\leq i \leq M}}\ q(i|\outC_r) \geq \threshIx$}{
    \Comment{Assign data to outer block}
    $q(\bfwi{\hat{i}}|\bfY)$ $\leftarrow$ \excdata{$q(\bfv|\outC_r)$};
    }
    \Else{
    \Comment{Save for post-assignment}
    $\poprRe$ $\leftarrow$ $\poprRe \cup \{\bfy : \bfy \in \outC_r\}$\;
    }
}
\Comment{Opt.\ post-assignment}
\Comment{(see Sec.~\ref{sec:postclust})}
\For{$\bfy \in \poprRe$}{
    \For{$i \in \poprTop_\bfy$}{
        \Comment{Compare payload dist.}
        \If{$\distPost(\bfwi{i},\bfy) < \threshClus$}{
        \Comment{Assign data to outer block}
        $q(\bfwi{\hat{i}}|\bfY)$ $\leftarrow$ \excdata{$p(\bfv|\bfy)$};}
        break\;
    }
}
\Comment{Outer decoding}
\Comment{(see Sec.~\ref{sec:dec-outer})}
\infhat $\leftarrow$ \outdec{$q(\bfwi{1}|\bfY),\ldots,q(\bfwi{\chM}|\bfY)$}\;

\end{algorithm}

\section{Experimental Setup}\label{sec:dataset}

The stored dataset consists of three files labeled \texttt{file1}, \texttt{file2}, and \texttt{file3}, comprised each of ${\sim}30\,600$ DNA strands. Each strand has a length of $150\, \text{nt}$, of which $110$ are dedicated to carrying data, and $40$ are reserved for two file-specific $20\, \text{nt}$ primer adapter sequences at both ends of the strand. The primer sequences were chosen among those designed by Organick \emph{et al.} in~\cite{organick_random_2018}. The $110\,\text{nt}$ data sequences were drawn uniformly at random and checked for collisions with the primers following~\cite{organick_random_2018}. The resulting ${\sim}92\,000$ strands were synthesized by \textit{GenScript} and stored jointly in a DNA pool. The pool was delivered in a volume of $80$ \textmu L at $17.5$ ng/\textmu L. (The synthesized pool contains a large number of physical copies of each strand.)

To retrieve a file, an aliquot of the pool is amplified via primer-targeted PCR following the amplification protocol in~\cite{organick_random_2018} and sequenced on a MinION Mk1C nanopore sequencing device produced by \textit{Oxford Nanopore Technologies} with Kit 14 chemistry and the corresponding standard Ligation Sequencing protocol. The resulting raw current signal readouts are basecalled (i.e., converted to sequences of nucleotides with associated quality scores) with both a fast and a high-accuracy basecaller available on the MinION device based on the \emph{Guppy} basecaller. We refer to the two different processes of basecalling as \emph{fast basecalling} (\fastbc) and \emph{accurate basecalling} (\accbc). The main difference is that \fastbc\ can be performed online (i.e., concurrently and keeping pace with the sequencing process) at the cost of higher error rates in the basecalled sequences, whereas \accbc\ is only available offline. In both cases, a quality score threshold parameter of $8$ (the default value for fast basecalling) is specified to split the reads into two groups\textemdash those that pass the quality score threshold and those that do not. To extract strand-specific parts of the basecalled reads, we align the reads against the six primer sequences used in our dataset (two primer sequences per file) using BLAST~\cite{BLAST} and crop the segments that correspond to adjacent alignments of correct primer pairs separated by $150\pm15\,\text{nt}$. It is important to note that PCR amplification yields double-stranded DNA, and subsequent nanopore sequencing produces readouts of a strand in both its forward and reverse-complemented (or backward) form. We separate the two during segmentation by keeping track of the orientation of BLAST-generated alignments. In summary, for each file, we split the reads according to \accbc\ versus \fastbc, passed versus failed quality score, and forward versus backward orientation, resulting in eight groups of reads per file. The number of segments in each group is shown in~\cref{tab:dataset}. 
\begin{table*}[t]
\begin{center}
\caption{Dataset Specifications$^a$ }
\label{tab:dataset}
\vspace{-2.3ex}
\begin{tabular}{c c c c c c c c c c }
    \toprule
    File & Num. stored strands  & \multicolumn{4}{c}{Accurate basecalling (\accbc)} & \multicolumn{4}{c}{Fast basecalling (\fastbc)} \\ 
                &     & \multicolumn{2}{c}{passed Q-score}       & \multicolumn{2}{c}{failed Q-score} & \multicolumn{2}{c}{passed Q-score} & \multicolumn{2}{c}{failed Q-score} \\
     & & forward & backward & forward & backward & forward & backward & forward & backward   \\[1ex] \hline \\[-0.5ex]
    \texttt{file1}     & $30\,589$ & $244\,510$ & $193\,079$ & $44\,411$ & $32\,826$ & $166\,713$ & $96\,385$ & $63\,160$ & $39\,254$ \\[0.75ex]
    \texttt{file2}$^b$ & $30\,589$ & $494\,026$ & $602\,463$ & $19\,658$ & $25\,987$ &   $242\,106$    &   $391\,019$   &   $113\,530$  &  148\,695 \\[0.75ex]
    \texttt{file3}$^c$ & $30\,588$ & $432\,501$ & $470\,052$ & $73\,106$ & $51\,346$ & $641\,832$ & $613\,096$ & $114\,602$ & $105\,714$ \\[0.5ex]
    \bottomrule
\end{tabular}
\end{center}
\footnotesize{$^a$Each stored strand consists of a $110\,\text{nt}$ random data part and on each side file-specific $20\,\text{nt}$ primers.}\\
\footnotesize{$^b$We used a smaller and cheaper flongle flow cell for nanopore sequencing; a larger number ($24$ versus $18$) of PCR cycles was performed.} \\
\footnotesize{$^c$We obtained a total $42$ million reads for \texttt{file3} but only clustered a subsample of it.}
\end{table*}

We remark that merging forward and backward reads by reverse-complementing the backward reads is a mistake if the channel model is not invariant under this operation; in particular, doing so for the \kmerCh\ channel model we describe in Section~\ref{sec:kmer-model} would erase the asymmetries in the substitution matrix and possible directional memory effects, whereas the i.i.d. IDS channel model is invariant under such transformation.

To estimate the parameters of the channel model, we need a set of input-output pairs, which in turn means that we need to assign one of the originally stored strands as channel input to each read. To do so, we calculated the Levenshtein distance between each read and each stored strand and assigned to each read the input strand that is closest to it.

In particular, a DNA storage dataset $\mathcal{D}$ contains $\chM^\mathcal{D}$ clusters, where each cluster consists of an input strand $\bfx^{(i)}$ of length $\chL$ and its corresponding, possibly empty, set of output strands $\bfy^{(i)}_1, \dots, \bfy^{(i)}_{\chM^{\mathcal{D}}_i}$. 

In the following, we show how to use the dataset for the channel estimation on which our scheme relies (\cref{sec:kmermodelestimation}). Moreover, we show how we use the dataset to obtain desired DNA storage channel parameters (\cref{sec:datasetsim}). In \cref{sec:datasetsim}, we present a method to conduct Monte-Carlo simulations on experimental data for coding schemes that use pseudo randomization of strands \cite{Srinivasavaradhan2021TrellisBMA}.

\subsection{Channel Estimation} \label{sec:model-estimation}

In this subsection, we describe the estimation of the data-dependent parameters for our approach and analysis.

\subsubsection{KMER Channel Transition Probability Estimation} \label{sec:kmermodelestimation}

The transition probabilities of the channel model are estimated using a DNA storage dataset containing labeled input (before synthesis) strands and the corresponding collection of output (after nanopore sequencing) strands split into forward and backward reads. We estimate a forward model \kmerChfo\ from the set of mapped input and forward read pairs $\datasetFo$ and a backward model \kmerChba\ from the set of mapped input and backward read pairs $\datasetBa$. In the following, we describe the estimation of the forward model since the backward model follows similarly, taking the reverse-complement effect into account.

For each pair $(\bfx^{(i)}, \bfy^{(i)}_j)$, we compute the edit distance matrix via the Wagner-Fischer algorithm \cite{wagner1974string}. By backtracking from the end and always choosing the minimum value (for ties deciding randomly), we can determine a chain of events $\bfe^{(i,j)}$ of length $L_e \geq \chL$, $e^{(i,j)}_l \in \evspac = \{\Ins, \Del, \Sub, \Tra\}$, that describes a possible outcome of how the strand $\bfx^{(i)}$ is edited to form $\bfy^{(i)}_j$. 
To each event, we add the information of the corresponding $\kmer$ %
and the preceding event $e^\prime \in \evspac$ to form a chain of $3$-tuples with entries $(e^{(i,j)}_l, \kmer^{(i,j)}_l, e^{(i,j)}_{l-1})$ and $e^{(i,j)}_0 = \tt{Beg}$. Note that by slight abuse of notation, we call here a \kmer\ also a substring of $\bfx^{(i)}$ shorter than $k$, since for $k > 1$, the actual $k'\text{mers}$ at the beginning and end of $\bfx^{(i)}$ are fitted.

Subsequently, we can estimate the transition probabilities $p( e | \kmer, e^\prime)$ by a simple counting argument. For a pair $(\bfx^{(i)}, \bfy^{(i)}_j)$, we define 
\begin{align*}
    S_{e,e',\kmer}^{(i,j)}= \Big\lvert \big\{ l :\, (e, \kmer, e') = (e_l^{(i,j)}, \kmer_l^{(i,j)}, e_{l-1}^{(i,j)}) \big\} \Big\rvert\,.
\end{align*}
Then, $p(e=\epsilon|\kmer = \boldsymbol{\kappa}, e'=\epsilon ')$ is estimated by 
\begin{align*}
\frac{\sum_{i=1}^{|\datasetFo|}\sum_{j=1}^{\chM^{\datasetFo}_i} S_{\epsilon,\epsilon',\boldsymbol{\kappa}}^{(i,j)} }{\sum_{e \in \{\Ins,\Del,\Sub,\Tra \}} \sum_{i=1}^{|\datasetFo|}\sum_{j=1}^{\chM^{\datasetFo}_i} S_{e,\epsilon',\boldsymbol{\kappa} }^{(i,j)}}\,.
\end{align*}
We remark that instead of the counting-based parameter estimation described above we have also tried using the Baum-Welch algorithm \cite{Baum1970AMT} for HMMs using our single-strand BCJR decoder (presented in \cref{sse:inner-decoding}) on a slightly different trellis allowing state transition at the same time $t$ (similar to the expansion of the trellis in \cite{Srinivasavaradhan2021TrellisBMA}). However, we have not observed any performance gain compared to the above counting method. Nonetheless, the Baum-Welch algorithm might become useful if the error rates are roughly known and the transition probabilities need to be estimated using only a small dataset (e.g., stored pilot strands). This is an interesting problem that would need further investigation and falls outside the scope of this paper.

We list the average IDS error rates of our dataset in \cref{tab:data-error-rates}.

\begin{table}[t]
\begin{center}
\caption{Average Error Rates for \texttt{File3} (Similar Across All Files) }
\label{tab:data-error-rates}
\vspace{-2.3ex}
\begin{tabular}{c c c c c c c c}
    \toprule
    Q-score & \multicolumn{3}{c}{\accbc$^a$ } & \vphantom{0} &\multicolumn{3}{c}{\fastbc$^a$ } \\ 
    & $\pI$  & $\pD$ & $\pS$ & \vphantom{0} & $\pI$  & $\pD$ & $\pS$   \\[0.75ex] \midrule
    passed & $0.009$ & $0.014$ & $0.020$ & \vphantom{0} & $0.014$ & $0.038$ & $0.043$ \\[0.75ex]
    failed & $0.037$ & $0.052$ & $0.094$ & \vphantom{0} & $0.023$ & $0.062$ & $0.080$ \\[0.1ex]
    \bottomrule
\end{tabular}
\end{center}
\footnotesize{$^a$We denote the average insertion, deletion, and substitution probability as $\pI$, $\pD$, and $\pS$, respectively.} 
\end{table}

\subsubsection{Draw Distribution Estimation}\label{sec:drawdist-est}

For each input strand $\bfx^{(i)}$, we calculate its drawing probability $p_i^\mathrm{d}$ by simply dividing the number of corresponding output strands by the total number of strands, i.e.,
\begin{align*}
    p_i^\mathrm{d} = \frac{\chM^{\mathcal{D}}_i}{\lvert \cD \rvert}\,.
\end{align*}
Note that we can generate a drawing distribution for any smaller number of input DNA strands $\chM \leq \chM^{\mathcal{D}}$ by uniformly at random subsampling the clusters from the dataset. The probability of reverse-complementing, i.e., observing a backward read, is simply
\begin{align*}
    p^\mathrm{rc} = \frac{\lvert \cD^\mathrm{b} \rvert}{\lvert \cD \rvert}\,.
\end{align*}

\subsection{Using Datasets for Simulation}\label{sec:datasetsim}

Since our coding scheme uses a random offset, i.e., a scrambling sequence, we can execute performance evaluations on real data as described in \cite{ARXIV_Srinivasavaradhan2021TrellisBMA}. In the following, we revisit their method shortly. Let us describe a dataset as pairs of stored uniformly at random strands and corresponding reads. Let $\bfw$ be the vector encoded by the index/inner code, \textsc{Enc} the encoding function, and $\bfz$ the random offset. Then, a stored sequence $\bfx$ can be described by $\bfx = \textsc{Enc}(\bfw) + \bfz$, where the symbols of $\bfx$ are thus uniformly distributed. Hence, we can use a random dataset for evaluation by fitting a $\bfz = \bfx -\textsc{Enc}(\bfw)$ for every strand of data. Since we conduct Monte-Carlo analyses, averaging over multiple simulations, we conclude that a general fixed $\bfz$ exists that must perform as well or better by the pigeon-hole principle. Hence, the Monte-Carlo results on a random dataset are ensemble averages of the proposed system. We remark that the random offset is not applied to the primer region; hence, we are restricted in the choice of primers by the dataset.

\section{Optimizing the Component Codes}\label{sec:codeopt}

The design of our coding scheme setup is flexible for any channel parameter $\beta$, i.e., any $\chM$ and $\chL$, and \kmerCh\ channel noise. Nonetheless, its performance depends on how well the system parameters are optimized for the specific channel. In \cref{sec:model-estimation}, we have already discussed how to infer the \kmerCh\ transition probabilities (we demonstrate in \cref{sec:simresults} that these parameters generalize over multiple experimental runs). This section discusses the criteria for optimizing the specific codes according to given system parameters and presents performance metrics for our coding scheme overall as well as for its components.

We give a brief overview of the optimization procedure in the following, where we always optimize based on experimental data (see~\cref{sec:dataset}). First, we compute AIRs for multiple inner code rates $\Rin$ on the extended trace reconstruction channel (see~\cref{sec:trace-recon-model}) for different numbers of traces $\trchM$ for each input strand (see~\cref{sec:codeopt-inner}). Within this computation, we directly optimize the scaling parameter $s(\trchM)$, which mitigates the overconfidence/underconfidence bias of the mismatched decoding approach. Given the estimated drawing distribution and the information rates, we calculate an artificial read/write cost trade-off for each inner code rate $\Rin$, assuming no permutation effect of the channel and a rate-achieving outer code. We pick the inner code rate $\Rin$ that shows the best artificial read/write cost trade-off across multiple coverage levels that we target. Then, we switch to the sampling channel, focusing on conducting the optimization based on average MIRs for the outer code, disregarding the finite-length phenomena due to computational complexity constraints. We choose the index code that maximizes the MIRs for the targeted coverage levels. Note that we do not consider clustering of reads here due to complexity in the considered parameter regime (see \cref{sse:clustering}). Subsequently, we directly optimize the index threshold $\threshIx$ and the number of top candidates considered in post-processing $\lvert \poprTop_{\bfy}\rvert$ using the same method (see~\cref{sec:codeopt-index,sec:codeopt-postproc}). For the best setup, we verify the MIRs with an outage analysis. The outage probability is a prominent performance metric in similar communication scenarios, e.g., the block-fading channel, and serves as a lower bound on the system's FER. Based on the outage analysis results, we construct an optimized SC-LDPC code such that its rate $\Rout$ is close to the outage bound for a targeted FER (see~\cref{sec:codeopt-outer}).

For the sake of presentation, we do not distinguish between RVs and their realizations in the following.

\subsection{Performance Metrics}\label{sec:performance-metric}

This subsection presents the main performance metrics applied in our analysis: the \emph{read/write cost trade-off} and \emph{outage probability}.

\subsubsection{Read/Write Cost Trade-Off}

For an overall performance measure of a DNA storage coding scheme, one can consider the read/write cost trade-off introduced in \cite{chandak_improved_2019}.\footnote{This definition is similar (inverse) to the \textit{storage-sequencing trade-off} definition described in \cite{HeckelShomorony_Foundations_2022}. Additionally, \cite{HeckelShomorony_Foundations_2022} also considers the difference in the synthesis and sequencing cost per base. We choose \cite{chandak_improved_2019} due to its better graphical visualization.} We define the writing cost $\wc$ as the fraction of the number of synthesized nucleotides and the number of information bits stored. Similarly, we define the reading cost $\rc$ as the fraction of the number of sequenced nucleotides by the number of information bits stored. Assuming a coverage depth $c \geq 1$, this  implies the following two formulations given the overall rate $\Rtot$: 
\begin{align*}
    \wc &= \Rtot^{-1} &\geq 0.5 \quad &\left[ \text{nt/bit} \right] \, ,\\
    \rc &= c \Rtot^{-1} &\geq 0.5  \quad &\left[ \text{nt/bit} \right] \,.
\end{align*}

\subsubsection{Outage Probability}

Due to its drawing effect \cite{weinberger_dna_2022}, the DNA storage channel seen by the outer code resembles a block-fading channel when considering a finite number of strands $\chM$ and finite strand length $\chL$. Hence, it also shares its afflictions, most notably its non-ergodic property. In particular, an \emph{outage} event occurs when an insufficient number of strands are drawn for a specific outer code block. Formally, an outage event occurs when the instantaneous mutual information between the input and output of the channel is lower than the transmission rate $R$. %
We consider the \textit{BCJR-once} \cite{Kavcic2003DE, muller_capacity_2004, soriaga_determining_2007} version of the information-outage probability adapted from \cite{buckingham_informationoutage_2008}, to which we refer as the \emph{mismatched information-outage probability} $\qout$. It incorporates our mismatched decoding approach and the dispersion due to the finite blocklength phenomena.

For that, we first introduce the \emph{instantaneous BCJR-once rate} $\rbcjr$, which is defined as the instantaneous symbolwise mutual information between the input and the MAP decoder output for a fixed drawing realization $\bfd$ and with uniform input distribution. This definition aligns with our applied outer decoding metric in \cref{sec:dec-outer}. Fixing the input distribution, multiplying symbolwise posteriors for multiple strands, impairments introduced in the reordering, and neglecting output correlations of the APPs introduce mismatches that only decrease the metric $q(\bfw|\chY)$ compared to the true---but infeasible to compute---metric $p(\bfw|\chY)$. In detail, the instantaneous BCJR-once rate $\rbcjr$ for the finite-length regime and fixed drawing realization $\bfd$ is computed as 
\begin{align*}
    \rbcjr 
    &= \frac{1}{\chM \chL} \sum_{i=1}^\chM \sum_{t=1} ^{\bleno} \left(- \log_2 p(w_t^{(i)}) + \log_2 q(w^{(i)}_t | \chY )\right)\,,
\end{align*}
where $p(w_t^{(i)})=\frac{1}{4}$ is the a priori probability and $q(w^{(i)}_t | \chY )$ is the decoder's mismatched metric for the symbols of the corresponding block $\bfwi{i}$ as computed in \cref{sec:dec-outer}.

Then, the mismatched information-outage probability $\qout$ is formally defined as
\begin{align*}
    \qout = {\rm Pr}\left( \rbcjr < R \right) \geq \pout\,,
\end{align*}
where $R$ is the overall system rate measured in bit/nt and $\pout$ is the true outage probability. The value $\qout$ gives a lower bound on the FER for a given encoder and mismatched decoder pair for a fixed finite number of strands, fixed finite blocklength, and fixed channel parameters. We approximate $\qout$ using the Monte-Carlo method by generating drawing realizations $\bfd$ for a given $R$.

\subsection{Inner Code}\label{sec:codeopt-inner}

As mentioned in \cref{sec:coding-scheme}, we consider MR codes as our inner codes due to their good performance in the high-rate/low-error probability regime in the case of i.i.d. errors and mismatched decoding, as shown in \cite{ARXIV_Srinivasavaradhan2021TrellisBMA}. Nevertheless, the rate $\Rin = \frac{\kin}{\nin} = \frac{\kin}{\kin+1}$ still needs to be optimized based on  experimental data. 

For the extended trace reconstruction channel with a fixed number of draws $\trchM$ per block $\bfwi{i}$ (see \cref{sec:trace-recon-model}), we can compute actual AIRs for a fixed joint-code mismatched MAP decoder on a dataset. This is due to the reintroduced ergodicity by fixing the drawing realization, fixing $\chL$, and $\chM \to \infty$. Specifically, we compute identically uniformly distributed (i.u.d.) \emph{BCJR-once} rates $\Rtrace$, measured in bit/nt, which are again defined as the symbolwise mutual information between the input and the MAP decoder output with uniform input distribution \cite{Kavcic2003DE, muller_capacity_2004, soriaga_determining_2007}. We also follow a mismatched decoding approach for the extended trace reconstruction channel due to the separate strand decoding technique and symbolwise multiplication combining, and neglecting symbolwise APP correlations at the MAP decoder output. Let the input to the inner encoder be $\bfw=(\bfwi{1},\dots,\bfwi{\chM})$ and the output be defined as $\trchY = (\trchY_1, \dots, \trchY_\chM)$, where $\trchY_i = (\bfy_{i,1}, \dots, \bfy_{i,\trchM})$. We let $\mi(\bfw;\trchY)$ denote the true mutual information between the input $\bfw$ and the output $\trchY$, $\entr(\bfw)$ the entropy of $\bfw$, $\entr(\bfw|\trchY)$ the conditional entropy of $\bfw$ given $\trchY$, and $\expect_{X}[\cdot]$  the expectation with respect to  RV $X$. It consequently holds that 
\begin{align*}
    &\mi(\bfw;\trchY) \\
    &= \entr(\bfw) - \entr(\bfw|\trchY) \nonumber \\
	&=  \expect_{(\bfw,\trchY)}\left[ -\log_2 p(\bfw) + \log_2 p(\bfw|\trchY) \right] \\
	&= \expect_{(\bfw,\trchY)}\left[ -\log_2 p(\bfw) \vphantom{ \log_2\frac{p(\bfw|\trchY)}{q(\bfw|\trchY)^s}} \right. \\
    &\hspace{6em}+ \left. \left(\log_2 q(\bfw|\trchY)^s + \log_2\frac{p(\bfw|\trchY)}{q(\bfw|\trchY)^s}\right) \right] \\
	&\geq\expect_{(\bfw,\trchY)}\left[ -\log_2 p(\bfw) +\log_2  q(\bfw|\trchY)^s \right] \\
    &=\expect_{(\bfw,\trchY)}\left[\sum_{i=1}^\chM \sum_{t=1} ^{\bleno} \left(-\log_2 p(w_t^{(i)}) +\log_2 q(w_t^{(i)}|\trchY)^s\right) \right] \,,
 \label{eq:mi-mismatched}
\end{align*}
where the expectation is over the true joint distribution $p(\bfw,\trchY)$ of the RVs. The inequality holds by identifying that the third summand corresponds to a non-negative Kullback-Leibler divergence. The last equality holds due to the mismatched memoryless metric assumed for the outer code.
In detail, for a fixed number of traces $\trchM$ and an inner code rate $\Rin$, we define the achievable BCJR-once information rate as
\begin{align*}
    &\Rtrace(\trchM,\Rin)  \triangleq\\
    & \max_{s\geq0} \lim_{\chM \to \infty} \frac{1}{\chM \chL} \sum_{i=1}^\chM \sum_{t=1} ^{\bleno} 
    \!\left(-\log_2 p(w_t^{(i)}) +\log_2 q(w_t^{(i)}|\trchY)^s\right)\!,
\end{align*}
using  for multi-read combining the decoder's mismatched metric
\begin{align*}
    q(w_t^{(i)}|\trchY)^{s(\trchM)} = \left[ \prod _{j=1}^{\trchM} p(w^{(i)}_t|\bfy_{i,j}) \right]^{s(\trchM)}\,.
\end{align*}

We approximate the achievable BCJR-once information rate using experimental data by averaging over a large number of data blocks ($\chM \approx 30\,000$) as
\begin{align*}
    &\Rtrace(\trchM,\Rin) \approx \\
    &\max_{s(\trchM)\geq0} \frac{1}{\chM \chL} \sum_{i=1}^\chM \sum_{t=1} ^{\bleno} \left(- \log_2 p(w_t^{(i)}) + \log_2 q(w^{(i)}_t | \trchY_i )^{s(\trchM)}\right) \,.
\end{align*}
Note that due to the maximization over $s(\trchM)$, we directly obtain the values for the scaling parameter for the decoding on the \multidraw\ channel.

Furthermore, let $\theta_{\beta,c}(d)$ denote the fraction of input reads drawn $d$ times, i.e., $\frac{Q_d}{\chM}$, for a \multidraw\ channel with parameter $\beta$, associated drawing distribution $\drawdist$, and a certain coverage depth $c$ (see \cref{fig:drawdist} in Appendix~\ref{sec:appendix-drawdist} depicting $\theta_{\beta,c}(d)$ for various coverages $c$ measured on our dataset).\footnote{For uniform drawing distributions, $\theta_{\beta,c}(d)$ asymptotically tends to a Poission distribution, $\theta_{\beta,c}(d) = \frac{c^d}{d!}{\mathrm{e}}^{-c}$ \cite{lenz_upperboundcapDNA_2019, weinberger_dna_2022}.} Then, for an i.i.d. uniform data input distribution, we can compute the achievable information rate $\Rtheo$ in alignment with \cite{shomorony_dna-based_2021,lenz_noisyDrawChan_2023,weinberger_dna_2022} as
\begin{align*}
    \Rtheo &= \sum_{d=1} ^\infty \theta_{\beta,c}(d) \Rtrace(d,\Rin) - \beta(1-\theta_{\beta,c}(0)) \,.
\end{align*}
The ergodic rate $\Rtheo$ represents the theoretical rate for a channel with no errors in indexing and an optimal rate-achieving outer code and, thus, large enough number of input strands $\chM$. We can draw a connection between   $\Rtheo$ and the instantaneous BCJR-once rate $\rbcjr$. In the limit $\chM \chL \to \infty$, we have
\begin{align*}
    \Rtheo \geq \expect_{\bfd} \left[ \rbcjr \right]\,,
\end{align*}
where the inequality solely comes from the fact that $\rbcjr$ incorporates the rate loss of the index code (still assuming perfect permutation recovery here).

As a final step, we choose the inner code rate $\Rin$ which minimizes the theoretical read/write cost trade-off, $(\wc,\rc)_\text{th} = (\Rtheo^{-1}, c \Rtheo^{-1})$, for the targeted coverage levels.

\subsection{Index Code}\label{sec:codeopt-index}

In general, using indices for regrouping reads reduces the overall rate of the system directly by $\beta$. Since coding the indices further reduces the overall rate but increases the correct decoding probability, one must carefully design the index rate $\Rix$ and, thus, the index code itself. We target index codes with a good Levenshtein distance spectrum.\footnote{Even though the random offset is also applied in the index region, which destroys the designed Levenshtein distance properties, we have observed better performance when using codes optimized for the Levenshtein distance instead of other metrics.} 

The cardinality of the index code needs to be at least $\chM$. Unfortunately, finding good codes becomes more difficult for larger $\chM$ since the theory of error correction in the Levenshtein metric is not as advanced as, e.g., in the Hamming metric. As explained in Section~\ref{sec:coding-scheme}, to circumvent this challenge,  we build the index code by concatenating shorter component codes of smaller cardinality. For very small cardinality, we have obtained codes via an exhaustive graph search algorithm optimizing the code's Levenshtein distance spectrum \cite{Sewell1998clique2, Maarouf2022ConcatenatedCF}. In \cref{tab:indexcode}, we give two codes of length $4$, cardinality $14$, and minimum Levenshtein distance $3$ which we alternate to build an index code. For larger cardinalities, we investigated the non-binary Varshamov-Tenegolts codes \cite{tenengolts_nonbinVT_1984} and $2$-fold repetition codes. While the latter two codes can correct a deletion or an insertion, they cannot uniquely correct a substitution. 
\begin{table}[t]
\begin{center}
\caption{Exhaustive Search Index Codes} \label{tab:indexcode}
\vspace{-1.8ex}
\setlength{\tabcolsep}{1pt}
\begin{tabular}{l c c l }
\toprule 
   Code & & & Codebook \\   \midrule
    Code I & & & $(0,0,3,3)$, $(0,1,0,1)$, $(0,2,2,0)$, $(0,3,1,2)$, $(1,1,2,2)$,  \\
    & & & $(1,2,0,3)$, $(1,3,3,1)$, $(2,0,0,2)$, $(2,1,3,0)$, $(2,2,1,1)$,   \\
    & & & $(3,0,2,1)$, $(3,1,1,3)$, $(3,2,3,2)$, $(3,3,0,0)$  \\
      \midrule
      Code II & & & $(0,0,2,0)$, $(0,2,3,2)$, $(0,3,0,1)$, $(1,1,3,1)$, $(1,2,1,0)$,  \\
      & & & $(1,3,2,3)$, $(2,0,1,1)$, $(2,1,2,2)$, $(2,2,0,3)$, $(2,3,3,0)$,  \\
      & & & $(3,0,3,3)$, $(3,1,0,0)$, $(3,2,2,1)$, $(3,3,1,2)$  \\
         \bottomrule
\end{tabular}
\end{center}
\vspace{-4.5ex}
\end{table}

We evaluate the index codes by simple average MIRs $\avMIR$. Formally, for a fixed coding and decoding scheme, for a feasible number of drawing realizations $\bfd_1, \ldots, \bfd_\Phi$, we compute
\begin{align*}
    \avMIR \triangleq \frac{1}{\Phi} \sum_{\phi=1}^\Phi r_{\bfd_\phi}\,.
\end{align*}
We emphasize that the computed value $\avMIR$ does not imply any achievability guarantee due to the non-ergodic property of the \multidraw\ channel for finite $\chM$ and $\chL$ (in contrast to the infinite assumption with $\Rtheo$).
However, we have observed that $\avMIR$ correlates well in our parameter regime with more elaborate performance metrics, such as outage probability (which also requires more computations for evaluation analysis).

For each index code $\cC_\mathrm{ix}$, we search for the decoding threshold $\threshIx$ by maximizing the average MIR, i.e., we choose 
\begin{align*}
    \threshIx^\text{opt} = \argmax_{0 \leq \threshIx \leq 1} \avMIR\,.
\end{align*}
We remark that optimizing the index threshold $\threshIx$ does not depend on the coverage when applying no reads clustering before index decoding. Finally, we choose the index code that maximizes $\avMIR$ with the optimized $\threshIx$ for the targeted coverage levels.

\subsection{Post-Assignment Optimization}\label{sec:codeopt-postproc}

In the post-assignment, we can control the cardinality of the top candidate list $\poprTop_\bfy$. This parameter controls the trade-off between complexity and performance, as discussed in \cref{sec:postclust}. We limit $\lvert \poprTop _\bfy\rvert$ such that it is constant with respect to $\chN$ and choose the cardinality where the MIR $\avMIR$ of the scheme empirically saturates for the targeted coverage levels. As seen in \cref{sec:simresults}, we choose $\lvert \poprTop_{\bfy} \rvert = 5$ independently of the scheme. However, we do not claim that this performance saturation holds in general. 

\subsection{Outer Code}\label{sec:codeopt-outer}

We employ protograph-based non-binary SC-LDPC codes as outer codes in our concatenated coding scheme due to their excellent performance in standard transmission models under belief-propagation (BP) window decoding, e.g., over erasure channels \cite{siegel2010windowBPSCLPDC}, block-fading channels \cite{hassan_diversityscldpc_2014}, etc. \cite{7112076}. We mitigate the block-fading effects of the channel due to the drawing effect through the interleaving operation since we scramble possible block errors and erasures randomly over the entire outer codeword. %

Formally, a protograph is a small multi-edge-type Tanner graph with $\protn$ variable-node (VN) types and $\protr$ check-node (CN) types. A protograph can be represented by a base matrix 
\begin{align*}
    \prot = \begin{pmatrix}
    b_{1,1} & b_{1,2} & \dots & b_{1,\protn} \\
    b_{2,1} & b_{2,2} & \dots & b_{2,\protn} \\
    \vdots & \vdots & \dots & \vdots \\
     b_{\protr,1} & b_{\protr,2} & \dots & b_{\protr,\protn}\\
    \end{pmatrix}\,,
\end{align*}
where entry $b_{i,j}$ is an integer representing the number of edge connections from a type-$i$ CN to a type-$j$ VN. A parity-check matrix $\bfH$ of an LDPC code can then be constructed by lifting the base matrix $\prot$ by replacing each non-zero (zero) $b_{i,j}$ with a $\protlift \times \protlift$ circulant (zero) matrix with row and column weight equal to $b_{i,j}$. The resulting lifted parity-check matrix, of dimensions $\protlift\protr \times \protlift\protn$,  defines an LDPC code of length $\protlift  \protn$ and dimension at least $\protlift(\protn-\protr)$. To construct a non-binary code from the lifted matrix, we randomly assign non-zero entries from $\field_{\outq}$ to the edges of the corresponding Tanner graph.

An SC-LDPC code is constructed from a series of $\sccoup$ (coupling length) Tanner graphs by interconnecting the neighboring $\scmem$ (coupling memory) graphs according to a predefined pattern.
Constructing an SC-LDPC protograph from a corresponding LDPC protograph $\bfB$ is done as follows. The protograph $\bfB$ of the LDPC code is copied $\sccoup$ times. Then, the edges of each protograph are interconnected to the $\scmem$ neighbors. The protograph matrix of an SC-LDPC code then has the  form
\begin{align*}
    \scprot\!=\!\begin{pmatrix}
    \prot_{1} & & & \\
    \prot_{2} & \prot_{1}&  & \\
    \vdots & \vdots & \ddots & \\
    \prot_{\scmem + 1} & \prot_{\scmem} & \dots & \prot_{1}\\
     &\prot_{\scmem + 1} &\dots  & \prot_{2}\\
     & &\ddots &\vdots \\
     & & &\prot_{\scmem + 1}
    \end{pmatrix}_{\!\!\!(\sccoup + \scmem)\protr \times \sccoup \protn}\,,
\end{align*}
where $\sum_{i = 1}^{\scmem + 1}\prot_i = \prot$ to maintain the same VN and CN degrees of the underlying block protograph $\bfB$. The resulting SC-LDPC code is then constructed by lifting and assigning edge weights in the same manner as mentioned earlier. 

In detail, we build our coupled codes from regular LDPC protographs with fixed VN degree $d_\mathrm{VN}$ and CN degree $d_\mathrm{CN}$ as the underlying base protograph $\bfB$ and fix the coupling memory $\scmem = d_\mathrm{VN}-1 = 2$ (i.e., implying $d_\mathrm{VN}=3$) such that each memory block $\bfB_i$ is connected to each of its neighbors exactly once, i.e.,  $\bfB = \sum_{i=1}^{d_\mathrm{VN}} [1,\ldots,1]$, where $[1,
\ldots,1]$ is of length $\frac{d_\mathrm{CN}}{d_\mathrm{VN}}$. Moreover, we always choose $\protlift$ such that one spatial position covers approximately $\protlift \frac{d_\mathrm{CN}}{d_\mathrm{VN}} \approx 10\,000$ VN positions and, when needed, shorten positions uniformly at random to fit the required blocklength $\leno$. We remark that we limit the rate optimization significantly to the possible coding rates $\Rout \approx \frac{1}{2}, \frac{2}{3}, \ldots, \frac{8}{9}$. Higher rates and/or better performance could possibly be achieved by considering different ways of coupling or using irregular protographs. We choose to couple regular protographs due to their simple design and promising decoding performance under BP decoding on classical transmission channels. For decoding, we consider sliding window decoding from the front and back simultaneously. We fix the window size to nine spatial positions and limit the BP iterations to five per window position, with ten iterations at the boundaries of the coupled chain when the window is not yet slid fully into the Tanner graph to kick-off decoding \cite{roman2023scalingSCLDPCLimIter}.

\section{Simulation Results} \label{sec:simresults}

We design our coding scheme to be viable for different modes of operation, with a specific focus on three primary coverages: $c = 5$, $c = 8$, and $c = 10$ for both \accbc\ and \fastbc\ (refer to \cref{tab:data-error-rates} for IDS error estimates obtained for the two basecalling methods) and fixed $\beta = \frac{\log_2(\chM)}{\chL} = 0.135$ ($\chM = 30\,588$, $\chL=110$). 

We perform code optimization on \texttt{file1} using MIRs following the steps and criteria discussed in \cref{sec:codeopt}. Then, we analyze our proposed coding scheme on the experimental data for \texttt{file3} by means of the information-outage probability for the outer code and FER simulations for the above channel setups. Moreover, the channel estimation for \kmerCh\ is performed using \texttt{file1}. Recall that the stored DNA strands in our dataset have a total length of $110\,\text{nt}$ plus additionally on each side a $20\,\text{nt}$ file-specific primer, and we stored $30\,589$ (\texttt{file3} $30\,588$) strands per file (more details in \cref{tab:dataset}). We subsample reads uniformly at random when targeting a specific $c$ such that we mimic the underlying sampling distribution due to possible synthesis, PCR, and/or sequencing bias. 

We focus only on using the passed Q-score reads. However, we believe cleverly utilizing the failed Q-score reads could increase the system's performance at no additional cost.

\subsection{Read/Write Cost Trade-Off} \label{sec:simresults-readwrite}

\begin{figure*}[t]
    \centering
     \begin{tikzpicture}    
    \begin{groupplot}[
    group style={
        group size=3 by 1,
        xticklabels at=all,
        horizontal sep=0pt
    },
    ymin=0.4, ymax=1.4,
    xmajorgrids,
    ytick={0.5, 0.6, 0.7, 0.8, 0.9, 1.0, 1.1, 1.2, 1.3, 1.3},
    height=0.35\textwidth,
    width=.99\textwidth
    ]

    \node[rotate=90] (ylabel) at (-1.2,2.3) {$\wc$ {\footnotesize (writing cost)  [nt/bit]}};

    \nextgroupplot[xmin=0.5,xmax=9.5,
    xtick={1, 2, 3, 4, 5, 6, 7, 8, 9, 10},
    axis x line*=middle,
    axis y line=middle,
    xlabel={$\rc$ {\footnotesize (reading cost) [nt/bit]}},
    height=0.35\textwidth,
    width=.45\textwidth,
    legend cell align={left},
    legend style={font=\footnotesize,at={(2.45,0.01)},anchor=south east},
	]

 \addlegendimage{blue, mark=*, only marks}
\addlegendentry{Nanopore}
 \addlegendimage{red, mark=*, only marks}
\addlegendentry{Illumina}
 \addlegendimage{teal, mark=square*, only marks}
\addlegendentry{This work (\accbc)}
 \addlegendimage{magenta, mark=square*, only marks}
\addlegendentry{This work (\fastbc)}

    \draw[gray,opacity=0.8] (axis cs: 0,0.5) -- (axis cs: 9.5,0.5);
    \draw[gray,opacity=0.8] (axis cs: 0,0.6) -- (axis cs: 9.5,0.6);
    \draw[gray,opacity=0.8] (axis cs: 0,0.7) -- (axis cs: 9.5,0.7);
    \draw[gray,opacity=0.8] (axis cs: 0,0.8) -- (axis cs: 9.5,0.8);
    \draw[gray,opacity=0.8] (axis cs: 0,0.9) -- (axis cs: 9.5,0.9);
    \draw[gray,opacity=0.8] (axis cs: 0,1.0) -- (axis cs: 9.5,1.0);
    \draw[gray,opacity=0.8] (axis cs: 0,1.1) -- (axis cs: 9.5,1.1);
    \draw[gray,opacity=0.8] (axis cs: 0,1.2) -- (axis cs: 9.5,1.2);
    \draw[gray,opacity=0.8] (axis cs: 0,1.3) -- (axis cs: 9.5,1.3);
   
    \addplot[
        scatter/classes={a={blue}, b={red}, c={teal}, d={magenta}},
        scatter, mark=*, only marks, 
        scatter src=explicit symbolic,
        nodes near coords*={\tiny \Label},
        visualization depends on={value \thisrow{label} \as \Label} %
    ] table [meta=class] {
        x y class label
        3.22 0.85 a Lau-2023
		6.59 1.07 a Chandak-2020
		6.77 0.65 b Erlich-2017
		4.46 0.91 b Organick-2019
		};
    \addplot[
        scatter/classes={a={blue}, b={red}, c={teal}, d={magenta}},
        scatter, mark=*, only marks, 
        scatter src=explicit symbolic,
        nodes near coords*={\tiny \Label},
                nodes near coords align={above right},
        visualization depends on={value \thisrow{label} \as \Label} %
    ] table [meta=class] {
        x y class label
		4.47 0.78 b Chandak-2019
		};

    \addplot[
        scatter/classes={a={blue}, b={red}, c={teal}, d={magenta}},
        scatter, mark=triangle, only marks, 
        scatter src=explicit symbolic,
        nodes near coords*={\tiny \Label},
        nodes near coords align={below},
        visualization depends on={value \thisrow{label} \as \Label} %
    ] table [meta=class] {
        x y class label
        3.587 0.7173 c { }
        5.161 0.645 c { }
        6.9565 0.8696 d { }
        8.175 0.8175 d { }
		};

    \addplot[
        scatter/classes={a={blue}, b={red}, c={teal}, d={magenta}, e={orange}},
        scatter, mark=square*, only marks, 
        scatter src=explicit symbolic,
        nodes near coords*={\tiny \Label},
        nodes near coords align={above},
        visualization depends on={value \thisrow{label} \as \Label} %
    ] table [meta=class] {
        x y class label
        5.51148 0.68893 c acc-c8
        8.75331 0.87533 d fast-c10
        3.81944 0.76388 c acc-c5
        7.85722 0.98215 d fast-c8
		};

    \draw[teal, densely dotted] (axis cs: 5.51148,0.68893) -- (axis cs: 5.161,0.645);
    \draw[magenta, densely dotted] (axis cs: 8.75331,0.87533) -- (axis cs: 8.175,0.8175);
    \draw[orange, densely dotted] (axis cs: 3.81944,0.76388) -- (axis cs: 3.587,0.7173);
    \draw[orange, densely dotted] (axis cs: 7.85722,0.98215) -- (axis cs: 6.9565,0.8696);

    \nextgroupplot[xmin=0,xmax=60,
    xtick={15,25,35,45,55},
    axis y line=none,
    axis x line*=middle,
    axis x discontinuity=crunch,
     height=0.35\textwidth,
    width=0.34\textwidth
    ]

    \draw[gray,opacity=0.8] (axis cs: 0,0.5) -- (axis cs: 60,0.5);
    \draw[gray,opacity=0.8] (axis cs: 0,0.6) -- (axis cs: 60,0.6);
    \draw[gray,opacity=0.8] (axis cs: 0,0.7) -- (axis cs: 60,0.7);
    \draw[gray,opacity=0.8] (axis cs: 0,0.8) -- (axis cs: 60,0.8);
    \draw[gray,opacity=0.8] (axis cs: 0,0.9) -- (axis cs: 60,0.9);
    \draw[gray,opacity=0.8] (axis cs: 0,1.0) -- (axis cs: 60,1.0);
    \draw[gray,opacity=0.8] (axis cs: 0,1.1) -- (axis cs: 60,1.1);
    \draw[gray,opacity=0.8] (axis cs: 0,1.2) -- (axis cs: 60,1.2);
    \draw[gray,opacity=0.8] (axis cs: 0,1.3) -- (axis cs: 60,1.3);
    
    \addplot[
        scatter/classes={a={blue}, b={red}},
        scatter, mark=*, only marks, 
        scatter src=explicit symbolic,
        nodes near coords*={\tiny \Label},
        visualization depends on={value \thisrow{label} \as \Label} %
    ] table [meta=class] {
        x y class label
        21.121 0.9183 a Lopez-2019
		18.27 0.62516 b Bar-Lev-2021
		42.098 0.62516 a Bar-Lev-2021
		15.03 0.512 b Anavy-2019
        47.06 1.176 b Bornholt-2016
		};

    \addplot[
        scatter/classes={a={blue}, b={red}},
        scatter, mark=*, only marks, 
        scatter src=explicit symbolic,
        nodes near coords align={below},
        nodes near coords*={\tiny \Label},
        visualization depends on={value \thisrow{label} \as \Label} %
    ] table [meta=class] {
        x y class label
		32.73 0.91 a Organick-2019
		};

	\nextgroupplot[xmin=-200,xmax=800,
	xtick={200,600},
	axis y line=none,
	axis x line=middle,
	axis x discontinuity=crunch,
	grid style = {line width=.1pt},
     height=0.35\textwidth,
	width=.2\textwidth
	]

     \draw[gray,opacity=0.8] (axis cs: -200,0.5) -- (axis cs: 800,0.5);
    \draw[gray,opacity=0.8] (axis cs: -200,0.6) -- (axis cs: 800,0.6);
    \draw[gray,opacity=0.8] (axis cs: -200,0.7) -- (axis cs: 800,0.7);
    \draw[gray,opacity=0.8] (axis cs: -200,0.8) -- (axis cs: 800,0.8);
    \draw[gray,opacity=0.8] (axis cs: -200,0.9) -- (axis cs: 800,0.9);
    \draw[gray,opacity=0.8] (axis cs: -200,1.0) -- (axis cs: 800,1.0);
    \draw[gray,opacity=0.8] (axis cs: -200,1.1) -- (axis cs: 800,1.1);
    \draw[gray,opacity=0.8] (axis cs: -200,1.2) -- (axis cs: 800,1.2);
    \draw[gray,opacity=0.8] (axis cs: -200,1.3) -- (axis cs: 800,1.3);

	\addplot[
		scatter/classes={a={blue}, b={red}},
		scatter, mark=*, only marks, 
		scatter src=explicit symbolic,
		nodes near coords*={\tiny \Label},
		visualization depends on={value \thisrow{label} \as \Label} %
	] table [meta=class] {
		x y class label
		171.62 0.59 a Yazdi-2017
		756.8 1.24 b Antkowiak-2020
		};

    \end{groupplot}

    \end{tikzpicture}
    \caption{Read/write cost trade-off for our setup and other experiments. Some experiments conducted storage trials with various parameters. We have selected the one that appears to be the best based on our knowledge. Label references: \emph{Lau-2023} \cite{stanford_magneticDNAstorage_2023}, \emph{Bar-Lev-2021} \cite{bar2021deep}, \emph{Antkowiak-2020} \cite{antkowiak_low_2020}, \emph{Chandak-2020} \cite{chandak_overcoming_2020}, \emph{Anavy-2019} \cite{Anavy2019composite}, \emph{Chandak-2019} \cite{chandak_improved_2019}, \emph{Lopez-2019} \cite{Lopez2019},  \emph{Organick-2019} \cite{organick_random_2018}, \emph{Erlich-2017} \cite{erlich_dna_2017}, \emph{Yazdi-2017} \cite{yazdi_portable_2017}, and \emph{Bornholt-2016} \cite{bornholt_dna-based_2016}.}
    \label{fig:readwrite}
\end{figure*}
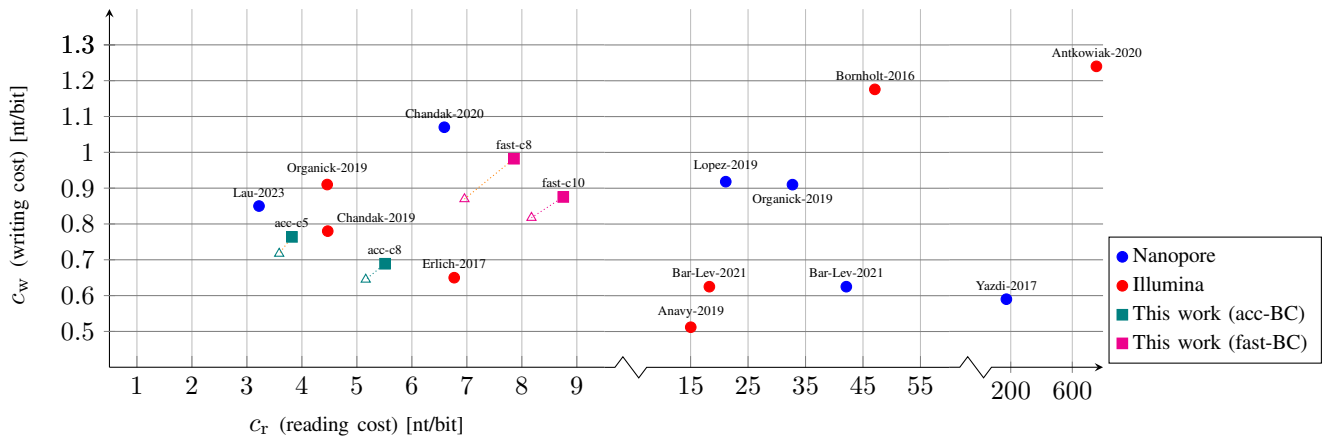

In \cref{fig:readwrite}, we show the read/write cost trade-off of our designed coding systems (optimized for different coverages $c$) and compare them to existing experiments. Note that we focus on the detailed construction and performance evaluation in the subsequent sections. With the setup for \accbc\ (green squared markers), we outperform other nanopore-based and even Illumina-based experiments while ensuring a failure rate of less than one percent (see \cref{fig:outage-fer}). We attribute our gain mostly to the clever combining/clustering techniques of the strands and the overall soft-decision-based decoding algorithms since our channel error rates, IDS errors combined roughly at $4\%$, are similar (or even worse) to some of the other experiments. Since it is expected that by optimizing the outer code better we could gain in rate, we have included the theoretical read/write cost trade-off of our inner system with an outage probability $\qout = 10^{-3}$ (empty triangle-shaped markers). We believe one could approach these points closer while ensuring a FER of approximately $10^{-3}$. 
Additionally, we have included the read/write cost trade-off for our \fastbc\ setups. Despite significantly higher IDS error rates (of roughly $10\%$), we remain competitive\textemdash but not better\textemdash than most other nanopore experiments. %

In our view, the read/write cost trade-off is the best available metric for experimental comparisons because it naturally incorporates the fundamental tension between synthesis and sequencing. However, the values reported in Fig.~\ref{fig:readwrite} should be interpreted with caution for two reasons: First, they do not reflect the differences in FER and the encoding/decoding complexity. Second, the read and write costs are affected by the details of the channel setup, which vary significantly between different approaches. For example, the choice of $\beta$, the IDS error rates, and different drawing distributions all naturally require different amounts of redundancy and thus massively influence the read-write cost performance of a system. For example, \emph{Lau-2023} \cite{stanford_magneticDNAstorage_2023} evaluates only with low $\beta = 0.0088$ ($\chM = 792$, $\chL = 114)$ or \emph{Erlich-2017} \cite{erlich_dna_2017} deals with IDS error rates of less than $1\%$. Additionally, the same can be seen by the different read/write cost trade-offs for our proposed system obtained for \accbc\ and \fastbc. 

\subsection{Extended Trace Reconstruction Channel: KMER Decoder }

\begin{figure}[t]
    \centering
    \begin{tikzpicture}
\begin{axis}[
width = 1.0\linewidth,
xmin=1,   xmax=15,
ymin=0.8,	ymax=2,
xticklabel style = {/pgf/number format/fixed, /pgf/number format/precision=6},
extra x ticks= {1, 15},
ylabel style={
	yshift=-1ex,
	name=label},
grid = both,
grid style = {line width=.1pt},
legend cell align={left},
legend style={font=\footnotesize,at={(0.99,0.01)},anchor=south east,legend columns=2},
xlabel = {$\trchM$ {\footnotesize (num. of traces)}},
ylabel = {$\Rtrace$\quad {\footnotesize[bit/nt]}},
cycle list name=color list
]

{\addplot+ [color=blue, mark=square, solid, mark options={solid}] table [col sep=comma,x=M,y=I] {Figures/TraceRecon/csv/k=0/acc_MR-108_IX=none.csv};}
{\addlegendentry{\accbc\ -- i.i.d.};}

{\addplot+ [color=red, mark=square, solid, mark options={solid}] table [col sep=comma,x=M,y=I] {Figures/TraceRecon/csv/k=0/nonacc_MR-100_IX=none.csv};}
{\addlegendentry{\fastbc\ -- i.i.d.};}

{\addplot+ [color=blue, mark=none, dashdotted, mark options={solid}] table [col sep=comma,x=M,y=I] {Figures/TraceRecon/csv/sone/acc_MR-108_IX=none.csv};}
{\addlegendentry{\accbc\ -- i.i.d. \cite{Maarouf2022ConcatenatedCF}};}

{\addplot+ [color=red, mark=none, dashdotted, mark options={solid}] table [col sep=comma,x=M,y=I] {Figures/TraceRecon/csv/sone/fast_MR-100_IX=none.csv};}
{\addlegendentry{\fastbc\ -- i.i.d. \cite{Maarouf2022ConcatenatedCF}};}

{\addplot+ [color=blue, mark=triangle, solid, mark options={solid}] table [col sep=comma,x=M,y=I] {Figures/TraceRecon/csv/k=1/acc_MR-108_IX=none.csv};}
{\addlegendentry{\accbc\ -- $k=1$};}

{\addplot+ [color=red, mark=triangle, solid, mark options={solid}] table [col sep=comma,x=M,y=I] {Figures/TraceRecon/csv/k=1/nonacc_MR-100_IX=none.csv};}
{\addlegendentry{\fastbc\ -- $k=1$};}

{\addplot+ [color=blue, mark=star, solid, mark options={solid}] table [col sep=comma,x=M,y=I] {Figures/TraceRecon/csv/k=3/acc-true_k-3_MR-108_IX=none.csv};}
{\addlegendentry{\accbc\ -- $k=3$};}

{\addplot+ [color=red, mark=star, solid, mark options={solid}] table [col sep=comma,x=M,y=I] {Figures/TraceRecon/csv/k=3/nonacc_MR-100_IX=none.csv};}
{\addlegendentry{\fastbc\ -- $k=3$};}

{\addplot+ [color=blue, mark=none, dotted, mark options={solid}] table [col sep=comma,x=M,y=I] {Figures/TraceRecon/csv/comparison/acc_MR-108_IX=none_k1_compare.csv};}
{\addlegendentry{\accbc\ -- $k=1$ \cite{frenchKmer_belaid_2023}};}

{\addplot+ [color=red, mark=none, dotted, mark options={solid}] table [col sep=comma,x=M,y=I] {Figures/TraceRecon/csv/comparison/nonacc_MR-100_IX=none_k1_compare.csv};}
{\addlegendentry{\fastbc\ -- $k=1$ \cite{frenchKmer_belaid_2023}};}

{\addplot+ [color=blue, mark=none, dashed, mark options={solid}] table [col sep=comma,x=M,y=I] {Figures/TraceRecon/csv/comparison/acc_MR-108_IX=none_k3_compare.csv};}
{\addlegendentry{\accbc\ -- $k=3$ \cite{frenchKmer_belaid_2023}};}

{\addplot+ [color=red, mark=none, dashed, mark options={solid}] table [col sep=comma,x=M,y=I] {Figures/TraceRecon/csv/comparison/nonacc_MR-100_IX=none_k3_compare.csv};}
{\addlegendentry{\fastbc\ --  $k=3$ \cite{frenchKmer_belaid_2023}};}

\addplot[forget plot, mark=none, color=blue, loosely dashdotdotted, opacity=1.0] coordinates {(1,1.96363636363636) (40,1.96363636363636)};

\addplot[forget plot, mark=none, color=red, loosely dashdotdotted, opacity=1.0] coordinates {(1,1.81818181818182) (40,1.81818181818182)};

\end{axis}%
\end{tikzpicture}%
    \caption{BCJR-once rates $\Rtrace$ versus number of traces $\trchM$ for the extended trace reconstruction channel with coded strands obtained on data of \texttt{file3}. We compare different single-strand decoders, i.e., different memory $k$ in the \kmerCh\ modeling in the decoder. For \accbc\, we use $\Rin = \nicefrac{216}{110}$ and for \fastbc\, we use $\Rin=\nicefrac{200}{110}$. Note that no index part is included here. The horizontal dash-dotted lines represent the rate limit imposed by the inner code rates.}
    \label{fig:sim-trace-recon-coded}
\end{figure}
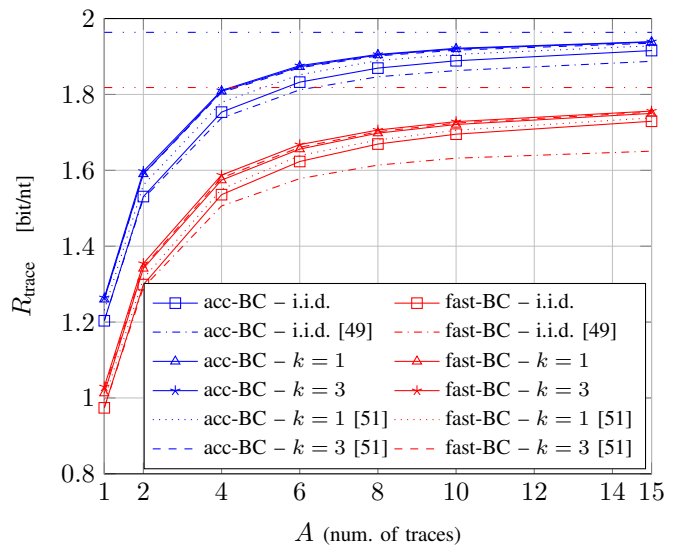

We analyze the proposed \kmerCh\ model for different values of $k$. In \cref{fig:sim-trace-recon-coded}, we illustrate the achievable rate $\Rtrace$ for the extended trace reconstruction channel (see \cref{sec:trace-recon-model}) as a function of the number of traces $\trchM$. We perform our analysis using \texttt{file3}, where in our computation, we average over all available clusters with at least size $\trchM$ for \accbc\ and \fastbc\ separately. A higher $\Rtrace$ for a fixed setup corresponds to a reduced mismatch with the data's true underlying strand error channel. We observe an increase in $\Rtrace$ once we introduce the \kmerCh\ model for both setups, \accbc\ and \fastbc. As expected, this gain diminishes slowly when increasing the number of traces $\trchM$ since the gain in extra physical redundancy dominates. Further, we observe that for the fast basecaller, increasing $k$ yields higher rates; however, this gain is marginal and is associated with a substantial increase in decoding complexity---e.g., changing from $k=1$ to $k=3$ increases decoding complexity $16$-fold. For the accurate basecaller, we can see no proper performance gain when increasing $k > 1$. In this case, we conjecture that the machine learning-based basecaller, which converts the electric current at the nanopore to nucleotides, already reduces the memory between the neighboring nucleotides. For completeness, we include the curves using the model from \cite{frenchKmer_belaid_2023} to show the performance gain of the slight adaption of the channel model at no additional cost of decoding complexity. Moreover, we have included the curves for i.i.d. decoding with fixed scaling parameter $s=1.0$ from \cite{Maarouf2022ConcatenatedCF}. The obtained scaling parameters of this work can be found in \cref{tab:scaling-logs} in Appendix~\ref{sec:ScalingParameter}. Furthermore, we present results for the uncoded traces in Appendix~\ref{sec:appendix-uncoded-tracerecon} in \cref{fig:trace-recon-uncoded}. Here, the main takeaway is that introducing the scaling parameter $s(\cdot)$ mitigates the APP mismatch significantly for increasing number of traces $\trchM$.

In our optimization procedure, we compute $\Rtrace(\trchM)$ for various inner rates $\Rin$ and select the rate that yields the best theoretical read/write cost trade-off $(\wc,\rc)_\text{th}$ as discussed in \cref{sec:codeopt-inner}. We have observed the best performance for $\Rin = \frac{216}{110}$ for \accbc\ and $\Rin=\frac{200}{110}$ for \fastbc. Due to the necessity of the coded address part in the strand of total length $110$, the data part is shorter than $110$. Thus, we fix the number of redundancy symbols for data protection: two symbols for \accbc\ and ten symbols for \fastbc. This effectively introduces a slight discrepancy in the inner code rates for the final setup relative to the values of $\Rin$ above.

\subsection{MIR Results: Optimization of the Joint-Code and Decoding Strategy} \label{sec:simresults-MIR}

For the choice of inner code and decoding optimization, we analyze MIRs $\avMIR$ versus coverage $c$ for different setups using \texttt{file1}. As a side-effect, the MIRs highlight the effects of the \multidraw\ channel, which we will include in the discussion. In this section, we focus on the results for \accbc; nonetheless, we have included the results for \fastbc\ in \cref{fig:mirs-indexcomp-fast,fig:mirs-fast} in Appendix~\ref{MIRFBase}, where similar conclusions (albeit different outcomes) can be obtained based on the same logic. Throughout this subsection, we fix $k=1$ in the single-strand joint-code decoder.

First, we compare the different index codes in \cref{fig:mirs-indexcomp-acc}. Recall that the overall index code is formed from concatenating an even number $a$ of component codes. Moreover, we split the index part equally to be at the strand's front and back. In detail, we compare  four index coding strategies: uncoded (UC), repetition (REP) coding, Varshamov-Tenengolts  (VT) coded, and exhaustive-search (ES) coded  (for the code details see \cref{tab:comp-ix-list} and \cref{sec:codeopt-index}). For all setups, we use two redundancy symbols for the inner data part coding; hence, we briefly discuss the inner code rate implications using the UC index coding strategy as an example. For UC we need $\lceil \log_4(30\,589)\rceil = 8\, \text{nt}$ of the total $110\, \text{nt}$ for the index. Hence, we have an inner data rate of $\Rin = \frac{200}{102}$; consequently, imposing a natural limit at $\frac{200}{110}$ (rate limits for each setup are indicated by the dash-dotted lines). Similarly, a VT-coded index needs $12\, \text{nt}$ of indexing; thus, combined with having two redundancy symbols for the data part, it imposes a natural limit of $\frac{192}{110}$. 

We begin by comparing the index codes using the decoding strategy where the decoder bases its regrouping decisions solely on the soft information of the decoded index (see \cref{sse:index-decoding}), i.e., no clustering or post-assignment is applied (squared markers). In a general view, we can conclude that although coding induces a rate loss, we obtain higher $\avMIR$ for the coded setups REP, ES, and VT than for simply UC. 
Further, for low coverages $c \leq 5$, we obtain a clear ranking in the performance of the studied index coding approaches. We conjecture that the respective Levenshtein distance spectrums of the index codes directly impact decoding performance. Codes with the spectrum shifted toward higher Levenshtein distances achieve higher $\avMIR$ even with a higher coding rate $\Rix$. For higher coverages, there is a slight change in the behavior of the codes, e.g., for $c=10$, UC overtakes REP. This can be explained by the fact that we operate closer to the respective rate limits of each system, e.g., REP's limit is 1.68 bit/nt, while UC's limit is 1.82 bit/nt. Consequently, the choice of the index code depends on the targeted coverage $c$. 

Additionally, for VT and ES index codes, we included in \cref{fig:mirs-indexcomp-acc} the curves for an alternative decoding strategy where we assign an index to a specific read if the decoded soft information exceeds a reliability threshold $\threshIx$ (triangle markers). After optimization of $\threshIx$, which turns out to be a convex optimization problem, we acquired for the  ES index code a reliability threshold of $\threshIx=0.985$ and for the VT index code a threshold of $\threshIx=0.995$. This corresponds to discarding $26\%$ of the reads for the ES strategy and $33.5\%$ for the VT strategy, independent of the coverage $c$. Due to the performance gain compared to the previous method, we conjecture that, most often, the discarded reads would have been grouped incorrectly since discarding is less detrimental than incorrect index assignments. Surprisingly, the benefit of this method, which comes at no additional decoding cost, remains relatively constant over different coverages (except $c=1$). We had anticipated that discarding incorrectly assigned reads at low coverage would be significantly more beneficial than at higher coverage. Moreover, we observe that the index code that performs best in the simple index grouping strategy is not the best choice when introducing the reliability threshold $\threshIx$, i.e., the VT index code performs better for higher coverages than the ES index code. Be that as it may, the VT index code has a higher rate limit than the ES index code, which might be responsible for the performance difference. In a final conclusion, we have picked the VT index code for the \accbc\ setup since it had the best performance for coverages $c=5$ until $c=10$ using the reliability threshold $\threshIx=0.995$.
\begin{table}[t]
\begin{center}
\caption{List of Compared Index Codes}
\label{tab:comp-ix-list}
\vspace{-2.3ex}
\begin{tabular}{c c c c}
    \toprule
    Name & $\Rix$ {\footnotesize [bit/nt]} & Component code & $a$ \\ \midrule
    Uncoded (UC) & $2$ & $[4,4]_4$ & $2$ \\ 
    Repetition (REP) & $1$ & $[8,4]_4$ & $2$  \\
    Varshamov-Tenengolts (VT) & $1.25$ & $[6,\log_4(178)]_4$ & $2$ \\
    Exhaustive-search (ES)$^a$ & $0.952$ & $[4,\log_4(14)]_4$ & $4$ \\
    \bottomrule
\end{tabular}
\end{center}
\footnotesize{$^a$For the ES index code, we alternate  Code~$(1)$ and Code~$(2)$ from \cref{tab:indexcode} to increase the inter minimum Levenshtein distance between two neighboring codebooks. } 
\end{table}
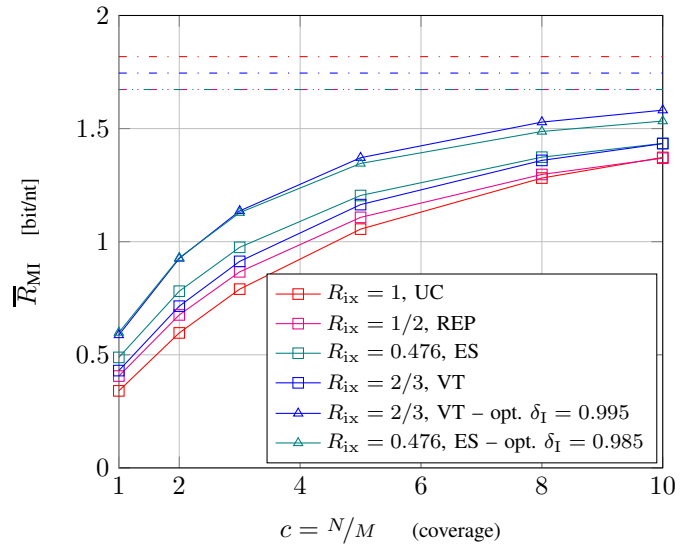
\begin{figure}[t]
    \centering
    \begin{tikzpicture}
\begin{axis}[
width = 0.995\linewidth,
xmin=1,   xmax=10,
ymin=0.0,	ymax=2,
xticklabel style = {/pgf/number format/fixed, /pgf/number format/precision=6},
extra x ticks= {1},
grid = both,
grid style = {line width=.1pt},
legend cell align={left},
legend style={font=\footnotesize,at={(0.99,0.01)},anchor=south east},
legend image post style={dash phase=0pt},
xlabel = {$c =\nicefrac{\chN}{\chM}$ \quad {\footnotesize(coverage)}},
ylabel = {$\avMIR$\quad {\footnotesize [bit/nt]}},
cycle list name=color list
]

{\addplot+ [color=red, mark=square, solid, mark options={solid}] table [col sep=comma,x=c,y=I] {Figures/MIR-index/acc-true/csv/Acc-true_MR-100_IX=one_IXDEC=ML.csv};}
{\addlegendentry{$R_\mathrm{ix}=1$, UC };}

\addplot[forget plot, mark=none, color=red, loosely dashdotted, opacity=1.0] coordinates {(1,1.81818181818182) (40,1.81818181818182)};

{\addplot+ [color=magenta, mark=square, solid, mark options={solid}] table [col sep=comma,x=c,y=I] {Figures/MIR-index/acc-true/csv/Acc-true_MR-92_IX=rep_IXDEC=ML.csv};}
{\addlegendentry{$R_\mathrm{ix}=1/2$, REP};}

\addplot[forget plot, mark=none, color=magenta, loosely dashdotted, opacity=1.0] coordinates {(1,1.67272727272727) (40,1.67272727272727)};

{\addplot+ [color=teal, mark=square, solid, mark options={solid}] table [col sep=comma,x=c,y=I] {Figures/MIR-index/acc-true/csv/Acc-true_MR-92_IX=ER_IXDEC=ML.csv};}
{\addlegendentry{$R_\mathrm{ix}=0.476$, ES };}

\addplot[forget plot, mark=none, color=teal, loosely dashdotted, opacity=1.0, dash phase=2pt] coordinates {(1,1.67272727272727) (40,1.67272727272727)};

{\addplot+ [color=blue, mark=square, solid, mark options={solid}] table [col sep=comma,x=c,y=I] {Figures/MIR-index/acc-true/csv/Acc-true_MR-96_IX=VT_IXDEC=ML.csv};}
{\addlegendentry{$R_\mathrm{ix}=2/3$, VT };}

\addplot[forget plot, mark=none, color=blue, loosely dashdotted, opacity=1.0] coordinates {(1,1.74545454545455) (40,1.74545454545455)};

{\addplot+ [color=blue, mark=triangle, solid, mark options={solid}] table [col sep=comma,x=c,y=I] {Figures/MIR-index/acc-true/csv/Acc-true_MR-96_IX=VT_IXDEC=ML_ixTh=0995.csv};}
{\addlegendentry{$R_\mathrm{ix}=2/3$, VT -- opt. $\threshIx=0.995$};}

{\addplot+ [color=teal, mark=triangle, solid, mark options={solid}] table [col sep=comma,x=c,y=I] {Figures/MIR-index/acc-true/csv/Acc-true_MR-92_IX=ER_IXDEC=ML_ixTh=0985.csv};}
{\addlegendentry{$R_\mathrm{ix}=0.476$, ES -- opt. $\threshIx=0.985$};}

\end{axis}%
\end{tikzpicture}%
    \caption{MIRs $\avMIR$ versus  coverage $c$ for the \multidraw\ channel on experimental data of \texttt{file1} ($\chM=30\,589$, $\chL=110$) for \accbc\ using $k=1$ in the single-strand decoder. We compare different index codes plus the index thresholding strategy. The horizontal dash-dotted lines represent the rate limit imposed by the combined respective index code rates and inner code rates (line colors match the index codes indicated in the legend). We use two redundancy symbols for the data part coding.}
    \label{fig:mirs-indexcomp-acc}
\end{figure}

\begin{figure}[t]
    \centering
    \begin{tikzpicture}
\begin{axis}[
width = 0.995\linewidth,
xmin=1,   xmax=10,
ymin=0.0,	ymax=2,
xticklabel style = {/pgf/number format/fixed, /pgf/number format/precision=6},
extra x ticks= {1},
grid = both,
grid style = {line width=.1pt},
legend cell align={left},
legend style={font=\footnotesize,at={(0.99,0.01)},anchor=south east,cells={align=left}},
xlabel = {$c =\nicefrac{\chN}{\chM}$ \quad {\footnotesize(coverage)}},
ylabel = {$\avMIR$\quad {\footnotesize [bit/nt]}},
cycle list name=color list
]

{\addplot+ [color=red, mark=square, solid, mark options={solid}] table [col sep=comma,x=c,y=I] {Figures/MIRs/acc/Acc-true_MR-100_IX=one_IXDEC=ML.csv};}
{\addlegendentry{UC, ix. dec.};}

{\addplot+ [color=red, mark=star, dashed, mark options={solid}] table [col sep=comma,x=c,y=I] {Figures/MIRs/acc/Acc-true_MR-100_IX=one_IXDEC=G.csv};}
{\addlegendentry{UC, genie ix. dec.};}

{\addplot+ [color=blue, mark=square, solid, mark options={solid}] table [col sep=comma,x=c,y=I] {Figures/MIRs/acc/Acc-true_MR-96_IX=VT_IXDEC=ML.csv};}
{\addlegendentry{VT, ix. dec.};}

{\addplot+ [color=blue, mark=star, dashed, mark options={solid}] table [col sep=comma,x=c,y=I] {Figures/MIRs/acc/Acc-true_MR-96_IX=VT_IXDEC=ML_clus=genie.csv};}
{\addlegendentry{VT, ix. dec., genie cluster};}

{\addplot+ [color=blue, mark=triangle, solid, mark options={solid}] table [col sep=comma,x=c,y=I] {Figures/MIRs/acc/Acc-true_MR-96_IX=VT_IXDEC=ML_ixTh=0995.csv};}
{\addlegendentry{VT, ix. dec. with $\threshIx$};}

{\addplot+ [color=blue, mark=diamond, solid, mark options={solid}] table [col sep=comma,x=c,y=I] {Figures/MIRs/acc/Acc-true_MR-96_IX=VT_IXDEC=ML_ixTh=0995_RomClus=5.csv};}
{\addlegendentry{VT, ix. thresh. dec., post-clust.};}

{\addplot+ [color=black, mark=star, dashed, mark options={solid}] table [col sep=comma,x=c,y=I] 
{
c,I
1,1.1026
2,1.4989
3,1.6668
4,1.7466
5,1.7881
6,1.8108
7,1.8245
8,1.8328
9,1.8384
10,1.8421
};}
{\addlegendentry{UC, noiseless};}

{\addplot+ [color=black, mark=star, dotted, mark options={solid}] table [col sep=comma,x=c,y=I] 
{
c,I
1,1.1786
2,1.6122
3,1.7717
4,1.8304
5,1.8520
6,1.8599
7,1.8628
8,1.8639
9,1.8643
10,1.8645
};}
{\addlegendentry{UC, noiseless, no. bias};}

\addplot[forget plot, mark=none, color=black, loosely dashdotted, opacity=1.0] coordinates {(1,1.86453886069668) (40,1.86453886069668)};

\end{axis}%
\end{tikzpicture}%
    \caption{MIRs $\avMIR$ versus coverage $c$ for the \multidraw\ channel on experimental data of \texttt{file1} ($\chM=30\,589$, $\chL=110$) for \accbc\ using $k=1$ in the single-strand decoder. We compare different decoding strategies and provide genie-based benchmarks for different scenarios. For the index reliability threshold, we use $\threshIx=0.995$, and for post-assignment, we use $\threshClus=44.0$ for the distance threshold and fix $|\poprTop|=5$.  For clarity, the horizontal black dash-dotted line represents the rate limit $(2-\beta)$ imposed by the index for error-free sampling channel transmission.}
    \label{fig:mirs-acc}
\end{figure}
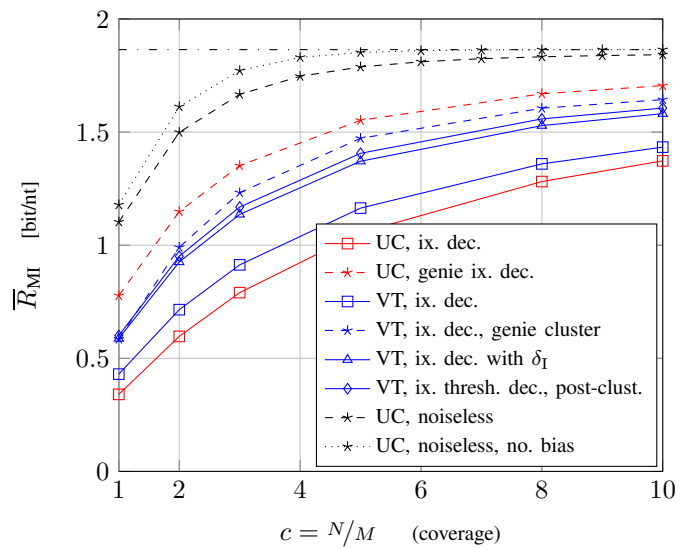

Next, we compare the decoding strategies by plotting $\avMIR$ versus coverage $c$ in \cref{fig:mirs-acc} using data simulations on \texttt{file1}. This plot highlights the different effects of the \multidraw\ \kmerCh\ channel---i.e., the full DNA storage channel---nicely.
The horizontal dash-dotted line marks the rate limit imposed by indexing and highlights the unavoidable rate loss due to the \textit{permutation} effect.
The next two curves (highest to lowest) correspond to the idealized case without strand-level noise (i.e., for which the rate loss is only due to the \textit{permutation} and \textit{sampling} effects). We observe a significant sampling-related rate loss at low coverage (caused primarily by the non-drawn strands); for increasing coverage $c$, this loss diminishes thanks to the multi-copy gain introduced by sampling.
More specifically, the two curves show the $\avMIR$ of an optimal index-based coding scheme over the \multidraw\ channel without strand noise for two drawing distributions: one curve (black, dashed, star markers) uses the drawing distribution based on our dataset, while the other (black, dotted, star markers) uses a uniform distribution $\frac{1}{\chM}$ with replacement, highlighting the rate loss due to the synthesis/PCR/sequencing bias in terms of drawing distribution (see \cref{fig:drawdist} in Appendix~\ref{sec:DrawDistribution} for a drawing distribution comparison). Comparing the two noiseless benchmarks, we conclude that without knowledge about the strand bias at the transmitter, we suffer a direct rate loss due to the sampling bias in any setup. With strand-dependent IDS noise, we include another benchmark for an index-based coding approach given an index genie in the decoding process, i.e., we artificially exclude the permutation loss of the channel (red, dashed, star markers). The rate gap to the noiseless scenario can be explained by the IDS noise and is also due to our chosen sub-optimal coding and mismatched decoding approach, i.e., simply multiplying the symbol APPs to combine reads in the decoding and memoryless outer decoding metric. This curve also functions as a true upper bound for our coding approach. The last benchmark we present uses the VT-coded index and has a clustering genie in the decoder, which groups reads that originate from the identical stored strands together before making index decisions (blue, dashed, star markers). This benchmark provides insight into the performance of our approach when leveraging the multi-copy gain, i.e., APP combining, for the index decisions as well. During optimization, we desire to approach this benchmark as closely as possible. In theory, we could surpass this benchmark, e.g., by trying to discard clusters that would group to the wrong index, which is the dominant effect responsible for the rate loss to the benchmark of the genie index decoding curve.

We can now move on to specific decoding approaches for the empirical channel (\cref{fig:mirs-acc}, solid curves) instead of the benchmark genie-aided or noise-free setups covered earlier. As discussed before, protecting the index via error-correction mechanisms produces a performance gain (VT, blue, solid, square markers versus UC, red, solid, square markers), whereas disregarding reads with unreliable index information (here, $35\%$ of all reads) boosts the possible system rate even higher (blue, solid, triangle markers). Using the post-assignment strategy of the discarded reads, i.e., assigning them to indices by comparing payload distances (see \cref{sec:postclust}) of discarded reads to the already-grouped clusters based on index decisions, provides an additional performance gain (blue, solid, diamond markers). While the fraction of discarded reads for the simple index threshold strategy is basically constant for all coverages, after the post-assignment step, we discard only $24\%$ of the reads for $c=1$ and even only $16\%$ for $c=10$. This could stem from the effect that there are fewer empty pre-grouped clusters for higher coverages, and empty pre-grouped clusters cannot benefit from the post-processing strategy. We have not observed valuable performance gains for top candidate lists of size $|\poprTop|>5$ for our parameter regimes. Overall, we find a coding/decoding system tackling and leveraging the \multidraw\ effect of the channel for the experimental data with \accbc, which approaches the genie clustering benchmark. We remark that we have deliberately not discussed any clustering approaches before index decoding since we could not find a clustering algorithm with a good complexity/performance trade-off for our number of reads. In particular, we have not found a clustering approach of DNA reads or APPs that did not dominate the system's decoding complexity while providing a performance gain for our coding/decoding setup. One possible approach that could be investigated is clustering exclusively based on the index APP distances.

For fast-BC, we observe the same effects of the channel for different coding/decoding strategies (see \cref{fig:mirs-indexcomp-fast,fig:mirs-fast} in Appendix~\ref{MIRFBase}). However, there are stronger changes in performance measured in $\avMIR$ for different setups. Moreover, there is still a bigger gap between the benchmark of the clustering genie curve and our best setup. The best setup uses an ES-coded index, an inner code rate of $\frac{168}{110}$, an index reliability threshold of $\threshIx=0.975$, a clustering threshold of $\threshClus=42.5$, and  $|\poprTop_\bfy|=5$ during post-assignment.

\subsection{Outage and FER Results} \label{sec:simresults-outageFER}

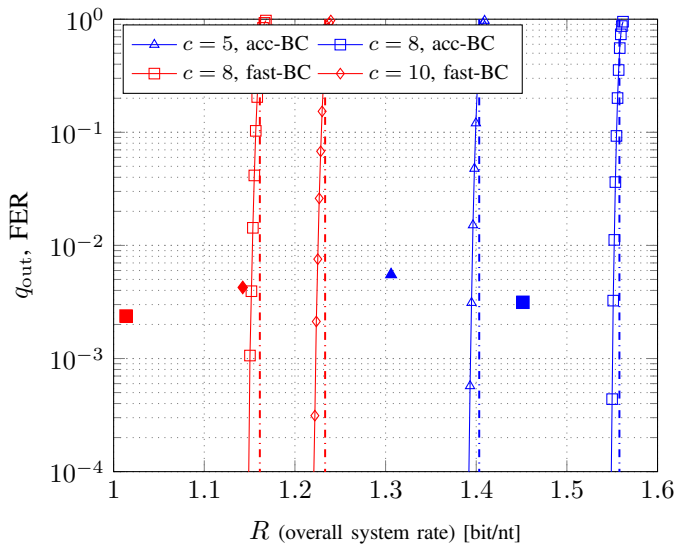
\begin{figure}[t]
    \centering
    \begin{tikzpicture}
\begin{axis}[
width = 0.995\linewidth,
xmin=1.0, xmax=1.6,
ymin=1e-04, ymax=1,
xlabel={$R$  {\footnotesize (overall system rate) [bit/nt]}},
ylabel={$\qout$,\ FER},
ylabel style={
	yshift=0ex,
	name=label},
grid style={gray,opacity=0.5,dotted},
xmajorgrids,
ymajorgrids, yminorgrids,
ymode=log,
axis background/.style={fill=white},
legend columns = 2,
legend style={font=\footnotesize},
legend cell align=left,
legend style={at={(axis cs: 1.01,9e-1)}, anchor=north west},
cycle list name=color list
]

{\addplot+ [color=blue, mark=triangle, solid, mark options={solid}, each nth point=4, filter discard warning=false, unbounded coords=discard] table [col sep=comma,x=R,y=qout] {Figures/outage/acc-true/outagefile-2_acc-true_c=5.000_.csv};}
{\addlegendentry{$c=5$, \accbc};}

\draw [ dashdotted, line width=0.75pt, color = blue] (axis cs:1.4032922177777971,1) -- (axis cs:1.4032922177777971,0.00000001);

{\addplot+ [color=blue, mark=square, solid, mark options={solid,fill},each nth point=4, filter discard warning=false, unbounded coords=discard] table [col sep=comma,x=R,y=qout] {Figures/outage/acc-true/outagefile-2_acc-true_c=8.000_.csv};}
{\addlegendentry{$c=8$, \accbc};}

\draw [ dashdotted, line width=0.75pt, color = blue] (axis cs:1.557954015625016,1) -- (axis cs:1.557954015625016,0.00000001);

{\addplot+ [color=red, mark=square, solid, mark options={solid},each nth point=4, filter discard warning=false, unbounded coords=discard] table [col sep=comma,x=R,y=qout] {Figures/outage/acc-false/outagefile-2_acc-false_c=8.000_.csv};}
{\addlegendentry{$c=8$, \fastbc};}

\draw [ dashdotted, line width=0.75pt, color = red] (axis cs:1.1612950968749929,1) -- (axis cs:1.1612950968749929,0.00000001);

{\addplot+ [color=red, mark=diamond, solid, mark options={solid,fill},each nth point=4, filter discard warning=false, unbounded coords=discard] table [col sep=comma,x=R,y=qout] {Figures/outage/acc-false/outagefile-2_acc-false_c=10.00_.csv};}
{\addlegendentry{$c=10$, \fastbc};}

\draw [ dashdotted, line width=0.75pt, color = red] (axis cs:1.233309496875003,1) -- (axis cs:1.233309496875003,0.00000001);

\addplot [forget plot, mark=triangle*, line width=1.0pt, color=blue, mark options={solid, line width = 0.5pt, fill=blue, mark size=2.3pt}] table[x=r,y=FER] {
r FER
1.30606060606061 0.005523809523809524
};
\addplot [forget plot, mark=square*, line width=1.0pt, color=blue, mark options={solid, line width = 0.5pt, fill=blue, mark size=2.3pt}] table[x=r,y=FER] {
r FER
1.45151515151515 0.0031417941298057048
};

\addplot [forget plot, mark=square*, line width=1.0pt, color=red, mark options={solid, line width = 0.5pt, fill=red, mark size=2.3pt}] table[x=r,y=FER] {
r FER
1.01369223819204 0.002375
};
\addplot [forget plot, mark=diamond*, line width=1.0pt, color=red, mark options={solid, line width = 0.5pt, fill=red, mark size=2.3pt}] table[x=r,y=FER] {
r FER
1.14242424242424 0.00425
};

\end{axis}%
\end{tikzpicture}%
    \caption{Information-outage probability $\qout$ and FER versus the overall system rate $R$ for \accbc\ and \fastbc\ and various coverages $c$ on experimental data of \texttt{file3} ($\chM=30\,588$, $\chL=110$). We apply the decoding strategy by pre-assigning reads with the index threshold and applying post-assignment. Parameters for \accbc: VT-coded index, an inner code with $\Rin=\nicefrac{192}{110}$, $\threshIx=0.995$, $\threshClus=44$, $\lvert \poprTop \rvert = 5$, and $k = 3$. Parameters for \fastbc: ES-coded index, an inner code with $\Rin=\nicefrac{168}{110}$, $\threshIx=0.975$, $\threshClus=42.5$, $\lvert \poprTop \rvert = 5$, and $k = 3$. The stand-alone markers correspond to the FER performance using an SC-LDPC code ensemble (see \cref{tab:scldpc-params} for parameters).}
    \label{fig:outage-fer}
\end{figure}

\cref{fig:outage-fer} shows the information-outage probability $\qout$ versus the overall system rate $R$ on our four channel setups using the best respective index/inner coding and decoding scheme as described in the previous subsections simulated on data of \texttt{file3} using $\sim\!\!16\, 000$ Monte-Carlo simulations. Recall that the mismatched information-outage probability $\qout$ represents a lower bound on the FER for the whole system at the given overall system rate $R$ for any outer code using the decoding metric from \cref{sec:dec-outer}. We included the value $\avMIR$ for each channel setup as a vertical line. We can see that for all setups, the waterfall of the outage is very steep. We attribute this to the long blocklengths of $\leno = 2\, 569\, 392\,\text{nt}$ for \fastbc\ and $\leno = 2\, 936\, 448\,\text{nt}$ for \accbc, the comparatively large numbers of blocks $\chM = 30\, 588$, and the fact that the sampling drawing distribution mimics very closely the population of the drawing distribution for our choices of $c$ and fixed $\beta$. Moreover, we infer that using $\avMIR$ as the code optimization criterion is reasonable for our parameters. For a fixed $\qout$, we achieve the highest possible system rate for the setup with $c=8$ and \accbc, since the IDS error rates are low and the coverage is relatively high. 

For the above channel setups, we also include FER results (solid shaped markers, \cref{fig:outage-fer}) for the protograph-based SC-LDPC code ensembles with parameters depicted in \cref{tab:scldpc-params}. We can ensure a FER  around $10^{-3}$ while the constructed outer codes have an outer code rate $\Rout$ of $5$--$10\%$ far from the value given by the outage bound for $\qout \approx 10^{-3}$.  (Note that in \cref{fig:outage-fer}, the overall system rate $R$ is used and not the outer code rate $\Rout$.) Formally, the results are ensemble averages of the outer code due to the random lifting and edge weight assignment, the random interleaver, and the offset of the joint code (see \cref{sec:datasetsim}).

Finally, the stored information amount for \texttt{file3} would evaluate for \accbc\ with $c=5$ to $549.3 \,\text{kByte}$ and with $c=8$ to $610.5 \,\text{kByte}$, and for \fastbc\ with $c=8$ to $426.5 \,\text{kByte}$ and with $c=10$ to $480.5 \,\text{kByte}$.

\begin{table}[t]
\begin{center}
\caption{SC-LDPC Code Parameters$^a$}
\label{tab:scldpc-params}
\vspace{-2.3ex}
\begin{tabular}{@{\hskip.05cm}ccccccc@{\hskip.05cm}}
\toprule 
   Basecalling & $c$ & $\Rout$ {\footnotesize [bit/nt]} & $\bfB_i$ & $\scmem$ & $\sccoup$ & $\protlift$ \\
      \midrule
    \accbc & $5$ & $1.497$ & $[1,1,1,1]$ & $2$ & $288$ & $2549$ \\    
    \accbc & $8$ & $1.663$ & $[1,1,1,1,1,1]$ & $2$ & $192$ & $2549$ \\   
    \fastbc & $8$ & $1.327$ & $[1,1,1]$ & 2 & $227$ & $3773$ \\
    \fastbc & $10$ & $1.496$ & $[1,1,1,1]$ & 2 & $252$ & $2549$ \\
    \bottomrule
\end{tabular}
\end{center}
\footnotesize{$^a$For all setups, sliding window decoding from front and end simultaneously with $5$ BP iterations per window position is performed. Window size is fixed to $9$ spatial positions. Double the BP iterations are performed at the front and back until the window is slid fully into the Tanner graph.} 
\end{table}

\section{Conclusion}
\label{sec:conclusion}

We proposed a competitive end-to-end coding system that can be optimized given the specific DNA storage channel requirements. Besides code design, we presented fitting low-complexity (only linear in the total number of reads $\chN$) decoding techniques that push the performance of the proposed system toward its fundamental limits. Further, we have shown that using strand-dependent error modeling in the decoder improves the performance of the system. Finally, we have obtained an explicit construction that outperforms state-of-the-art techniques in terms of read/write cost trade-off with FER of magnitude $10^{-3}$, where the results are obtained by simulations on true DNA storage data with a nanopore sequencer.

There are several future interesting research directions. To name a few, we could try to use neural network decoders (similar as in \cite{stanford_magneticDNAstorage_2023} directly on the signal) or try to reduce the model mismatch at the decoder's side to the true DNA storage error channel by replacing the BCJR decoder with a clever neural network architecture (similar as in \cite{bar2021deep}). A more comprehensive future step would be to design the system for the idea of composite DNA \cite{Anavy2019composite} since this approach seems promising to reduce the writing cost even further by a (possibly) slight increase of the reading cost.

\bibliographystyle{IEEEtran}
\bibliography{bib/biblio.bib}

\begin{thebibliography}{10}
\providecommand{\url}[1]{#1}
\csname url@samestyle\endcsname
\providecommand{\newblock}{\relax}
\providecommand{\bibinfo}[2]{#2}
\providecommand{\BIBentrySTDinterwordspacing}{\spaceskip=0pt\relax}
\providecommand{\BIBentryALTinterwordstretchfactor}{4}
\providecommand{\BIBentryALTinterwordspacing}{\spaceskip=\fontdimen2\font plus
\BIBentryALTinterwordstretchfactor\fontdimen3\font minus \fontdimen4\font\relax}
\providecommand{\BIBforeignlanguage}[2]{{%
\expandafter\ifx\csname l@#1\endcsname\relax
\typeout{** WARNING: IEEEtran.bst: No hyphenation pattern has been}%
\typeout{** loaded for the language `#1'. Using the pattern for}%
\typeout{** the default language instead.}%
\else
\language=\csname l@#1\endcsname
\fi
#2}}
\providecommand{\BIBdecl}{\relax}
\BIBdecl

\bibitem{multidrawConCat_ITW_2023}
L.~Welter, I.~Maarouf, A.~Lenz, A.~Wachter-Zeh, E.~Rosnes, and A.~Graell~i Amat, ``Index-based concatenated codes for the multi-draw {DNA} storage channel,'' in \emph{Proc. IEEE Inf. Theory Workshop}, Saint-Malo, France, Apr. 2023, pp. 383--388.

\bibitem{IssamEirikAlex_AIRkmer_23}
I.~Maarouf, E.~Rosnes, and A.~Graell~i Amat, ``Achievable information rates and concatenated codes for the {DNA} nanopore sequencing channel,'' in \emph{Proc. IEEE Inf. Theory Workshop}, Saint-Malo, France, Apr. 2023, pp. 377--382.

\bibitem{feynman1959}
R.~P. Feynman, ``There's plenty of room at the bottom,'' \emph{Talk at the Meeting of the American Physical Society at the California Institute of Technology}, 1959.

\bibitem{davis1996microvenus}
J.~Davis, ``Microvenus,'' \emph{Art J.}, vol.~55, no.~1, pp. 70--74, 1996.

\bibitem{clelland1999hiding}
C.~T. Clelland, V.~Risca, and C.~Bancroft, ``Hiding messages in {DNA} microdots,'' \emph{Nature}, vol. 399, no. 6736, pp. 533--534, Jun. 1999.

\bibitem{leier2000cryptography}
A.~Leier, C.~Richter, W.~Banzhaf, and H.~Rauhe, ``Cryptography with {DNA} binary strands,'' \emph{Biosystems}, vol.~57, no.~1, pp. 13--22, Jun. 2000.

\bibitem{bancroft2001long}
C.~Bancroft, T.~Bowler, B.~Bloom, and C.~T. Clelland, ``Long-term storage of information in {DNA},'' \emph{Science}, vol. 293, no. 5536, pp. 1763--1765, Sep. 2001.

\bibitem{gibson2010creation}
D.~G. Gibson, J.~I. Glass, C.~Lartigue, V.~N. Noskov, R.-Y. Chuang, M.~A. Algire, G.~A. Benders, M.~G. Montague, L.~Ma, M.~M. Moodie \emph{et~al.}, ``Creation of a bacterial cell controlled by a chemically synthesized genome,'' \emph{Science}, vol. 329, no. 5987, pp. 52--56, May 2010.

\bibitem{church_next-generation_2012}
G.~M. Church, Y.~Gao, and S.~Kosuri, ``\BIBforeignlanguage{en}{Next-generation digital information storage in {DNA}},'' \emph{\BIBforeignlanguage{en}{Science}}, vol. 337, no. 6102, p. 1628, Aug. 2012.

\bibitem{goldman_towards_2013}
N.~Goldman, P.~Bertone, S.~Chen, C.~Dessimoz, E.~M. LeProust, B.~Sipos, and E.~Birney, ``Towards practical, high-capacity, low-maintenance information storage in synthesized {DNA},'' \emph{Nature}, vol. 494, no. 7435, pp. 77--80, Jan. 2013.

\bibitem{grass_robust_2015}
R.~N. Grass, R.~Heckel, M.~Puddu, D.~Paunescu, and W.~J. Stark, ``Robust chemical preservation of digital information on {DNA} in silica with error-correcting codes,'' \emph{Angew. Chem. Int. Ed.}, vol.~54, no.~8, pp. 2552--2555, Feb. 2015.

\bibitem{yazdi_rewritable_2015}
S.~M. H.~T. Yazdi, Y.~Yuan, J.~Ma, H.~Zhao, and O.~Milenkovic, ``A rewritable, random-access {{DNA}}-based storage system,'' \emph{Sci. Rep.}, vol.~5, no. 14138, pp. 1--10, Sep. 2015.

\bibitem{erlich_dna_2017}
Y.~Erlich and D.~Zielinski, ``\BIBforeignlanguage{en}{{DNA} fountain enables a robust and efficient storage architecture},'' \emph{\BIBforeignlanguage{en}{Science}}, vol. 355, no. 6328, pp. 950--954, Mar. 2017.

\bibitem{yazdi_portable_2017}
S.~M. H.~T. Yazdi, R.~Gabrys, and O.~Milenkovic, ``\BIBforeignlanguage{en}{Portable and error-free {DNA}-based data storage},'' \emph{\BIBforeignlanguage{en}{Sci. Rep.}}, vol.~7, no. 5011, pp. 1--6, Jul. 2017.

\bibitem{organick_random_2018}
L.~Organick, S.~D. Ang, Y.-J. Chen, R.~Lopez, S.~Yekhanin, K.~Makarychev, M.~Z. Racz, G.~Kamath, P.~Gopalan, B.~Nguyen, C.~N. Takahashi, S.~Newman, H.-Y. Parker, C.~Rashtchian, K.~Stewart, G.~Gupta, R.~Carlson, J.~Mulligan, D.~Carmean, G.~Seelig, L.~Ceze, and K.~Strauss, ``\BIBforeignlanguage{en}{Random access in large-scale {DNA} data storage},'' \emph{\BIBforeignlanguage{en}{Nature Biotechnol.}}, vol.~36, no.~3, pp. 242--248, Mar. 2018.

\bibitem{antkowiak_low_2020}
P.~L. Antkowiak, J.~Lietard, M.~Z. Darestani, M.~M. Somoza, W.~J. Stark, R.~Heckel, and R.~N. Grass, ``Low cost {{DNA}} data storage using photolithographic synthesis and advanced information reconstruction and error correction,'' \emph{Nature Commun.}, vol.~11, no. 5345, pp. 1--10, Oct. 2020.

\bibitem{bornholt_dna-based_2016}
J.~Bornholt, R.~Lopez, D.~M. Carmean, L.~Ceze, G.~Seelig, and K.~Strauss, ``\BIBforeignlanguage{en}{A {DNA}-based archival storage system},'' in \emph{\BIBforeignlanguage{en}{Proc. ACM {Int}. {Conf}. {Architect.} {Supp.} {Program.} {Lang.} {Operat.} {Syst.}}}, Atlanta, GA, USA, Apr. 2016, pp. 637--649.

\bibitem{Lopez2019}
R.~Lopez, Y.-J. Chen, S.~D. Ang, S.~Yekhanin, K.~Makarychev, M.~Z. Racz, G.~Seelig, K.~Strauss, and L.~Ceze, ``{DNA} assembly for nanopore data storage readout,'' \emph{Nature Commun.}, vol.~10, no. 2933, pp. 1--9, Jul. 2019.

\bibitem{Anavy2019composite}
L.~Anavy, I.~Vaknin, O.~Atar, R.~Amit, and Z.~Yakhini, ``Data storage in {DNA} with fewer synthesis cycles using composite {DNA} letters,'' \emph{Nature Biotechnology}, vol.~37, no.~10, pp. 1229--1236, Oct. 2019.

\bibitem{stanford_magneticDNAstorage_2023}
B.~Lau, S.~Chandak, S.~Roy, K.~Tatwawadi, M.~Wootters, T.~Weissman, and H.~P. Ji, ``Magnetic {DNA} random access memory with nanopore readouts and exponentially-scaled combinatorial addressing,'' \emph{Sci. Rep.}, vol.~13, no. 8514, pp. 1--15, May 2023.

\bibitem{heckel_characterization_2019}
R.~Heckel, G.~Mikutis, and R.~N. Grass, ``\BIBforeignlanguage{en}{A characterization of the {DNA} data storage channel},'' \emph{\BIBforeignlanguage{en}{Sci. Rep.}}, vol.~9, no. 9663, pp. 1--12, Jul. 2019.

\bibitem{HeckelShomorony_Foundations_2022}
I.~Shomorony and R.~Heckel, ``Information-theoretic foundations of {DNA} data storage,'' \emph{Found. Trends Commun. Inf. Theory}, vol.~19, no.~1, pp. 1--106, Feb. 2022.

\bibitem{heckel_fundamental_ISIT_2017}
R.~Heckel, I.~Shomorony, K.~Ramchandran, and D.~N.~C. Tse, ``Fundamental limits of {DNA} storage systems,'' in \emph{Proc. IEEE Int. Symp. Inf. Theory}, Aachen, Germany, Jun. 2017, pp. 3130--3134.

\bibitem{lenz_upperboundcapDNA_2019}
A.~Lenz, P.~H. Siegel, A.~Wachter-Zeh, and E.~Yaakobi, ``An upper bound on the capacity of the {DNA} storage channel,'' in \emph{Proc. IEEE Inf. Theory Workshop}, Visby, Sweden, Aug. 2019, pp. 1--5.

\bibitem{lenz_achieving_2020}
------, ``Achieving the capacity of the {DNA} storage channel,'' in \emph{Proc. IEEE {Int}. {Conf}. {Acoust}., {Speech}, {Sig}. {Process}.}, Barcelona, Spain, May 2020, pp. 8846--8850.

\bibitem{lenz_noisyDrawChan_2023}
------, ``The noisy drawing channel: Reliable data storage in {DNA} sequences,'' \emph{IEEE Trans. Inf. Theory}, vol.~69, no.~5, pp. 2757--2778, May 2023.

\bibitem{weinberger_dna_2022}
N.~Weinberger and N.~Merhav, ``The {DNA} storage channel: Capacity and error probability bounds,'' \emph{IEEE Trans. Inf. Theory}, vol.~68, no.~9, pp. 5657--5700, Sep. 2022.

\bibitem{weinberger_codedindex_2022}
N.~Weinberger, ``Error probability bounds for coded-index {DNA} storage systems,'' \emph{IEEE Trans. Inf. Theory}, vol.~68, no.~11, pp. 7005--7022, Nov. 2022.

\bibitem{levick2022achieving}
K.~Levick, R.~Heckel, and I.~Shomorony, ``Achieving the capacity of a {DNA} storage channel with linear coding schemes,'' in \emph{Proc. 56th Annu. Conf. Inf. Sci. Syst.}, Princeton, NJ, USA, Mar. 2022, pp. 218--223.

\bibitem{lenz_concatachieve_2020}
A.~Lenz, L.~Welter, and S.~Puchinger, ``Achievable rates of concatenated codes in {{DNA}} storage under substitution errors,'' in \emph{Proc. IEEE Int. Symp. Inf. Theory Appl.}, Kapolei, HI, USA, Oct. 2020, pp. 269--273.

\bibitem{polarDNAshuffle2023haghighat}
J.~Haghighat and T.~M. Duman, ``A practical concatenated coding scheme for noisy shuffling channels with coset-based indexing,'' in \emph{Proc. IEEE Global Commun. Conf.}, Kuala Lumpur, Malaysia, Dec. 2023, pp. 1842--1847.

\bibitem{davey_reliable_2001}
M.~C. Davey and D.~J.~C. MacKay, ``Reliable communication over channels with insertions, deletions, and substitutions,'' \emph{IEEE Trans. Inf. Theory}, vol.~47, no.~2, pp. 687--698, Feb. 2001.

\bibitem{ratzer_marker_2005}
E.~A. Ratzer, ``\BIBforeignlanguage{en}{Marker codes for channels with insertions and deletions},'' \emph{\BIBforeignlanguage{en}{Annales Télécommun.}}, vol.~60, no. 1-2, pp. 29--44, Feb. 2005.

\bibitem{briffa_improved_2010}
J.~A. Briffa, H.~G. Schaathun, and S.~Wesemeyer, ``An improved decoding algorithm for the {Davey}-{MacKay} construction,'' in \emph{Proc. IEEE {Int}. {Conf}. {Commun}.}, Cape Town, South Africa, May 2010, pp. 1--5.

\bibitem{buttigieg_improved_2015}
V.~Buttigieg and N.~Farrugia, ``Improved bit error rate performance of convolutional codes with synchronization errors,'' in \emph{Proc. IEEE Int. Conf. Commun.}, London, U.K., Jun. 2015, pp. 4077--4082.

\bibitem{banerjee_sequential_2022}
A.~Banerjee, A.~Lenz, and A.~Wachter-Zeh, ``Sequential decoding of convolutional codes for synchronization errors,'' in \emph{Proc. IEEE Inf. Theory Workshop}, Mumbai, India, Nov. 2022, pp. 630--635.

\bibitem{maarouf2023finite}
I.~Maarouf, G.~Liva, E.~Rosnes, and A.~Graell~i Amat, ``Finite blocklength performance bound for the {DNA} storage channel,'' in \emph{Proc. 12th Int. Symp. Topics Coding}, Brest, France, Sep. 2023, pp. 1--5.

\bibitem{carrillo1988multiple}
H.~Carrillo and D.~Lipman, ``The multiple sequence alignment problem in biology,'' \emph{SIAM J. Appl. Math.}, vol.~48, no.~5, pp. 1073--1082, Oct. 1988.

\bibitem{levenshtein2001efficient}
V.~I. Levenshtein, ``Efficient reconstruction of sequences,'' \emph{IEEE Trans. Inf. Theory}, vol.~47, no.~1, pp. 2--22, Jan. 2001.

\bibitem{batu2004reconstructing}
T.~Batu, S.~Kannan, S.~Khanna, and A.~McGregor, ``Reconstructing strings from random traces,'' in \emph{Proc. 15th Annu. ACM-SIAM Symp. Discrete Algorithms}, New Orleans, LA, USA, Jan. 2004, pp. 910--918.

\bibitem{holenstein2008trace}
T.~Holenstein, M.~Mitzenmacher, R.~Panigrahy, and U.~Wieder, ``Trace reconstruction with constant deletion probability and related results,'' in \emph{Proc. 19th Annu. ACM-SIAM Symp. Discrete Algorithms}, San Francisco, CA, USA, Jan. 2008, pp. 389--398.

\bibitem{peres2017average}
Y.~Peres and A.~Zhai, ``Average-case reconstruction for the deletion channel: Subpolynomially many traces suffice,'' in \emph{Proc. 58th IEEE Annu. Symp. Found. Comput. Sci. (FOCS)}, Berkeley, CA, USA, Oct. 2017, pp. 228--239.

\bibitem{holden2020subpolynomial}
N.~Holden, R.~Pemantle, Y.~Peres, and A.~Zhai, ``Subpolynomial trace reconstruction for random strings and arbitrary deletion probability,'' \emph{Math. Stat. Learn.}, vol.~2, no. 3/4, pp. 275--309, Oct. 2020.

\bibitem{Srinivasavaradhan_2021}
S.~R. Srinivasavaradhan, M.~Du, S.~N. Diggavi, and C.~Fragouli, ``Algorithms for reconstruction over single and multiple deletion channels,'' \emph{IEEE Trans. Inf. Theory}, vol.~67, no.~6, pp. 3389--3410, Jun. 2021.

\bibitem{sakogawa2020symbolwise}
R.~Sakogawa and H.~Kaneko, ``Symbolwise {MAP} estimation for multiple-trace insertion/deletion/substitution channels,'' in \emph{Proc. Int. Symp. Inf. Theory}.\hskip 1em plus 0.5em minus 0.4em\relax Los Angeles, USA: {IEEE}, Jun. 2020, pp. 781--785.

\bibitem{press2020hedges}
W.~H. Press, J.~A. Hawkins, S.~K. {Jones Jr}, J.~M. Schaub, and I.~J. Finkelstein, ``{HEDGES} error-correcting code for {DNA} storage corrects indels and allows sequence constraints,'' \emph{Proc. Nat. Acad. Sci.}, vol. 117, no.~31, pp. 18\,489--18\,496, Aug. 2020.

\bibitem{sabary2024reconstruction}
O.~Sabary, A.~Yucovich, G.~Shapira, and E.~Yaakobi, ``Reconstruction algorithms for {DNA}-storage systems,'' \emph{Sci. Rep.}, vol.~14, no. 1951, pp. 1--15, Jan. 2024.

\bibitem{Srinivasavaradhan2021TrellisBMA}
S.~R. Srinivasavaradhan, S.~Gopi, H.~D. Pfister, and S.~Yekhanin, ``Trellis {BMA}: Coded trace reconstruction on {IDS} channels for {DNA} storage,'' in \emph{Proc. IEEE Int. Symp. Inf. Theory}, Melbourne, Australia, Jul. 2021, pp. 2493--2498.

\bibitem{Maarouf2022ConcatenatedCF}
I.~Maarouf, A.~Lenz, L.~Welter, A.~Wachter-Zeh, E.~Rosnes, and A.~{Graell i Amat}, ``Concatenated codes for multiple reads of a {DNA} sequence,'' \emph{IEEE Trans. Inf. Theory}, vol.~69, no.~2, pp. 910--927, Feb. 2023.

\bibitem{hamoum_conf_aplanmer_2021}
B.~Hamoum, E.~Dupraz, L.~Conde-Canencia, and D.~Lavenier, ``Channel model with memory for {DNA} data storage with nanopore sequencing,'' in \emph{Proc. 11th Int. Symp. Topics Coding}, Montreal, QC, Canada, Aug./Sep. 2021, pp. 1--5.

\bibitem{frenchKmer_belaid_2023}
B.~Hamoum and E.~Dupraz, ``Channel model and decoder with memory for {DNA} data storage with nanopore sequencing,'' \emph{IEEE Access}, vol.~11, pp. 52\,075--52\,087, 2023.

\bibitem{mao2018models}
W.~Mao, S.~N. Diggavi, and S.~Kannan, ``Models and information-theoretic bounds for nanopore sequencing,'' \emph{IEEE Trans. Inf. Theory}, vol.~64, no.~4, pp. 3216--3236, Apr. 2018.

\bibitem{hulett2021coding}
R.~Hulett, S.~Chandak, and M.~Wootters, ``On coding for an abstracted nanopore channel for {DNA} storage,'' in \emph{Proc. IEEE Int. Symp. Inf. Theory}, Melbourne, Australia, Jul. 2021, pp. 2465--2470.

\bibitem{mcbain2022finite}
B.~McBain, E.~Viterbo, and J.~Saunderson, ``Information rates of the noisy nanopore channel,'' \emph{IEEE Trans. Inf. Theory}, 2024, early access.

\bibitem{banerjee2023error}
A.~Banerjee, Y.~Yehezkeally, A.~Wachter-Zeh, and E.~Yaakobi, ``Error-correcting codes for nanopore sequencing,'' \emph{IEEE Trans. Inf. Theory}, 2024, early access.

\bibitem{concatViterbo2024}
A.~Vidal, V.~B. Wijekoon, and E.~Viterbo, ``Concatenated nanopore {DNA} codes,'' \emph{IEEE Trans. NanoBioscience}, vol.~23, no.~2, pp. 310--318, Apr. 2024.

\bibitem{chandak_overcoming_2020}
S.~Chandak, J.~Neu, K.~Tatwawadi, J.~Mardia, B.~Lau, M.~Kubit, R.~Hulett, P.~Grifﬁn, M.~Wootters, T.~Weissman, and H.~Ji, ``\BIBforeignlanguage{en}{Overcoming high nanopore basecaller error rates for {DNA} storage via basecaller-decoder integration and convolutional codes},'' in \emph{\BIBforeignlanguage{en}{Proc. IEEE {Int}. {Conf}. {Acoust}., {Speech}, {Sig}. {Process}.}}, Barcelona, Spain, May 2020, pp. 8822--8826.

\bibitem{rashtchian_clustering_2017}
C.~Rashtchian, K.~Makarychev, M.~R{\'a}cz, S.~D. Ang, D.~Jevdjic, S.~Yekhanin, L.~Ceze, and K.~Strauss, ``\BIBforeignlanguage{en}{Clustering billions of reads for {DNA} data storage},'' in \emph{\BIBforeignlanguage{en}{Proc. Adv. Neural Inf. Process. Syst.}}, Long Beach, CA, USA, Dec. 2017, pp. 3360--3371.

\bibitem{roman2023experimentaldata}
\BIBentryALTinterwordspacing
R.~Sokolovskii and L.~Welter, ``Experimental data for ``{An} end-to-end coding scheme for {DNA}-based data storage with nanopore sequenced reads'','' 2024. [Online]. Available: \url{https://zenodo.org/doi/10.5281/zenodo.10943281}
\BIBentrySTDinterwordspacing

\bibitem{del21troubles}
C.~Delahaye and J.~Nicolas, ``Sequencing {DNA} with nanopores: Troubles and biases,'' \emph{PloS ONE}, vol.~16, no.~10, pp. 1--29, Oct. 2021.

\bibitem{milenkovic_2ndStru_2005}
O.~Milenkovic and N.~Kashyap, ``{DNA} codes that avoid secondary structures,'' in \emph{Proc. IEEE Int. Symp. Inf. Theory}, Adelaide, Australia, Sep. 2005, pp. 288--292.

\bibitem{felstrom1999time}
A.~J. Felstr{\"o}m and K.~S. Zigangirov, ``Time-varying periodic convolutional codes with low-density parity-check matrix,'' \emph{IEEE Trans. Inf. Theory}, vol.~45, no.~6, pp. 2181--2191, Sep. 1999.

\bibitem{lentmaier_SCLDPC_2010}
M.~Lentmaier, A.~Sridharan, D.~J. {Costello, Jr.}, and K.~S. Zigangirov, ``Iterative decoding threshold analysis for {LDPC} convolutional codes,'' \emph{IEEE Trans. Inf. Theory}, vol.~56, no.~10, pp. 5274--5289, Oct. 2010.

\bibitem{inoue_adaptivemarker2012}
M.~Inoue and H.~Kaneko, ``Adaptive synchronization marker for insertion/deletion/substitution error correction,'' in \emph{Proc. IEEE Int. Symp. Inf. Theory}, Cambridge, MA, USA, Jul. 2012, pp. 508--512.

\bibitem{BCJR_1974}
L.~R. Bahl, J.~Cocke, F.~Jelinek, and J.~Raviv, ``Optimal decoding of linear codes for minimizing symbol error rate,'' \emph{IEEE Trans. Inf. Theory}, vol.~20, no.~2, pp. 284--287, Mar. 1974.

\bibitem{bahl_decoding_1975}
L.~R. Bahl and F.~Jelinek, ``\BIBforeignlanguage{en}{Decoding for channels with insertions, deletions, and substitutions with applications to speech recognition},'' \emph{\BIBforeignlanguage{en}{IEEE Trans. Inf. Theory}}, vol.~21, no.~4, pp. 404--411, Jul. 1975.

\bibitem{kaplan1993information}
G.~Kaplan and S.~Shamai, ``Information rates and error exponents of compound channels with application to antipodal signaling in a fading environment,'' \emph{AEU Int. J. Electron. Commun.}, vol.~47, no.~4, pp. 228--239, Jul. 1993.

\bibitem{ARXIV_Srinivasavaradhan2021TrellisBMA}
S.~R. Srinivasavaradhan, S.~Gopi, H.~D. Pfister, and S.~Yekhanin, ``Trellis {BMA}: Coded trace reconstruction on {IDS} channels for {DNA} storage,'' Jul. 2021, arXiv preprint arXiv:2107:06440.

\bibitem{BLAST}
S.~F. Altschul, W.~Gish, W.~Miller, E.~W. Myers, and D.~J. Lipman, ``Basic local alignment search tool,'' \emph{J. Mol. Biol.}, vol. 215, no.~3, pp. 403--410, Oct. 1990.

\bibitem{wagner1974string}
R.~A. Wagner and M.~J. Fischer, ``The string-to-string correction problem,'' \emph{J. ACM}, vol.~21, no.~1, pp. 168--173, Jan. 1974.

\bibitem{Baum1970AMT}
L.~E. Baum, T.~Petrie, G.~Soules, and N.~Weiss, ``A maximization technique occurring in the statistical analysis of probabilistic functions of {Markov} chains,'' \emph{Annals Math. Stats.}, vol.~41, no.~1, pp. 164--171, Feb. 1970.

\bibitem{chandak_improved_2019}
S.~Chandak, K.~Tatwawadi, B.~Lau, M.~Kubit, J.~Mardia, J.~Neu, P.~Griffin, M.~Wootters, T.~Weissman, and H.~Ji, ``\BIBforeignlanguage{en}{Improved read/write cost tradeoff in {DNA}-based data storage using {LDPC} codes},'' in \emph{\BIBforeignlanguage{en}{Proc. {Annu}. {Allerton} {Conf}. {Commun}., {Control}, {Comput}.}}, Monticello, IL, USA, Sep. 2019, pp. 147--156.

\bibitem{Kavcic2003DE}
A.~Kav{\vv c}i{\'c}, X.~Ma, and M.~Mitzenmacher, ``Binary intersymbol interference channels: Gallager codes, density evolution, and code performance bounds,'' \emph{IEEE Trans. Inf. Theory}, vol.~49, no.~7, pp. 1636--1652, Jul. 2003.

\bibitem{muller_capacity_2004}
R.~R. M{\"u}ller and W.~H. Gerstacker, ``On the capacity loss due to separation of detection and decoding,'' \emph{IEEE Trans. Inf. Theory}, vol.~50, no.~8, pp. 1769--1778, Aug. 2004.

\bibitem{soriaga_determining_2007}
J.~B. Soriaga, H.~D. Pfister, and P.~H. Siegel, ``Determining and approaching achievable rates of binary intersymbol interference channels using multistage decoding,'' \emph{IEEE Trans. Inf. Theory}, vol.~53, no.~4, pp. 1416--1429, Apr. 2007.

\bibitem{buckingham_informationoutage_2008}
D.~Buckingham and M.~C. Valenti, ``The information-outage probability of finite-length codes over {AWGN} channels,'' in \emph{Proc. 42nd Annu. Conf. Inf. Sci. Syst.}, Princeton, NJ, USA, Mar. 2008, pp. 390--395.

\bibitem{shomorony_dna-based_2021}
I.~Shomorony and R.~Heckel, ``{DNA}-based storage: Models and fundamental limits,'' \emph{IEEE Trans. Inf. Theory}, vol.~67, no.~6, pp. 3675--3689, Jun. 2021.

\bibitem{Sewell1998clique2}
E.~C. Sewell, ``A branch and bound algorithm for the stability number of a sparse graph,'' \emph{INFORMS J. Comput.}, vol.~10, no.~4, pp. 438--447, Nov. 1998.

\bibitem{tenengolts_nonbinVT_1984}
G.~Tenengolts, ``Nonbinary codes, correcting single deletion or insertion,'' \emph{IEEE Trans. Inf. Theory}, vol.~30, no.~5, pp. 766--769, Sep. 1984.

\bibitem{siegel2010windowBPSCLPDC}
M.~Papaleo, A.~R. Iyengar, P.~H. Siegel, J.~K. Wolf, and G.~E. Corazza, ``Windowed erasure decoding of {LDPC} convolutional codes,'' in \emph{Proc. IEEE Inf. Theory Workshop}, Cairo, Eqypt, Jan. 2010, pp. 1--5.

\bibitem{hassan_diversityscldpc_2014}
N.~ul~Hassan, M.~Lentmaier, I.~Andriyanova, and G.~P. Fettweis, ``Improving code diversity on block-fading channels by spatial coupling,'' in \emph{Proc. IEEE Int. Symp. Inf. Theory}, Honolulu, HI, USA, Jun./Jul. 2014, pp. 2311--2315.

\bibitem{7112076}
Y.~Fang, G.~Bi, Y.~L. Guan, and F.~C.~M. Lau, ``A survey on protograph {LDPC} codes and their applications,'' \emph{IEEE Commun. Surv. Tut.}, vol.~17, no.~4, pp. 1989--2016, 2015.

\bibitem{roman2023scalingSCLDPCLimIter}
R.~Sokolovskii, A.~Graell~i Amat, and F.~Brännström, ``Finite-length scaling of {SC-LDPC} codes with a limited number of decoding iterations,'' \emph{IEEE Trans. Inf. Theory}, vol.~69, no.~8, pp. 4869--4888, Aug. 2023.

\bibitem{bar2021deep}
D.~Bar-Lev, I.~Orr, O.~Sabary, T.~Etzion, and E.~Yaakobi, ``Deep {DNA} storage: Scalable and robust {DNA} storage via coding theory and deep learning,'' Sep. 2021, arXiv preprint arXiv:2109.00031.

\end{thebibliography}

\appendices
\renewcommand\thefigure{\thesection.\arabic{figure}}
\setcounter{figure}{0}    
\renewcommand\thetable{\thesection.\arabic{table}}
\setcounter{table}{0}    
\section{Draw Distribution}
\label{sec:appendix-drawdist}

\begin{figure}[H]
    \centering
    \begin{tikzpicture}
\begin{axis}[
width = 0.995\linewidth,
xmin=0,   xmax=30,
ymin=0.0,	ymax=0.45,
xticklabel style = {/pgf/number format/fixed, /pgf/number format/precision=6},
grid = both,
grid style = {line width=.1pt},
legend cell align={left},
legend style={font=\footnotesize,at={(0.99,0.99)},anchor=north east,cells={align=left}},
xlabel = {$d$ \quad {\footnotesize(draws)}},
ylabel = {$\theta_{\beta,c}(d)$ },
cycle list name=color list
]

{\addplot+ [color=red] table [col sep=comma,x=d,y=qd-acc] {Figures/drawdist/data/drawdist_c-1.csv};}
{\addlegendentry{$c=1$};}

{\addplot+ [forget plot, color=red, dashed] table [col sep=comma,x=d,y=poi] {Figures/drawdist/data/drawdist_c-1.csv};}

{\addplot+ [color=cyan] table [col sep=comma,x=d,y=qd-acc] {Figures/drawdist/data/drawdist_c-3.csv};}
{\addlegendentry{$c=3$};}

{\addplot+ [forget plot,color=cyan, dashed] table [col sep=comma,x=d,y=poi] {Figures/drawdist/data/drawdist_c-3.csv};}

{\addplot+ [color=magenta] table [col sep=comma,x=d,y=qd-acc] {Figures/drawdist/data/drawdist_c-5.csv};}
{\addlegendentry{$c=5$};}

{\addplot+ [forget plot,color=magenta, dashed] table [col sep=comma,x=d,y=poi] {Figures/drawdist/data/drawdist_c-5.csv};}

{\addplot+ [color=teal] table [col sep=comma,x=d,y=qd-acc] {Figures/drawdist/data/drawdist_c-10.csv};}
{\addlegendentry{$c=10$};}

{\addplot+ [forget plot, color=teal, dashed] table [col sep=comma,x=d,y=poi] {Figures/drawdist/data/drawdist_c-10.csv};}

{\addplot+ [color=blue] table [col sep=comma,x=d,y=qd-acc] {Figures/drawdist/data/drawdist_c-20.csv};}
{\addlegendentry{$c=20$};}

{\addplot+ [forget plot,color=blue, dashed] table [col sep=comma,x=d,y=poi] {Figures/drawdist/data/drawdist_c-20.csv};}

\end{axis}%
\end{tikzpicture}%
    \caption{The fraction of reads $\theta_{\beta,c}(d)$ exactly drawn $d$ times for a fixed $\beta = 0.13546$ and various coverages $c$ obtained via subsampling from dataset \texttt{file3} using \accbc. The dashed lines show the Poisson distribution for the respective coverage, which is obtained for uniform drawing distributions $\bfP^\mathrm{d}=[\nicefrac{1}{\chM},\ldots, \nicefrac{1}{\chM}]$. The skew in comparison to the Poisson curve indicates a drawing bias due to synthesis, PCR, or sequencing effects. The drawing distributions do not differ for \accbc\ and \fastbc, hence \fastbc\ is not included.}
    \label{fig:drawdist}
\end{figure}
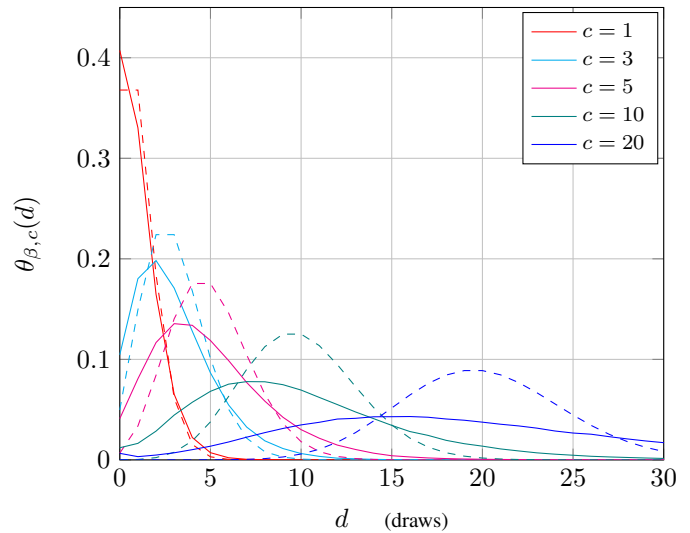

\section{Scaling Parameter $s(\trchM)$}
\label{sec:ScalingParameter}

\begin{table}[H]
\begin{center}
\caption{Scaling Parameter $s(\trchM)$ for Acc-BC With $\Rin = \nicefrac{216}{110}$ and Fast-BC With $\Rin = \nicefrac{200}{110}$ When Combining $\trchM$ Traces. Intermediate (Non-Depicted) Values of $s(\trchM)$ Are Obtained Via Interpolation}
\label{tab:scaling-logs}
\vspace{-0.75ex}
\begin{tabular}{c c c }
    \toprule
     $\trchM$ & $s(\trchM)$ for \accbc & $s(\trchM)$ for \fastbc \\[1ex] \hline \\[-0.75ex] 
     $1$  & $1.00$  & $1.01$ \\
     $2$  & $0.90$  & $0.88$ \\
     $4$  & $0.75$ & $0.69$ \\
     $6$  & $0.65$ & $0.57$ \\
     $8$  & $0.57$ & $0.49$ \\
     $10$ & $0.52$ & $0.43$ \\
     $15$ & $0.42$ & $0.33$ \\
     $20$ & $0.35$ & $0.27$ \\
     $30$ & $0.26$ & $0.20$ \\
     $50$ & $0.18$ & $0.13$ \\     
    \bottomrule
\end{tabular}
\end{center}
\end{table}

\section{Non-I.i.d. Decoder Improvement for the Uncoded Case}
\label{sec:appendix-uncoded-tracerecon}

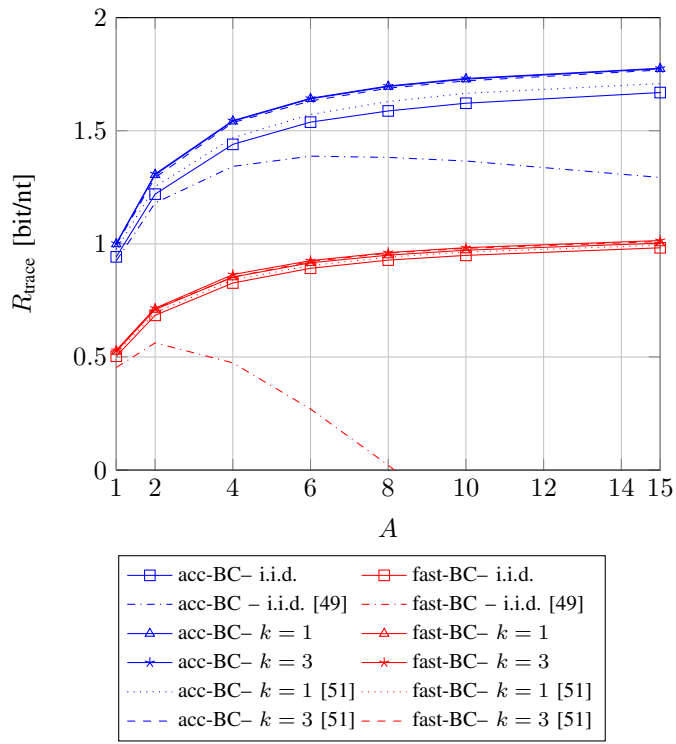
\begin{figure}[H]
    \centering
    \begin{tikzpicture}
\begin{axis}[
width = 0.995\linewidth,
xmin=1,   xmax=15,
ymin=0.0,	ymax=2,
xticklabel style = {/pgf/number format/fixed, /pgf/number format/precision=6},
extra x ticks= {1, 15},
grid = both,
grid style = {line width=.1pt},
legend cell align={left},
legend columns=2,
legend style={font=\footnotesize,at={(0.9,-0.6)},anchor=south east},
xlabel = {$\trchM$},
ylabel = {$\Rtrace$ [bit/nt]},
cycle list name=color list
]

{\addplot+ [color=blue, mark=square, solid, mark options={solid}] table [col sep=comma,x=M,y=I] {Figures/TraceRecon/csv/uncoded/k=0/acc-true_k-0_MR-110_IX=none.csv};}
{\addlegendentry{\accbc -- i.i.d.};}

{\addplot+ [color=red, mark=square, solid, mark options={solid}] table [col sep=comma,x=M,y=I] {Figures/TraceRecon/csv/uncoded/k=0/acc-false_k-0_MR-110_IX=none.csv};}
{\addlegendentry{\fastbc -- i.i.d.};}

{\addplot+ [color=blue, mark=none, dashdotted, mark options={solid}] table [col sep=comma,x=M,y=I] {Figures/TraceRecon/csv/uncoded/sone/acc_MR-110_IX=none.csv};}
{\addlegendentry{\accbc\ -- i.i.d. \cite{Maarouf2022ConcatenatedCF}};}

{\addplot+ [color=red, mark=none, dashdotted, mark options={solid}] table [col sep=comma,x=M,y=I] {Figures/TraceRecon/csv/uncoded/sone/fast_MR-110_IX=none.csv};}
{\addlegendentry{\fastbc\ -- i.i.d. \cite{Maarouf2022ConcatenatedCF}};}

{\addplot+ [color=blue, mark=triangle, solid, mark options={solid}] table [col sep=comma,x=M,y=I] {Figures/TraceRecon/csv/uncoded/k=1/acc-true_k-1_MR-110_IX=none.csv};}
{\addlegendentry{\accbc -- $k=1$};}

{\addplot+ [color=red, mark=triangle, solid, mark options={solid}] table [col sep=comma,x=M,y=I] {Figures/TraceRecon/csv/uncoded/k=1/acc-false_k-1_MR-110_IX=none.csv};}
{\addlegendentry{\fastbc -- $k=1$};}

{\addplot+ [color=blue, mark=star, solid, mark options={solid}] table [col sep=comma,x=M,y=I] {Figures/TraceRecon/csv/uncoded/k=3/acc-true_k-3_MR-110_IX=none.csv};}
{\addlegendentry{\accbc -- $k=3$};}

{\addplot+ [color=red, mark=star, solid, mark options={solid}] table [col sep=comma,x=M,y=I] {Figures/TraceRecon/csv/uncoded/k=3/acc-false_k-3_MR-110_IX=none.csv};}
{\addlegendentry{\fastbc -- $k=3$};}

{\addplot+ [color=blue, mark=none, dotted, mark options={solid}] table [col sep=comma,x=M,y=I] {Figures/TraceRecon/csv/comparison/uncoded/acc-true_k-1_uncoded_compare.csv};}
{\addlegendentry{\accbc -- $k=1$ \cite{frenchKmer_belaid_2023}};}

{\addplot+ [color=red, mark=none, dotted, mark options={solid}] table [col sep=comma,x=M,y=I] {Figures/TraceRecon/csv/comparison/uncoded/acc-false_k-1_uncoded_compare.csv};}
{\addlegendentry{\fastbc -- $k=1$ \cite{frenchKmer_belaid_2023}};}

{\addplot+ [color=blue, mark=none, dashed, mark options={solid}] table [col sep=comma,x=M,y=I] {Figures/TraceRecon/csv/comparison/uncoded/acc-true_k-3_uncoded_compare.csv};}
{\addlegendentry{\accbc -- $k=3$ \cite{frenchKmer_belaid_2023}};}

{\addplot+ [color=red, mark=none, dashed, mark options={solid}] table [col sep=comma,x=M,y=I] {Figures/TraceRecon/csv/comparison/uncoded/acc-false_k-3_uncoded_compare.csv};}
{\addlegendentry{\fastbc -- $k=3$ \cite{frenchKmer_belaid_2023}};}

\end{axis}%
\end{tikzpicture}%
    \caption{BCJR-once rates $\Rtrace$ versus number of traces $A$ with for the extended
trace reconstruction channel with uncoded strands obtained on data of \texttt{file3}.}
    \label{fig:trace-recon-uncoded}
\end{figure}

\newpage

\section{MIR Results: Fast Basecalling}
\label{MIRFBase}

\begin{figure}[H]
    \centering
    \begin{tikzpicture}
\begin{axis}[
width = 0.995\linewidth,
xmin=1,   xmax=8,
ymin=0.0,	ymax=2,
xticklabel style = {/pgf/number format/fixed, /pgf/number format/precision=6},
extra x ticks= {1},
grid = both,
grid style = {line width=.1pt},
legend cell align={left},
legend style={font=\footnotesize,at={(0.01,0.99)},anchor=north west},
legend image post style={dash phase=0pt},
xlabel = {$c =\nicefrac{\chN}{\chM}$ \quad {\footnotesize(coverage)}},
ylabel = {$\avMIR$\quad {\footnotesize [bit/nt]}},
cycle list name=color list
]

{\addplot+ [color=red, mark=square, solid, mark options={solid}] table [col sep=comma,x=c,y=I] {Figures/MIR-index/acc-false/csv/Acc-false_MR-92_IX=one_IXDEC=ML.csv};}
{\addlegendentry{$R_\mathrm{ix}=1$, UC };}

\addplot[forget plot, mark=none, color=red, loosely dashdotted, opacity=1.0] coordinates {(1,1.81818181818182) (40,1.81818181818182)};

{\addplot+ [color=magenta, mark=square, solid, mark options={solid}] table [col sep=comma,x=c,y=I] {Figures/MIR-index/acc-false/csv/Acc-false_MR-84_IX=rep_IXDEC=ML.csv};}
{\addlegendentry{$R_\mathrm{ix}=1/2$, REP};}

\addplot[forget plot, mark=none, color=magenta, loosely dashdotted, opacity=1.0] coordinates {(1,1.67272727272727) (40,1.67272727272727)};

{\addplot+ [color=teal, mark=square, solid, mark options={solid}] table [col sep=comma,x=c,y=I] {Figures/MIR-index/acc-false/csv/Acc-false_MR-84_IX=ER_IXDEC=ML.csv};}
{\addlegendentry{$R_\mathrm{ix}=0.476$, ES };}

\addplot[forget plot, mark=none, color=teal, loosely dashdotted, opacity=1.0, dash phase=2pt] coordinates {(1,1.67272727272727) (40,1.67272727272727)};

{\addplot+ [color=blue, mark=square, solid, mark options={solid}] table [col sep=comma,x=c,y=I] {Figures/MIR-index/acc-false/csv/Acc-false_MR-88_IX=VT_IXDEC=ML.csv};}
{\addlegendentry{$R_\mathrm{ix}=2/3$, VT };}

\addplot[forget plot, mark=none, color=blue, loosely dashdotted, opacity=1.0] coordinates {(1,1.74545454545455) (40,1.74545454545455)};

{\addplot+ [color=blue, mark=triangle, solid, mark options={solid}] table [col sep=comma,x=c,y=I] {Figures/MIR-index/acc-false/csv/Acc-false_MR-88_IX=VT_IXDEC=ML_ixTh=0985.csv};}
{\addlegendentry{$R_\mathrm{ix}=2/3$, VT -- opt. $\threshIx=0.985$};}

{\addplot+ [color=teal, mark=triangle, solid, mark options={solid}] table [col sep=comma,x=c,y=I] {Figures/MIR-index/acc-false/csv/Acc-false_MR-84_IX=ER_IXDEC=ML_ixTh=0975.csv};}
{\addlegendentry{$R_\mathrm{ix}=0.476$, ES -- opt. $\threshIx=0.975$};}

\end{axis}%
\end{tikzpicture}%
    \caption{MIRs $\avMIR$ versus coverage $c$ for the \multidraw\ channel on experimental data of \texttt{file1} ($\chM=30\,589$, $\chL=110$) for \fastbc\ using $k=1$ in the single-strand decoder. We compare different index codes plus the index thresholding strategy. The horizontal dash-dotted lines represent the rate limit imposed by the combined respective index code rates and inner code rates (line colors match the index codes indicated in the legend). We use ten redundancy symbols for the data part coding.}
    \label{fig:mirs-indexcomp-fast}
    \vspace{-2.3ex}
\end{figure}

\begin{figure}[H]
    \centering
    \begin{tikzpicture}
\begin{axis}[
width = 0.995\linewidth,
xmin=1,   xmax=8,
ymin=0.0,	ymax=2,
xticklabel style = {/pgf/number format/fixed, /pgf/number format/precision=6},
extra x ticks= {1},
grid = both,
grid style = {line width=.1pt},
legend cell align={left},
legend style={font=\footnotesize,at={(-0.10,-0.19)}, legend columns=2,  anchor=north west,cells={align=left}},
xlabel = {$c =\nicefrac{\chN}{\chM}$ \quad {\footnotesize(coverage)}},
ylabel = {$\avMIR$\quad {\footnotesize [bit/nt]}},
cycle list name=color list
]

{\addplot+ [color=red, mark=square, solid, mark options={solid}] table [col sep=comma,x=c,y=I] {Figures/MIRs/non-acc/Acc-false_MR-92_IX=one_IXDEC=ML.csv};}
{\addlegendentry{UC, ix. dec.};}

{\addplot+ [color=red, mark=star, dashed, mark options={solid}] table [col sep=comma,x=c,y=I] {Figures/MIRs/non-acc/Acc-false_MR-92_IX=one_IXDEC=G.csv};}
{\addlegendentry{UC, genie ix. dec.};}

{\addplot+ [color=blue, mark=square, solid, mark options={solid}] table [col sep=comma,x=c,y=I] {Figures/MIRs/non-acc/Acc-false_MR-84_IX=ER_IXDEC=ML.csv};}
{\addlegendentry{ES, ix. dec.};}

{\addplot+ [color=blue, mark=star, dashed, mark options={solid}] table [col sep=comma,x=c,y=I] {Figures/MIRs/non-acc/Acc-false_MR-84_IX=ER_IXDEC=ML_clus=genie.csv};}
{\addlegendentry{ES, ix. dec., genie cluster};}

{\addplot+ [color=blue, mark=triangle, solid, mark options={solid}] table [col sep=comma,x=c,y=I] {Figures/MIRs/non-acc/Acc-false_MR-84_IX=ER_IXDEC=ML_ixTh=0975.csv};}
{\addlegendentry{ES, ix. dec. with $\threshIx$};}

{\addplot+ [color=blue, mark=diamond, solid, mark options={solid}] table [col sep=comma,x=c,y=I] {Figures/MIRs/non-acc/Acc-false_MR-84_IX=ER_IXDEC=ML__ixTh=0975_RomClus=5.csv};}
{\addlegendentry{ES, ix. thresh. dec., post-clust.};}

{\addplot+ [color=black, mark=star, dashed, mark options={solid}] table [col sep=comma,x=c,y=I] 
{
c,I
1,1.1026
2,1.4989
3,1.6668
4,1.7466
5,1.7881
6,1.8108
7,1.8245
8,1.8328
9,1.8384
10,1.8421
};}
{\addlegendentry{UC, noiseless};}

{\addplot+ [color=black, mark=star, dotted, mark options={solid}] table [col sep=comma,x=c,y=I] 
{
c,I
1,1.1786
2,1.6122
3,1.7717
4,1.8304
5,1.8520
6,1.8599
7,1.8628
8,1.8639
9,1.8643
10,1.8645
};}
{\addlegendentry{UC, noiseless, no. bias};}

\end{axis}%
\end{tikzpicture}%
    \caption{MIRs $\avMIR$ versus coverage $c$ for the \multidraw\ channel on experimental data of \texttt{file1} ($\chM=30\,589$, $\chL=110$) for \fastbc\ using $k=1$ in the single-strand decoder. We compare different decoding strategies and provide genie-based benchmarks for different scenarios. For the index reliability threshold, we use $\threshIx=0.975$, and for post-clustering, we use $\threshClus=42.5$ for the distance threshold and fix $|\poprTop|=5$.}
    \label{fig:mirs-fast}
    \vspace{-2.3ex}
\end{figure}

\end{document}